\documentclass[twocolumn]{aastex62}

\usepackage{natbib}
\usepackage{url}
\usepackage{academicons}
\usepackage{xcolor}
\usepackage{amsmath}
\usepackage{hyperref}
\usepackage{comment}
\usepackage{fixltx2e}
\usepackage{floatrow}
\usepackage{lineno}

\newcommand{\angstrom}{\mbox{\normalfont\AA}}

\newfloatcommand{capbtabbox}{table}[][\FBwidth]

\ifxetex
  \usepackage{letltxmacro}
  \LetLtxMacro\SavedIncludeGraphics\includegraphics
  \def\includegraphics#1#{
    \IncludeGraphicsAux{#1}%
  }%
  \newcommand*{\IncludeGraphicsAux}[2]{%
    \XeTeXLinkBox{%
      \SavedIncludeGraphics#1{#2}%
    }%
  }%
\fi

\definecolor{orcidlogocol}{HTML}{A6CE39}

\shorttitle{Altair in Fizeau Mode}
\shortauthors{Spalding et al.}
\begin{document}

\title{High contrast imaging with Fizeau interferometry: The case of Altair\footnote{The LBT is an international collaboration among institutions in the United States, Italy and Germany. LBT Corporation partners are: The University of Arizona on behalf of the Arizona university system; Istituto Nazionale di Astrofisica, Italy; LBT Beteiligungsgesellschaft, Germany, representing the Max-Planck Society, the Astrophysical Institute Potsdam, and Heidelberg University; The Ohio State University, and The Research Corporation, on behalf of The University of Notre Dame, University of Minnesota and University of Virginia.}}

\author[0000-0003-3819-0076]{E. Spalding \href{https://orcid.org/0000-0003-3819-0076}{\includegraphics[scale=0.1]{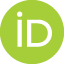}}}
\email{espaldin@nd.edu}
\affil{Steward Observatory, University of Arizona, Tucson, AZ 85719, USA}
\affil{Center for Astronomical Adaptive Optics, University of Arizona, Tucson, AZ 85719, USA}
\affil{Department of Physics, University of Notre Dame, Notre Dame, IN 46556, USA}

\author[0000-0002-1384-0063]{K.M. Morzinski \href{https://orcid.org/0000-0002-1384-0063}{\includegraphics[scale=0.1]{orcid_64x64.png}}}
\affil{Steward Observatory, University of Arizona, Tucson, AZ 85719, USA}
\affil{Center for Astronomical Adaptive Optics, University of Arizona, Tucson, AZ 85719, USA}

\author[0000-0002-1954-4564]{P. Hinz \href{https://orcid.org/0000-0002-1954-4564}{\includegraphics[scale=0.1]{orcid_64x64.png}}}
\affil{Department of Astronomy and Astrophysics, University of California Santa Cruz, Santa Cruz, CA 95064, USA}
\affil{Laboratory for Adaptive Optics, Center for Adaptive Optics, Santa Cruz, CA 95064, USA}

\author[0000-0002-2346-3441]{J. Males \href{https://orcid.org/0000-0002-2346-3441}{\includegraphics[scale=0.1]{orcid_64x64.png}}}
\affil{Steward Observatory, University of Arizona, Tucson, AZ 85719, USA}
\affil{Center for Astronomical Adaptive Optics, University of Arizona, Tucson, AZ 85719, USA}

\author[0000-0003-1227-3084]{M. Meyer \href{https://orcid.org/0000-0003-1227-3084}{\includegraphics[scale=0.1]{orcid_64x64.png}}}
\affil{Department of Astronomy, University of Michigan, Ann Arbor, MI 48109, USA}

\author[0000-0003-3829-7412]{S.P. Quanz \href{https://orcid.org/0000-0003-3829-7412}{\includegraphics[scale=0.1]{orcid_64x64.png}}}
\affil{Institute for Particle Physics and Astrophysics, ETH Z\"{u}rich, Z\"{u}rich, Switzerland}

\author[0000-0002-0834-6140]{J. Leisenring \href{https://orcid.org/0000-0002-0834-6140}{\includegraphics[scale=0.1]{orcid_64x64.png}}}
\affil{Steward Observatory, University of Arizona, Tucson, AZ 85719, USA}

\author[0000-0000-0000-0000]{J. Power}
\affil{Large Binocular Telescope Observatory, Tucson, AZ 85719, USA}

\begin{abstract}
The Large Binocular Telescope (LBT) has two 8.4-m primary mirrors that produce beams that can be combined coherently in a ``Fizeau'' interferometric mode. In principle, the Fizeau PSF enables the probing of structure at a resolution up to three times better than that of the adaptive-optics-corrected PSF of a single 8.4-m telescope. In this work, we examined the nearby star Altair ($5.13$ pc, type A7V, $\sim$100s Myr to $\approx$1.4 Gyr) in the Fizeau mode with the LBT at Br-$\alpha$ (4.05 $\mu$m) and carried out angular differential imaging to search for companions. This work presents the first filled-aperture LBT Fizeau science dataset to benefit from a correcting mirror which provides active phase control. In the analysis of the $\lambda/D$ angular regime, the sensitivity of the dataset is down to $\approx$0.5 $M_{\odot}$ at 1'' for a 1.0 Gyr system. This sensitivity remains limited by the small amount of integration time, which is in turn limited by the instability of the Fizeau PSF. However, in the Fizeau fringe regime we attain sensitivities of $\Delta m \approx 5$ at 0.2'' and put constraints to companions of 1.3 $M_{\odot}$ down to an inner angle of $\approx$0.15'', closer than any previously published direct imaging of Altair. This analysis is a pathfinder for future datasets of this type, and represents some of the first steps to unlocking the potential of the first ELT. Fizeau observations will be able to reach dimmer targets with upgrades to the instrument, in particular the phase detector.
\end{abstract}


\section{Introduction}
\label{sec:intro}

The technique of direct imaging offers the possibility of studying the emitted spectra of massive exoplanets which lie at large angular separations from their host stars (e.g., \citet{thalmann2011piercing,quanz2012searching,vigan2015high,rajan2017characterizing,greenbaum2018gpi,boehle2019combining,nielsen2019gemini,hunziker2020refplanets,vigan2021sphere}). To overcome the severe planet-to-star contrast, the first generation of direct imaging surveys preferentially targeted younger stars with ages ranging from a few Myr to $\approx$500 Myr, because massive planets in such systems would still be glowing with the heat of formation---a glow that  drops by orders of magnitude as the system ages from $\sim$1 Myr to $\sim$1 Gyr \citep{mordasini2012characterization,marleau2013constraining,bowler2016imaging}.

However, circumstances could also favor the imaging of planets around older stars, if they are particularly close to Earth. Within an arbitrarily-chosen 10 pc radius from Earth, there are 68 A- through K-type stars \citep{reyle202110}. One of the four A-type stars in that sample, Altair, is a $5.130\pm0.015$ pc distant $\delta$ Scuti variable \citep{buzasi2005altair,van2007validation}. At $m_{V}=0.8$ mag on the Vega scale, it is the 12th-brightest star to the human eye in the night sky \citep{hoffleit1995vizier}. The sheer brightness of Altair quickly saturates detectors and makes it a challenging candidate for transit or precise astrometric studies, and it will be technically challenging for JWST even with a defocused PSF and shorter subarray readout times \citep{beichman2014observations,talens2017multi}.

The tightest astrometric constraints are from HIPPARCOS observations over a baseline of 2.7 years, and are consistent with a single star. The parallax precision of $\approx$0.6 mas \citep{1997perrymannhipparcos,van2007validation} is equivalent to the astrometric semi-amplitude of a 10M$_{J}$ planet at $\approx$0.6 AU. Ground-based relative astrometry by \citet{gatewood1995map} also rule out $>$10M$_{J}$ companions with periods of 1.2 to 5 years, based on five years of relative astrometry observations of Altair, with a precision of $\approx$0.9 mas.

Altair is a challenging target for precise radial velocity studies, because it has few absorption lines due to its early type, and its lines are broadened due to fast rotation \citep{howard2016limits} and Stark broadening. The systemic velocity of Altair is constrainted with high-resolution spectroscopy to within an uncertainty of 3.5 km/sec \citep{prieto2004sn}. With repeated sampling, this would be enough to detect the 18 km/sec semi-amplitude of a ``dark'' solar-mass companion in an edge-on orbit at 1 AU, but the inclination of any companions to Altair remains unknown.

Direct imaging can help fill a niche here, even though Altair may be relatively old. It has isochrone-based age estimates of 0.7 Gyr \citep{vican2012age} to 1.2-1.4 Gyr \citep{kervella2005gravitational}, though some evidence from hydrogen mass fractions suggests that Altair has just left the zero-age main sequence, and may be as young as $\sim$100 Myr \citep{peterson2006resolving,bouchaud2020realistic}. In any event, Altair is close enough that planets on orbits of 1 AU would go out to a maximum of 0.2'', or 3.4$\lambda/D$ in $K$-band and 2.0$\lambda/D$ in $L'$-band of an 8 m adaptive optics (AO)-equipped telescope. Since Altair is an early-type star, planets would also re-emit thermally at wider orbits than around cooler stars. In addition, in thermal equilibrium the level of re-emission is less affected by the system's age. It can be expected that thermal re-emission alone will cause a system with an ``old'' Jovian planet on a 2 or 3 AU orbit around an A-type star to have a planet-to-star contrast of $\sim10^{-8}$ or $10^{-9}$ in $L'$-band.

For these reasons, Altair warrants close investigation in its own right with direct imaging in the thermal infrared, as a compliment to imaging surveys which put constraints on gas giant planets around young stars on wide orbits of $\sim$10s AU (e.g., \citet{males2014direct,quanz2015confirmation,meyer2018finding,wang2019new}). Currently, however, no companions of any mass are known to orbit Altair, even though the star has been included in a number of direct imaging surveys \citep{kuchner2000search,schroeder2000search,oppenheimer2001coronagraphic,leconte2010lyot,roberts2011astrometric,janson2011high,dieterich2012solar,stone2018leech}. The current tightest constraints on companions around Altair from direct imaging have essentially ruled out $\gtrsim$10 $M_{J}$ planets for orbits with projected radii of $\gtrsim$10 AU, and $\gtrsim$5 $M_{J}$ planets at $\gtrsim$20 AU, for a stellar age set to 500 Myr \citep{janson2011high}. 

In the meantime, other aspects of Altair have been characterized and have set the stage for detailed study of whatever exoplanet system it may host. Near-infared and optical long-baseline interferometry (OLBI) have revealed the star's angular width, high inclination, asymmetric surface brightness, an oblate spheroid shape due to the star's spin, and variable $K$-band-emissive exozodiacal dust in tight orbits of $\sim$0.1 AU, though no emission has been detected from a cold debris disk or $N$-band-emissive habitable-zone dust \citep{kuchner199811,ohishi2004asymmetric,reiners2004altair,kervella2005gravitational,suarez2005modelling,peterson2006resolving,monnier2007imaging,richichi2009list,lara2011gravity,millan2011exozodiacal,van2012interferometric,absil2013near,gaspar2013collisional,hadjara2014beyond,mennesson2014constraining,thureau2014unbiased,van2014near,baines2017fundamental,kirchschlager2017constraints,nunez2017near,ertel2018hosts,bouchaud2020realistic,ertel2020hosts}. \citet{van2001altair} also note that visibility data from the Palomar Testbed Interferometer (PTI) consistent with an elliptical shape for Altair is probably not an artifact of a tight binary companion, as Altair's proper motion and the sensitivity of the PTI do not bring apparent binary companions within range that could masquerade as an elliptical star.

Further constraints on the presence of companions to Altair will inform efficient survey design of space-based missions, for which observing time will be at a particular premium. Indeed, Altair is on the target lists of the Exo-C, Exo-S, and Nancy Grace Roman Space Telescope\footnote{formerly WFIRST-AFTA} study teams \citep{howard2016limits}, and is among the best candidates for detecting and characterizing planets with a starshade coupled to Roman \citep{romero2021starshade}. Simulations suggest that Altair is also one of the best candidates for finding Jovian companions from the ground in the thermal infrared, with the METIS instrument of the future 39-m European-ELT \citep{bowens2021exoplanets}. 

\begin{centering}
\begin{figure}
\includegraphics[clip, trim=20cm 13cm 20cm 12cm, width=0.95\linewidth]{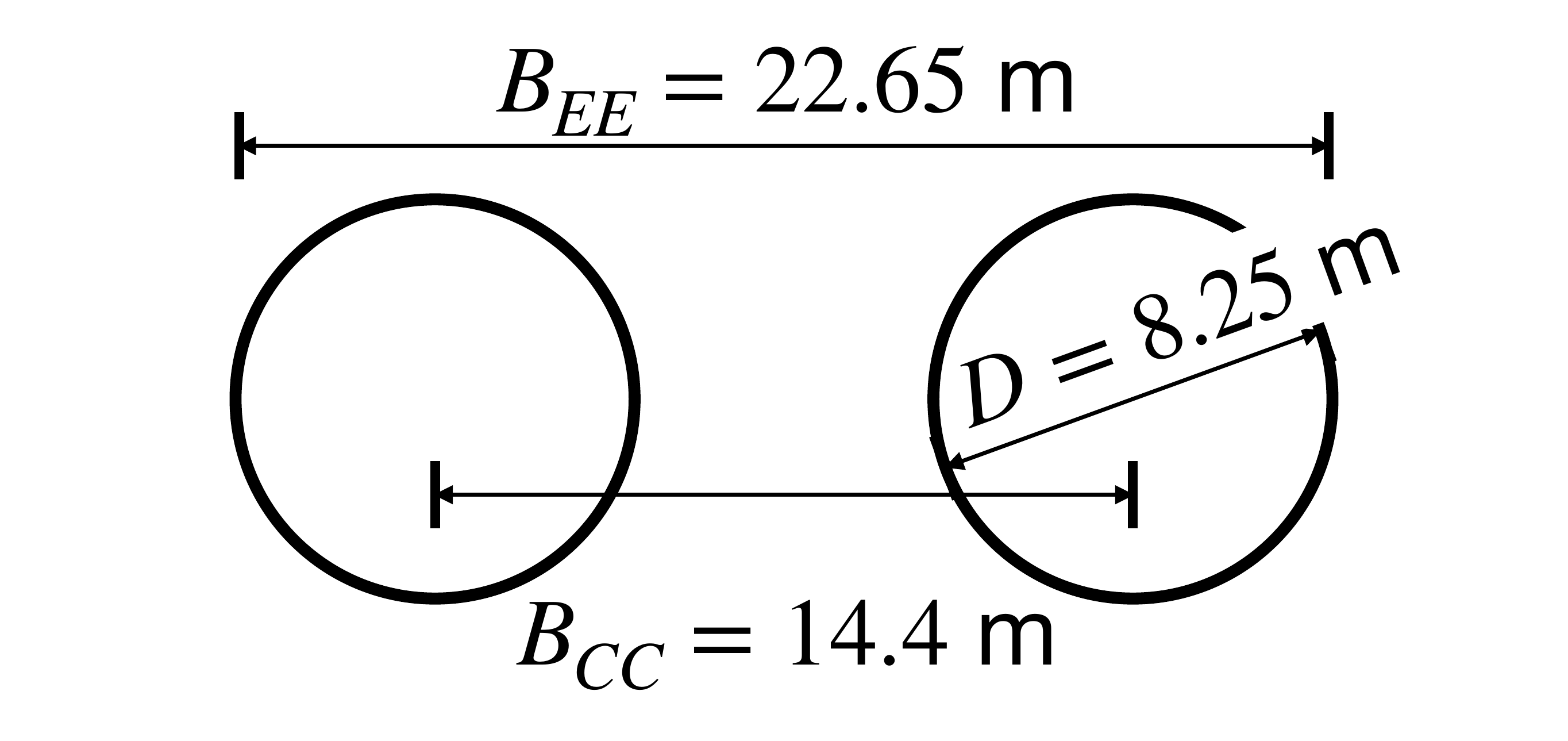}
\caption{Dimensions of the LBT aperture. ($EE:$ edge-to-edge; $CC:$ center-to-center.)} 
\label{fig:lbt_dims}
\end{figure}
\end{centering}

The Large Binocular Telescope (LBT), on Mt. Graham in southern Arizona, is in a unique position for obtaining high-angular-resolution, thermal infrared imaging akin to that of ELTs. The LBT has twin 8.4 m telescopes equipped with adaptive optics (AO) (e.g., \citet{hill2010large,rothberg2018current}), with a 14.4 m center-to-center separation that provides a maximum 22.65 m baseline (Fig. \ref{fig:lbt_dims}). The beams from both telescopes can be coherently combined, in a ``Fizeau'' mode which preserves the coherence envelope across the field of view. 

In this work, we carried out high-contrast imaging of the star Altair in the LBT Fizeau mode. Our work represents the first filled-aperture LBT Fizeau dataset with some degree of active phase control with a correcting mirror to counteract time-dependent differential piston between the two telescopes, and freeze the Fizeau fringe pattern. We developed a new code base for reducing these unique data, using a PCA-based decomposition to tolerate greater PSF phase diversity. In this analysis we increase the typical proportion of used frames from 2-20\% from the handful of past LBT Fizeau observations for static image deconvolution \citep{leisenring2014fizeau,conrad2015spatially,conrad2016role} to 57\%, and we place constraints on companions to Altair closer in to the star than in any previously-published direct imaging dataset. (Works by \cite{de2017multi} and \cite{de2021resolving} selected frames based on rapid-cadence changes in the observed object.) 

In Sec. \ref{subsec:observations} we describe the observations, and in Sec. \ref{sec:analysis} we describe the unique systematics of the PSF and the reduction process. In Sec. \ref{sec:results} we discuss the results in the angular regimes of both classical direct imaging and Fizeau imaging. We discuss the results in Sec. \ref{sec:discussion}, describe strategies for future improvement in Sec. \ref{sec:future_improvements}, and conclude in Sec. \ref{sec:consclusions}.

\section{Observations and Methods}

\subsection{LBT in Fizeau mode}

\begin{centering}
\begin{figure}
\includegraphics[width=\linewidth,trim=25cm 8cm 23cm 0cm, clip=True]{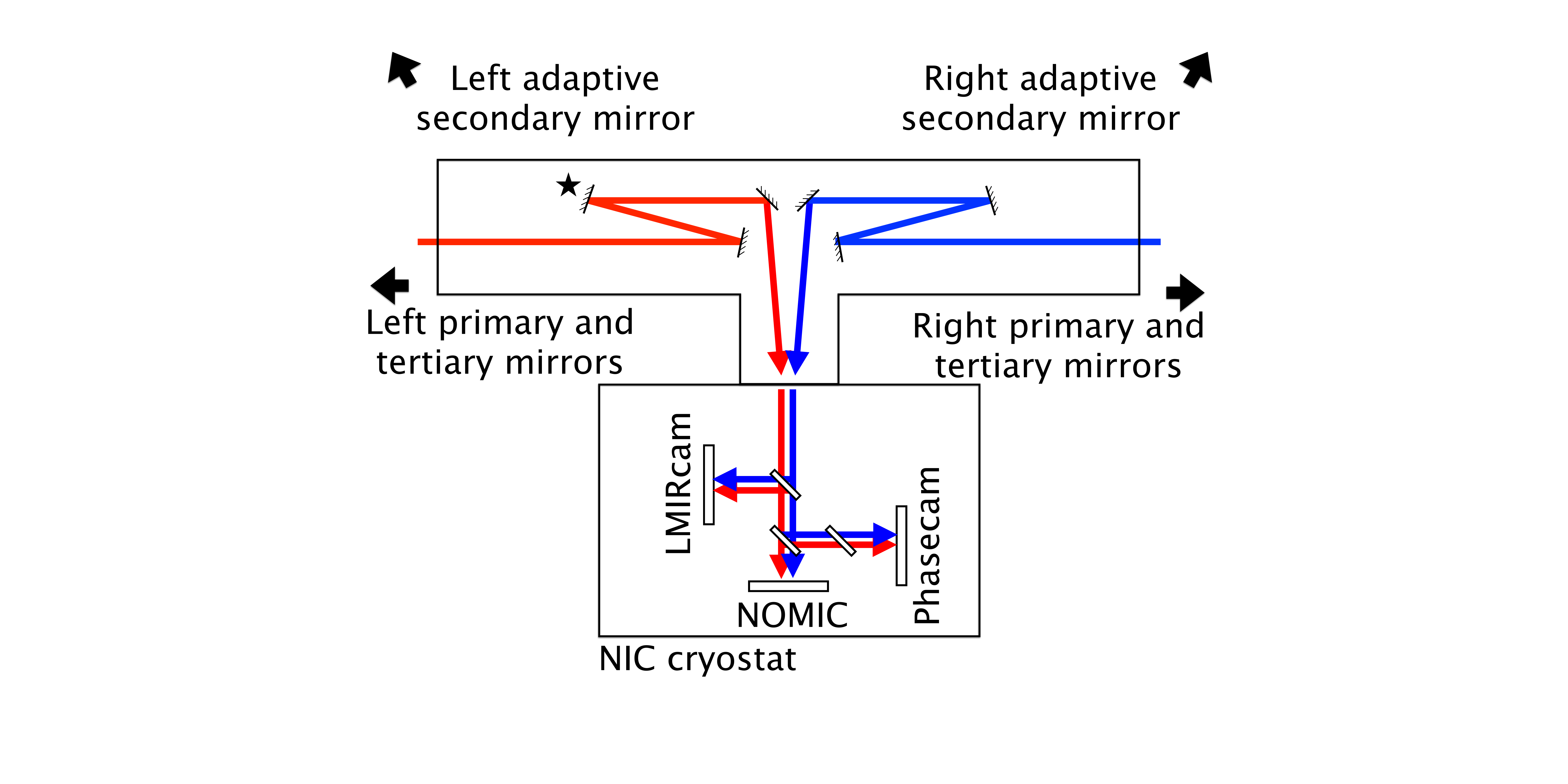}
\caption{A schematic layout of the LBTI instrument, which receives the beams from both sub-telescopes, combines them, and directs them through a trichroic and beamsplitter elements onto the science and phase detectors. The star indicates the fast pathlength corrector mirror, which is used to correct the phase.} 
\label{fig:overall_layout}
\end{figure}
\end{centering}

The interferometer LBTI receives the telescope beams from the LBT tertiary mirrors, feeds $<$1 $\mu$m light to AO wavefront sensors, and sends $>$1 $\mu$m light into the Nulling and Imaging Camera (NIC) cryostat \citep{hinz2008nic,bailey2014large,hinz2016overview}. There, each beam branches into a shorter-wavelength beam to the phase camera (Phasecam; \citet{defrere2014co}), and a longer-wavelength beam to the 1.2-5 $\mu$m LMIRCam \citep{skrutskie2010large,leisenring2012sky} and/or the 8-12 $\mu$m Nulling Optimized Mid-Infrared Camera (NOMIC; \citet{hoffmann2014operation}). (See Fig. \ref{fig:overall_layout} for a schematic.)

To combine the twin beams coherently in a  Fizeau configuration, the entrance pupil (the twin primary mirrors) and the exit pupil (a stop internal to the instrument) share the same spatial proportions, which effectively makes the twin telescopes act as one large telescope aperture with a mask that passes two circular subapertures. This configuration allows the center of the coherence envelope to exist across the field of view. The illumination on the detector is still a simple convolution of the object with the telescope PSF, but the PSF is now a multiplication of an Airy function with a corrugation pattern due to the separation of the telescopes.

\begin{centering}
\begin{figure}
\includegraphics[width=\linewidth,trim=1cm 1cm 1cm 0cm, clip=True]{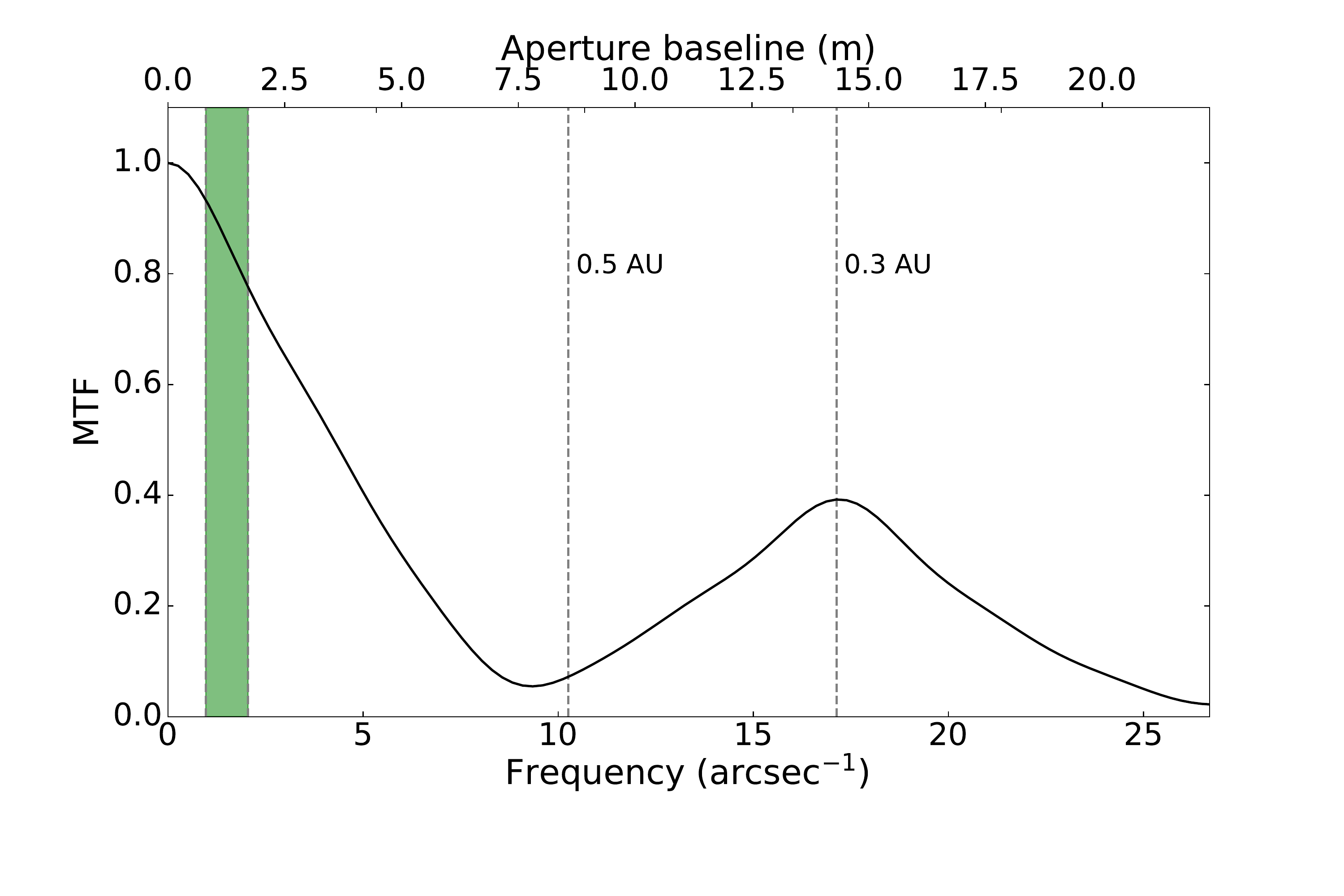}
\caption{A cross-section through the model LBT modulation transfer function (MTF) for the Br-$\alpha$ filter, with perfect wavefront correction and phase control. The MTF of a single 8.25-m aperture would consist only of the low-frequency peak, and go to zero power at approximately the equivalent of 0.5 AU around the Altair system. The secondary maximum of the LBT MTF is at a baseline of $B_{CC}$, equivalent to 0.3 AU. Green indicates the frequencies equivalent to radii of a face-on habitable zone as defined by \citet{cantrell2013solar}.} 
\label{fig:mtfs_altair}
\end{figure}
\end{centering}

The Fizeau PSF samples Fourier space at frequencies higher than from a single filled aperture. Fig. \ref{fig:mtfs_altair} shows the theoretical resolving power of the LBT based on the modulation transfer function (MTF), or the amplitude of the Fourier transform of the LBT PSF. The Fizeau PSF is also predicted to have contrast gains when searching for low-mass companions in the additional dark regions of the PSF, if detector integration times are shorter than the atmospheric coherence timescale $\tau_{0}$ \citep{patru2017_i}. 

\subsection{Observations}
\label{subsec:observations}

\begin{deluxetable}{l | l}
\tabletypesize{\scriptsize}
\tablecaption{Observing log, UT 2018 May 7}
\tablewidth{0pt}
\tablehead{
\colhead{Parameter} & \colhead{Value}
}
\startdata
Epoch (HH:MM UT) & 10:15 to 12:16    \\
Seeing (line-of-sight) &  mostly 0.9'' to 1.3''\\
Precipitable water vapor$^{a}$ & mostly 8 to 13 mm H$_{2}$O\\
Parallactic angle $q$& $-43^{\circ} \leq q\leq+3^{\circ}$ \\
Airmass & 1.1 to 1.2 \\
Wind & 0.5 to 4.5 m/s \\
AO frequency$^{b}$ & 1 kHz \\
AO deformation modes & 300 \\
AO control radius$^{c}$ $r_{c}$ & $4.05\mu \textrm{m} / (2\cdot d ) \approx$1.4''
\enddata
\tablenotetext{a}{Measured at the Heinrich Hertz Submillimeter Telescope}
\vspace{-0.2cm}
\tablenotetext{b}{Listed AO parameters are for both telescopes}
\vspace{-0.2cm}
\tablenotetext{c}{$d$: subaperture spacing}
\label{table:conds}
\end{deluxetable}

Our observations were carried out with the LBT in Fizeau mode on UT 2018 May 7, with conditions as tabulated in Table \ref{table:conds}. We followed our standard procedure for setting up for Fizeau observations. After closing both AO loops, a grism was inserted into the science beam. The two grism illuminations (one from each telescope) were manually and incoherently overlapped on the LMIRcam science detector by moving tip-tilt mirrors internal to the instrument beam combiner \citep{hinz2004large,leisenring2012sky}. 

An internal mirror was manually moved in piston to seek the coherence envelope of the two overlapping grism PSFs. This coherent combination of the beams leads to a ``barber pole'' or ``candy cane'' fringing pattern along the wavelength-dependent length of the combined grism illumination. Vertical fringes indicated an optical path difference of zero. The grism was then removed, leaving a Fizeau-Airy PSF (that is, a Fizeau PSF which is not chromatically dispersed) on the science detector. $Ks$-band fringes from a separate channel were placed on the phase detector by moving a beam combiner optic, and then the phase loop was closed (e.g., \citet{spalding2019status}).

The closed phase loop acts to make another mirror upstream of the science and phase detectors to move in piston, tip, and tilt at a rate of 100s of Hz to 1 kHz. Movements along each of these three degrees of freedom are centered around tip-tilt and piston setpoints which can be set manually to remove residual aberrations from the PSF.

\begin{centering}
\begin{figure}
\includegraphics[trim= 0cm 0.5cm 0cm 0cm, width=1\linewidth, clip=True]{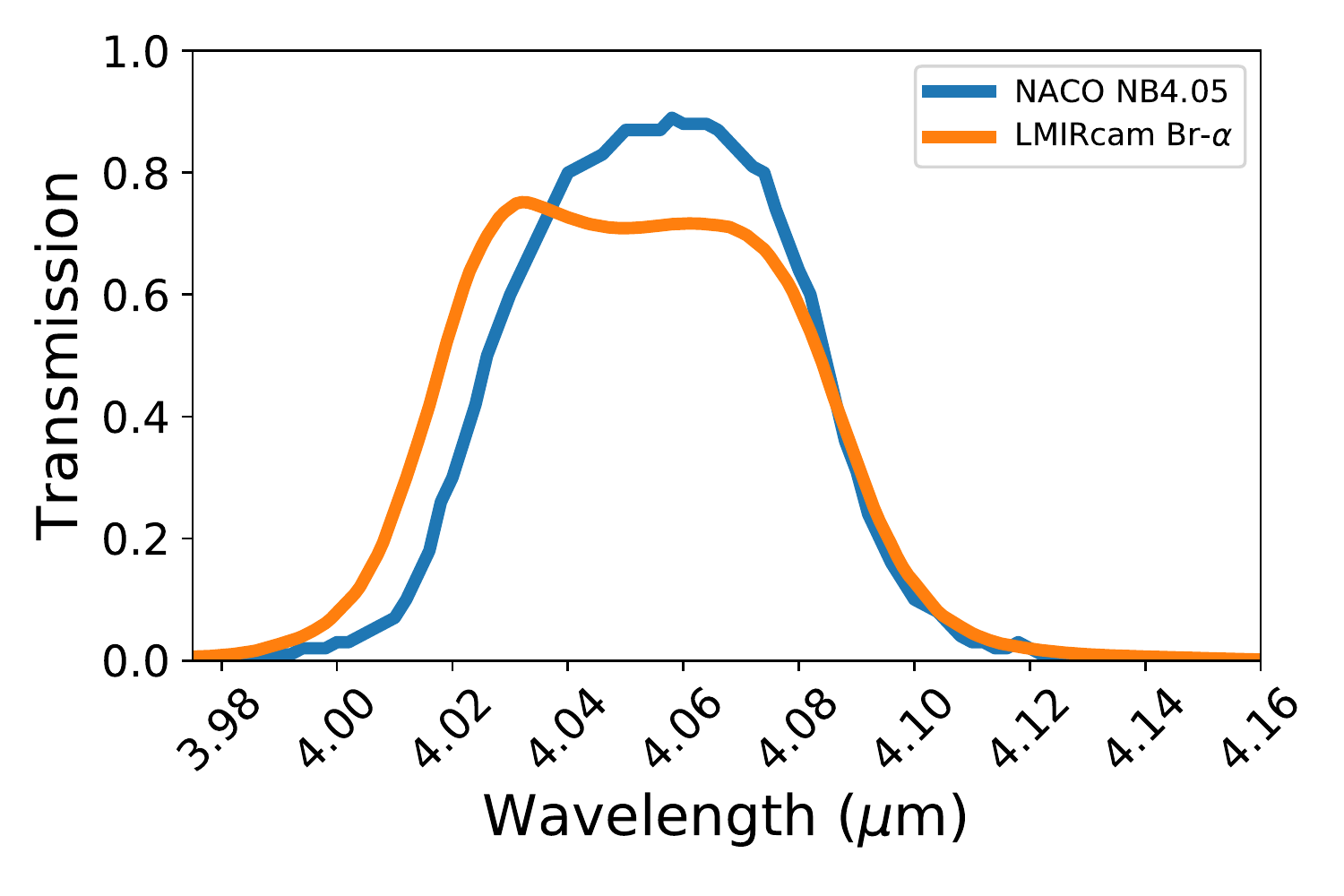}
\caption{A comparison of the NACO 4.05 $\mu$m filter with the cryogenic transmission of the LMIRCam Br-$\alpha$. Magnitude-to-mass conversions described in Sec. \ref{sec:results} are based on models as seen though the NACO filter.} 
\label{fig:filter_comparison}
\end{figure}
\end{centering}

Given the target brightness, we observed through a narrowband Brackett-$\alpha$ filter ($4.01\leq\lambda/\mu$m $\leq4.09$ at cryogenic temperature) and a 10\% neutral density (ND) transmission filter, so as to saturate only the central bright Fizeau fringe peaks. Our science filter is similar to that of the NACO NB4.05 filter \citep{rodrigo2012svo} (Fig. \ref{fig:filter_comparison}). Unsaturated PSF frames were taken by substituting the 10\% transmission filter with a 1\% transmission filter. LMIRcam detector readouts of 2048$\times$512 pixels, or a quarter of the array, were read out with an ISDEC controller backend \citep{burse2016isdec}. All frames had integration times of 0.146 sec, with one detector reset for each read.

\begin{deluxetable}{l | l | l}
\tabletypesize{\scriptsize}
\tablecaption{Blocks of frames by filter combination}
\tablewidth{0pt}
\tablehead{
\colhead{Frames} & \colhead{Filters (in series)} & \colhead{Parallactic angle $q$}
}
\startdata
\begin{tabular}{@{}l@{}}``A'' block \\ (saturated science frames) \end{tabular}& \begin{tabular}{@{}l@{}}ND 10\% transm. \\  Br-$\alpha$ (4.05 $\mu$m) \end{tabular}& -43.3$^{\circ}$ to -36.0$^{\circ}$\\ 
\hline
\begin{tabular}{@{}l@{}}``B'' block \\ (unsaturated) \end{tabular}& \begin{tabular}{@{}l@{}}ND 1\% transm. \\ Br-$\alpha$ (4.05 $\mu$m) \end{tabular}& -36.0$^{\circ}$ to -33.1$^{\circ}$ \\
\hline
\begin{tabular}{@{}l@{}}``C'' block \\ (unsaturated) \end{tabular}& \begin{tabular}{@{}l@{}}ND 1\% transm. \\ Wide 2.8-4.3 $\mu$m\\ Br-$\alpha$ (4.05 $\mu$m) \end{tabular}& -32.1$^{\circ}$ to -28.8$^{\circ}$\\
\hline
\begin{tabular}{@{}l@{}}``D'' block \\ (saturated science frames) \end{tabular}& \begin{tabular}{@{}l@{}}ND 10\% transm. \\Wide 2.8-4.3 $\mu$m \\ Br-$\alpha$ (4.05 $\mu$m) \end{tabular}& -22.0$^{\circ}$ to $+$3.3$^{\circ}$\\
\hline
\begin{tabular}{@{}l@{}}Total $\Delta q$, \\ saturated science frames \end{tabular}& - & 32.6$^{\circ}$
\enddata
\end{deluxetable}
\label{table:blocks}

Four different permutations of filters were combined in series upstream of the science detector. We denote the sets of detector readouts taken with each filter combination as ``blocks'' of frames, all four of which are listed in Table \ref{table:blocks}. Two of those blocks (A and D) were the saturated science frames. The other two (blocks B and C) were the unsaturated frames for reconstructing the PSFs. In the C and D blocks, we added a wideband 2.8-4.3 $\mu$m filter. This filter, frequently used for spectroscopy, removes some of the strong water vapor emissivity in the $M$-band.

In the stochastic seeing conditions, the phase loop repeatedly opened and had to be manually re-closed. The longest duration for which the phase loop was continuously closed was 4.4 minutes. In addition, there were occasional fringe ``jumps'' on Phasecam by 1$\lambda_{K_{s}}$ wavelength, which caused a jump of 0.53$\lambda_{Br\alpha}$ on LMIRCam (\citet{maier2018two,maier2020implementing}; see also Table 2 in \citet{spalding2018fizeau}). The telescopes were nodded only once during the entire observation, so as to keep the PSF stable and reduce overheads. 

\begin{centering}
\begin{figure}
\includegraphics[trim= 0cm 0cm 0cm 0cm, width=1\linewidth, clip=True]{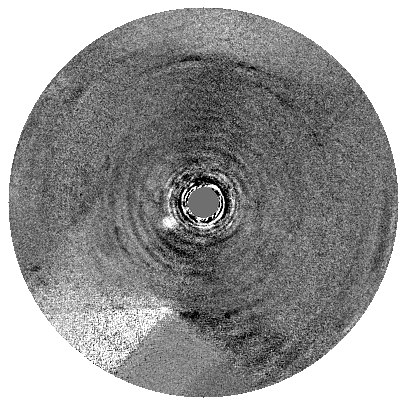}
\caption{An example ADI image in the classical $\lambda/D$ regime, where North is up and East is left. A $\Delta m=9.4$ injected companion corresponding to a detection with S/N=5 is at $\rho=0.42$ arcsec, 120$^{o}$ E of N. The residuals to the southeast at $\rho\gtrsim1$ arcsec are from detector bias variation noise.} 
\label{fig:adi_example}
\end{figure}
\end{centering}

No new companions to Altair were found in the data, though the analysis described in Sec. \ref{sec:analysis} characterizes the sensitivity of this novel mode of observation. Fig. \ref{fig:adi_example} shows an example of a reduced, classical ADI image with an injected companion. The region of the readout used in the ADI reduction was restricted to a circle for computational expediency, and the central region of the star was masked after the subtraction of the stellar PSF. Since the PSF also contained Fizeau fringes, an additional analysis was performed for the angular regime of Fizeau fringes.

\subsection{Analysis}
\label{sec:analysis}

Traditional host star subtraction routines based on angular differential imaging (ADI) make use of the fact that a faint astrophysical signal near the host star will rotate with the sky. The PSF is fixed relative to the pupil while the sky (and any potential companion) rotates beneath it, which allows decorrelation of the astrophysical signal from systematic effects like quasistatic speckles.

Code bases have already been developed to reduce AO-corrected direct imaging data to search for low-mass companions. However, these routines have tended to be applied to stable, AO-corrected Airy PSFs which are, to first order, axisymmetric. By contrast, the LBT Fizeau PSF is non-axisymmetric, and exhibits degrees of freedom that are unique to the coherent combination of two Airy PSFs. These include differential tip, tilt, and phase \citep{spalding2019status}, all three of which can vary with time.

To deal with this unique and time-variable PSF, we constructed a new reduction pipeline which treats each detector readout individually, under the assumption that the morphology of the PSF can change significantly from frame to frame. This pipeline also treats large-angle, classical ADI regimes ($\rho\gtrsim\lambda/D$) and small-angle Fizeau regimes ($\lambda/D\gtrsim\rho\gtrsim\lambda/B_{EE}$) separately. We describe the pipeline in the following sections, with more details in Appendix \ref{sec:app_pca}.

\subsubsection{Regime of classical ADI imaging: $\rho>\lambda/D$}

This part of the analysis is similar to that of classical direct imaging, with an AO-corrected PSF of size scale $\lambda/D=0.101''$. Modeled values of the atmospheric coherence timescale at the top of Mt. Graham can be $\approx$1.5-2 milliseconds towards the zenith at 0.5 $\mu$m\footnote{\url{http://alta.arcetri.astro.it}} (see \citet{turchi2016forecasts}). For 4 $\mu$m observations at zenith angles $\gamma$ of 25-35 degrees, the scaling law $\tau_{0}\propto\lambda^{6/5}($cos$\gamma)^{3/5}$ (e.g., \citet{roddier1999adaptive}) yields coherence timescales of $\approx$20 milliseconds. With no phase control, the integration time of 146 milliseconds will lead to smearing of the fringes. Thus, out of saturated frames representing 12.9 minutes of integration, we selected only frames which had phase control. The closed phase loop brought the phase rms values as low as $\approx$0.3 $\mu$m. This satisfies the condition of piston rms $\ll\lambda$ which maximizes fringe contrast in a given frame \citep{patru2017_ii}, barring phase jumps. The science data which passed these quality checks represented a total of 7.4 minutes of integration (Table \ref{table:strip_frame_subsets}). 

\begin{deluxetable*}{l | l | l | l | l | l }
\tabletypesize{\scriptsize}
\tablecaption{Subsets of science frames used for ADI}
\tablewidth{0pt}
\tablehead{
\colhead{Frames} &  \colhead{Relevant regime} & \colhead{Parallactic angle $q$} & \colhead{Angle E of N of long baselines$^{a}$} & \colhead{Integration time$^{b}$ $t_{int}$} & \colhead{Efficiency$^{c}$} 
}
\startdata
Blocks A and D & $\lambda/D$ & \begin{tabular}{@{}l@{}}-43.3$^{\circ}$ to -36.0$^{\circ}$,\\ -22.0$^{\circ}$ to +3.3$^{\circ}$\end{tabular} & -- & 445 sec$^{d}$ & 0.093 / 0.57 \\ 
\hline
Block A, strip 0 & $\lambda/B_{EE}$ & -43.3$^{\circ}$ to -36.0$^{\circ}$ & 50.32$^{\circ}$ (0E) and 230.32$^{\circ}$ (0W)& 124 sec & 0.072 / 0.45 \\ 
\hline
Block D, strip 1 & $\lambda/B_{EE}$ & -21.5$^{\circ}$ to -15.7$^{\circ}$ & 70.782$^{\circ}$ (1E) and 250.782$^{\circ}$ (1W)& 103 sec & 0.060 / 0.82 \\ 
\hline
Block D, strip 2 & $\lambda/B_{EE}$ & -15.7$^{\circ}$ to -9.4$^{\circ}$ & 76.57$^{\circ}$ (2E) and 256.57$^{\circ}$  (2W)& 75 sec & 0.098 / 0.60 \\ 
\hline
Block D, strip 3 & $\lambda/B_{EE}$ & -9.3$^{\circ}$ to -3.0$^{\circ}$ & 83.37$^{\circ}$ (3E) and 263.37$^{\circ}$  (3W)& 82 sec & 0.111 / 0.68 \\ 
\hline
Block D, strip 4 & $\lambda/B_{EE}$ & -3.0$^{\circ}$ to +3.3$^{\circ}$ & 90.04$^{\circ}$ (4E) and 270.04$^{\circ}$  (4W)& 61 sec & 0.084 / 0.52 
\enddata
\tablenotetext{a}{The naming convention of the baselines is a chronological number and a letter to indicate whether a half-baseline is the eastern or western half (see Fig. \ref{fig:strips}).}
\tablenotetext{b}{The total integration time of frames in this subset which were ultimately used in the reduction.}
\tablenotetext{c}{The first decimal is the used integration time $t_{int}$ divided by wall-time elapsed during that section(s) of the observing sequence. This ratio is heavily affected by science data transfer and management overheads. The second number is the ratio of used science frames to all the science frames (or $t_{int}$ out of the total  integration time). The lost efficiency of this number is primarily due to phase loop openings. The cumulative frames used in the $\lambda/B_{EE}$ reductions are the same as those in the $\lambda/D$ reduction, so the total percentage of used frames in the $\lambda/B_{EE}$ reductions is also 57\%.}
\tablenotetext{d}{Note that the borders of $\approx$70\% of the science frame cutouts centered on the PSF extend beyond the edge of the readout region by 1''. This brings the effective integration time of the science frames down to $\approx$130 sec for a region $>$1'' to the southeast of Altair.}
\label{table:strip_frame_subsets}
\end{deluxetable*}

Each host star PSF in a frame was reconstructed twice: once with a PSF PCA basis $S$ set made from saturated frames (i.e., those in blocks A and D), and once with a basis set $U$ made from unsaturated frames (those in blocks B and C). Unsaturated reconstructions of saturated frames were made by projecting the science frames onto $U$, with saturated pixels in the innermost bright fringes reconstructed based on the projection of the rest of the PSF on the basis set. 

\begin{figure}
\begin{centering}
\includegraphics[width=0.48\linewidth,trim=4.75cm 5cm 5.25cm 5cm,clip=True]{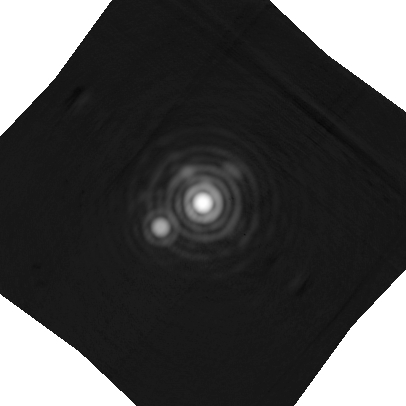}
\includegraphics[width=0.48\linewidth,trim=4.75cm 5cm 5.25cm 5cm,clip=True]{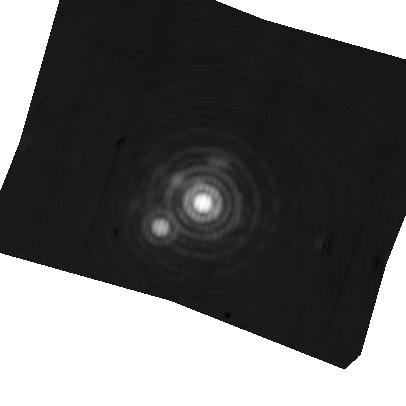}
\caption{Examples of derotated and medianed science frames, with $10^{-1}$ amplitude fake companions, using subsets of frames corresponding to those used in reductions of strips 0 (left) and strips 1 (right). These are the same as ADI frames, but without subtraction of the host star (and without imposing a tesselation pattern). The PCA-based reconstruction of the host star PSF reproduces in the companion PSF the same tip, tilt, phase (modulo 2$\pi$), and fringe visibility in every subsidiary frame. In these ADI frames, the fringes have different amounts of visibility, which can happen due to phase diversity among the subsidiary frames or different amounts of on-sky rotation.} 
\label{fig:median_sci}
\end{centering}
\end{figure}

The reconstruction of the unsaturated PSF for a given frame was injected into the frame as the fake planet, whose PSF shares the host star's Fizeau aberrations of differential tip/tilt and phase.  (See Fig. \ref{fig:median_sci}.) In this way we minimize the introduction of systematics stemming from the forcing of the fake planet PSF to fit a fixed PSF model, such as a median across a stack of frames.

Fake companions were injected at $\phi=\{0^{\circ},120^{\circ},240^{\circ}\}$ East of True North, and at distances from the host star corresponding to $\rho=\textrm{FWHM}\cdot\{2,3,4,5,7,9,12,15,19\}$, where $\textrm{FWHM}=0.104''$ is the full-width-at-half-maximum of a perfect Airy PSF. An entire reduction was done for every single combination of ($\phi,\rho$), one fake companion in a dataset at a time. The inner cutoff of $2\textrm{FWHM}$ is somewhat arbitrary, but was made to allow overlap with the treatment of the $\lambda/B_{EE}$ regime (Section \ref{subsubsec:regime_fizeau}).

After the host star was subtracted from each frame with the reconstruction based on a projection onto the basis set $S$, the images were derotated and medianed to produce a single ADI frame. This frame was convolved with a 2D Gaussian of size $\lambda/D$ to remove the effects of pixel-to-pixel noise. The signal was taken to be the maximum value of counts within a circular cutout centered on the fake companion, with a width of one FWHM of the Airy function. The noise was calculated as the standard deviation of counts in a series of ``necklace-bead'' patches along the same annulus as that containing the fake companion. The centers of the patches are circumferentially separated from each other by one FWHM.

The pipeline repeated the reductions iteratively, perturbing the fake planet amplitudes until convergence to S/N$=5$ at a given location ($\phi,\rho$). An azimuthal median of the companion amplitudes was made to generate a 1D ``classical'' contrast curve, which was then  modified to account for diminishing degrees of freedom at small radii, following the framework of \citet{mawet2014fundamental}. This framework also allows a false positive fraction (FPF) which varies as a function of radius, and a true positive fraction (TPF) which is fixed at 0.95. More details of this part of the reduction are described in Appendix \ref{subsec:lambdaDregime}, and Appendix \ref{appendix:contrast_curves} provides more details on the calculation of the contrast curves.

\subsubsection{Regime of Fizeau fringes: $\lambda/D\gtrsim\rho\gtrsim\lambda/B_{EE}$}
\label{subsubsec:regime_fizeau}

At 4.05 $\mu$m, the Fizeau regime can in principle carry information down to $\lambda/B_{EE}=0.037"$. However, the Fizeau fringes effectively rotate around the object during the observation, because the fringes are stationary on the detector while the sky rotates. ADI applied to all the frames at once would then only have the effect of washing out the fringes. We therefore subdivided the science frames into subsets listed in the rows corresponding to $\lambda/B_{EE}$ in Table \ref{table:strip_frame_subsets}. It should be noted from Table \ref{table:strip_frame_subsets} that preserving the Fizeau fringes by subdividing the dataset comes at the cost of total integration time for each subset of frames.

\begin{centering}
\begin{figure}
\includegraphics[width=\linewidth,trim=0.5cm 5cm 0cm 4cm,clip=True]{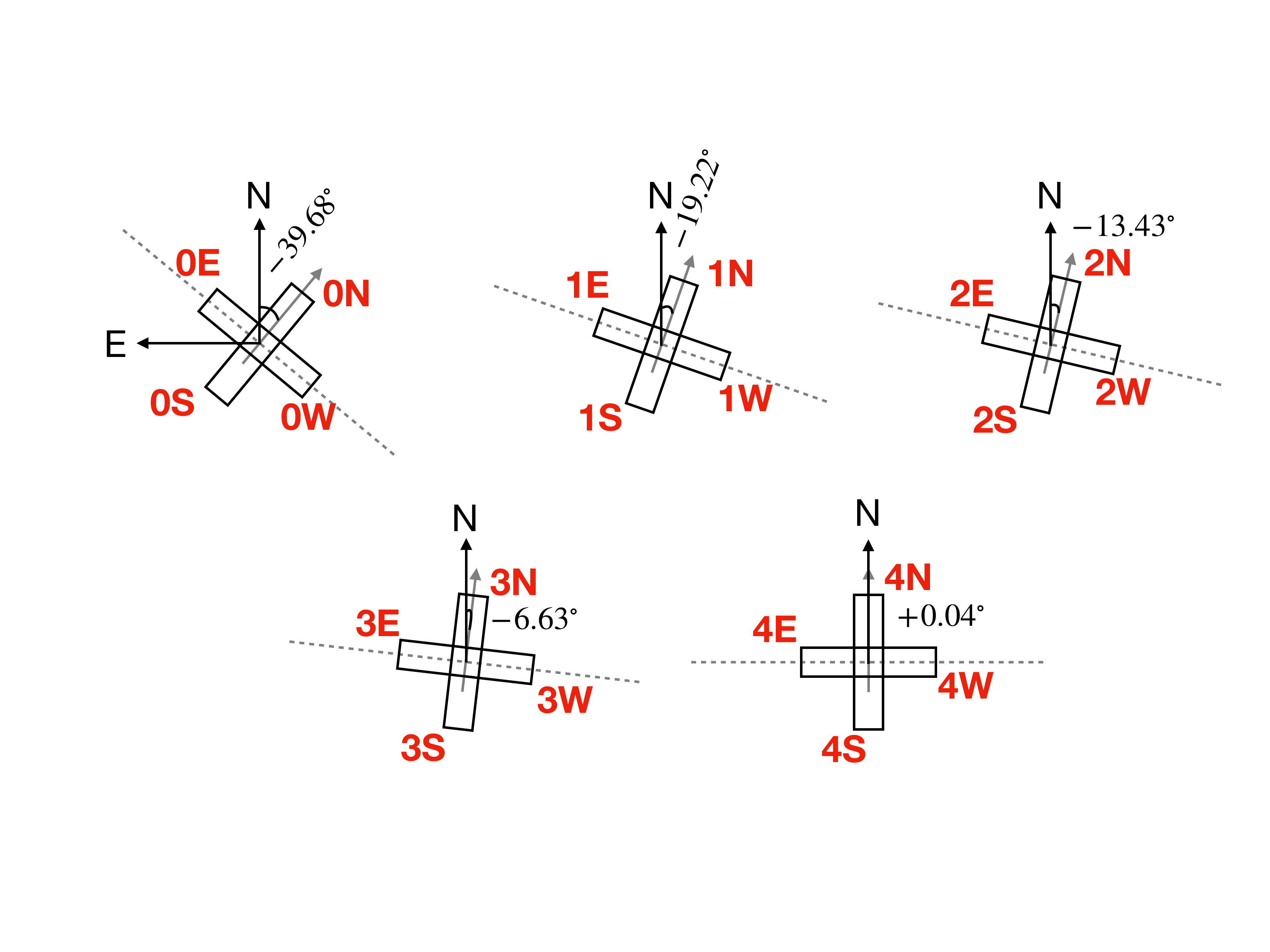}
\caption{Illustration of the strips taken across the PSF for the $\lambda/B_{EE}$ analysis. Angles are those between the short baseline direction and True North. Each half-strip is named based on the subset of frames (`0', `1', `2', `3', `4'; see Table \ref{table:strip_frame_subsets}), and the closest cardinal direction (`N', `E', `S', `W'). The dashed line is the long baseline.} 
\label{fig:strips}
\end{figure}
\end{centering}

To subtract the host star, we use PCA-reconstructed regions along rectangular strips. In one set of reductions, a strip was set with the long axis along the long (Fizeau) LBT baseline. In another set of reductions, the reconstructed region is along the short LBT baseline, which would most closely approximate classical imaging. Those regions are overlapped for illustration in Fig. \ref{fig:strips} to show how they rotate on the sky. These perpendicular orientations were chosen to compare the results based on the longest and shortest LBT baselines, though baselines could be chosen for any arbitrary orientation.

To determine sensitivity to a companion along a given baseline, frames are reduced with injected fake companions ranging across a grid of amplitudes equivalent to the star-to-planet contrast $0 < \Delta m < 10$, and separations equivalent to $0.28 < \rho/(\lambda/B_{EE}) < 14.1$. Again, reductions are always performed with one fake companion at a time. This is done for all frames, including those for strips which have angles different from the baseline whose sensitivity we want to determine. This allows for a companion PSF that has a center outside a given strip to ``bleed'' into the other strips on the other baselines, as a true PSF would. 

\begin{centering}
\begin{figure}
\includegraphics[width=0.49\linewidth,trim=5cm 6cm 5cm 6cm,clip=True]{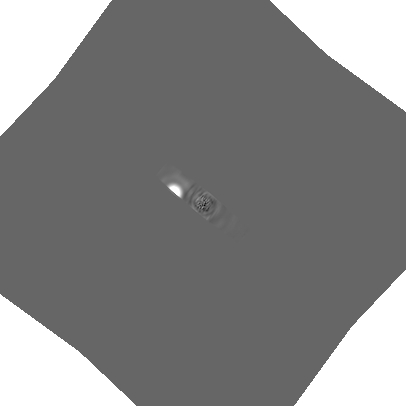}
\includegraphics[width=0.49\linewidth,trim=5cm 6cm 5cm 6cm,clip=True]{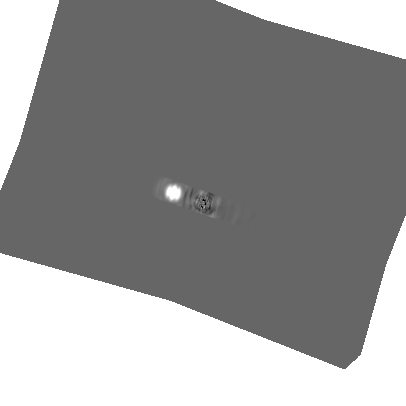} \\
\includegraphics[width=0.49\linewidth,trim=5cm 6cm 5cm 6cm,clip=True]{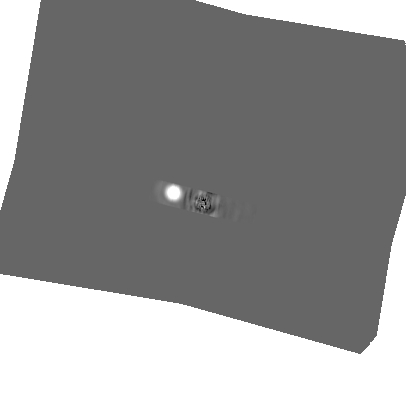} 
\includegraphics[width=0.49\linewidth,trim=5cm 6cm 5cm 6cm,clip=True]{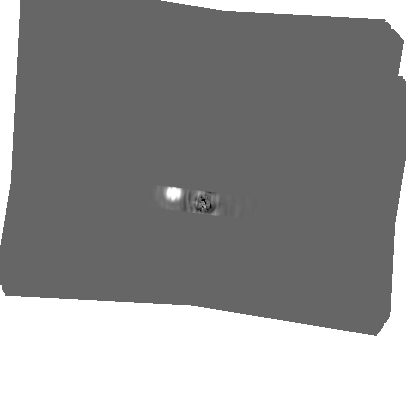} \\
\includegraphics[width=0.49\linewidth,trim=5cm 6cm 5cm 6cm,clip=True]{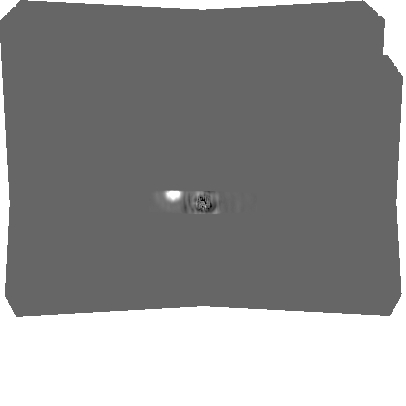}
\includegraphics[width=0.49\linewidth,trim=5cm 6cm 5cm 6cm,clip=True]{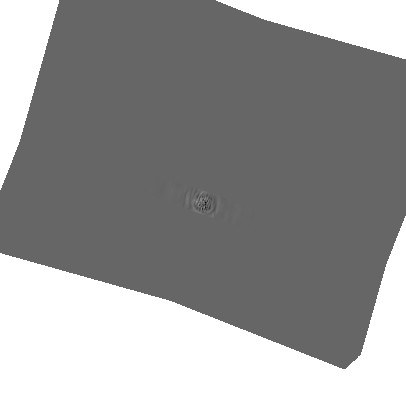}
\caption{Example ADI images of strips involving a $10^{-1}$ companion along the average angle of half-strip 1E (at 70.8$^{\circ}$ East of North), where North is up and East is left. Left to right, in the top row: strip 0 and strip 1. Middle row: strip 2 and strip 3. Bottom row: strip 4 and the baseline strip 1 with no injected companions. Greyscale is square-root scaled, and NaNs outside the strips have been replaced with zeros for display.}
\label{fig:sweep}
\end{figure}
\end{centering}

After subtraction of the PCA-reconstructed host star, a series of two-sample Kolmogorov-Smirnov (KS) tests is done with pairs of 1D residuals of strips. We use the two-sample KS test to determine the acceptance or rejection of the null hypothesis $H_{0}$ which states that two empirical distributions come from the same parent sample, with a confidence of 95\%. This is effectively a method of detecting azimuthal anomalies which betray the presence of a companion as the long baseline of the LBT sweeps out the area around the host star (Fig. \ref{fig:sweep}).

The residuals of the strip with planets injected along it are compared with  the residuals of other strips pointed in the nearest cardinal direction. These KS grids in ($\rho,\Delta m$)-space are then averaged into one grid corresponding to the closest cardinal direction N, S, E, or W. The critical value of the contour plot of the KS landscape is that which divides detections with 95\% confidence and non-detections. Appendix \ref{subsec:lambdaBregime} provides more details of this stage of the pipeline and shows that the critical dividing line corresponds to the contour value 0.2716.


\section{Results}
\label{sec:results}

This section presents the results of the data analysis, in terms of contrast and sensitivity. We show a 1D contrast curve for the $\lambda/D$ regime in Fig. \ref{fig:cont_curve_1d}. The contrast curve was converted into absolute magnitudes (see Appendix \ref{sec:abs_mag}) and then to mass amplitudes with the use of four different models derived from the \texttt{PHOENIX} radiative transfer code \citep{phoenix} for standalone, self-luminous objects. They include AMES-Cond \citep{allard2001limiting,baraffe2003evolutionary}; BT-Cond, which incorporates updated molecular opacities \citep{plez1998new,ferguson2005low,barber2006high}; and BT-Dusty and BT-Settl, which allow for suspended particulates \citep{allard2012models}. 

\begin{centering}
\begin{figure}
\includegraphics[width=0.95\linewidth]{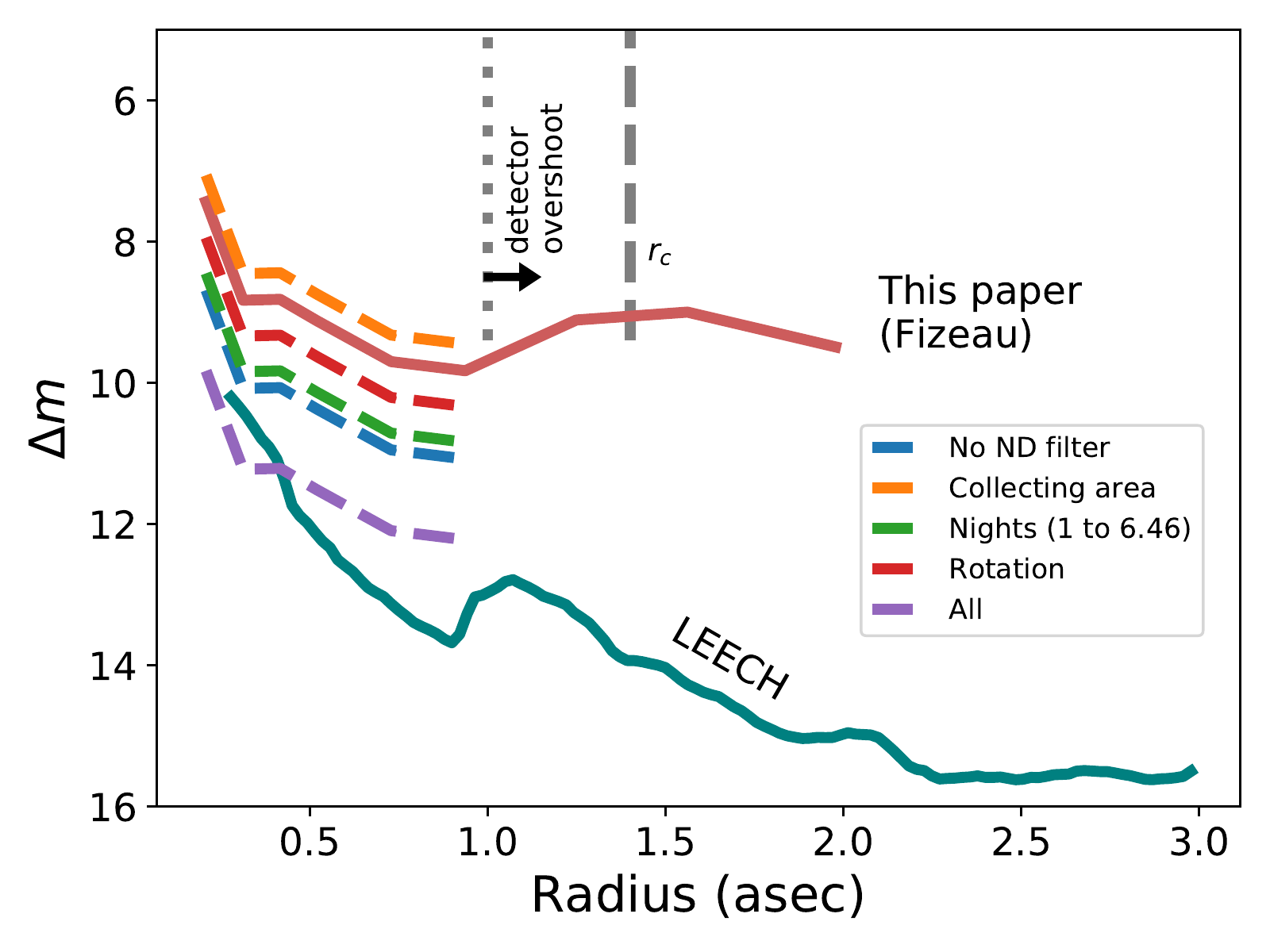}
\caption{The $\Delta m$ contrast curve for Altair in the regime $\rho\gtrsim\lambda /D$ with total integration time in Fizeau mode of $t_{int}=$445 sec for most azimuths around Altair. The dotted line indicates the radius corresponding to the edge of the detector readout for a subset of the frames, effectively reducing the integration time to $\approx$130 sec for a region to the southeast of Altair. The dashed line indicates the AO control radius, beyond which the AO correction is expected to deteriorate due to Nyquist spatial sampling limitations. Seeing was 0.9-1.4'', and $\Delta \textrm{PA}=33^{\circ}$. (See Tables \ref{table:conds} and \ref{table:blocks}). A contrast curve for Altair from the direct imaging LEECH survey is also shown, obtained with one of the two LBT sub-telescopes, with $\lambda_{C}=3.8$ $\mu$m, $t_{int}=2873$ sec, seeing $\sim$0.9'', and $\Delta \textrm{PA}=83^{\circ}$. (J. Stone, pers. comm.)}
\label{fig:cont_curve_1d}
\end{figure}
\end{centering}

\begin{centering}
\begin{figure}
\includegraphics[width=0.95\linewidth]{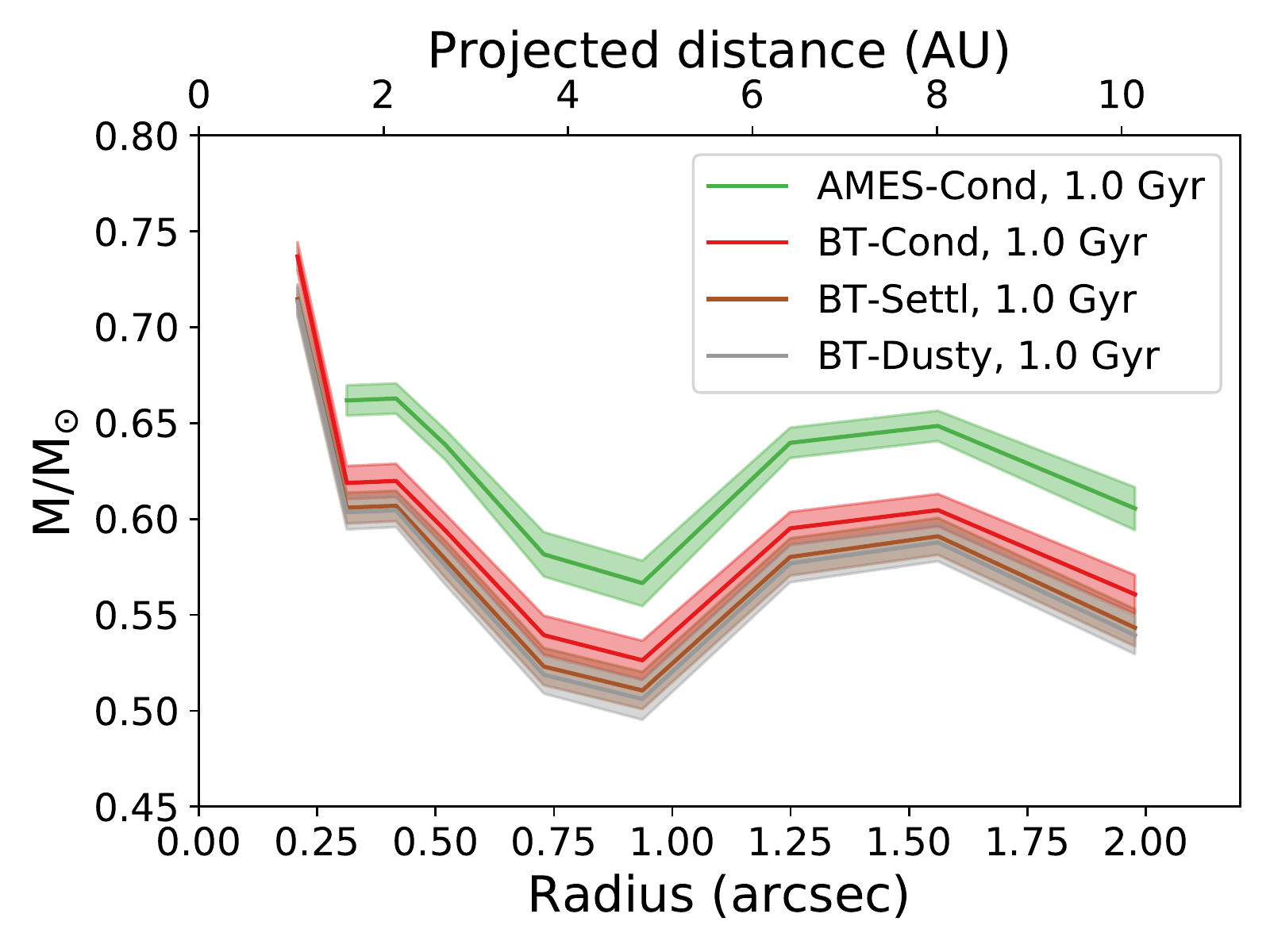}
\caption{The mass sensitivity curve of the $\lambda/D$ regime, based on fake planet injections. Note the effective integration time decreases at radii $>$1'' (see note $d$ in Table \ref{table:strip_frame_subsets}).}
\label{fig:mass_sensitivity_curve_1d}
\end{figure}
\end{centering}

Fig. \ref{fig:mass_sensitivity_curve_1d} shows the mass sensitivity, which is limited to objects $\gtrsim0.5M_{\odot}$, well above the hydrogen-burning limit and the brown dwarf mass range of $13\lesssim M/M_{J} \lesssim80$ \citep{spiegel2011deuterium}. At these sensitivities the differences in the models between ages 1.0 and 0.7 Gyr were negligible, and so we consider only 1.0 Gyr in what follows. 

\begin{centering}
\begin{figure}
\includegraphics[width=1\linewidth]{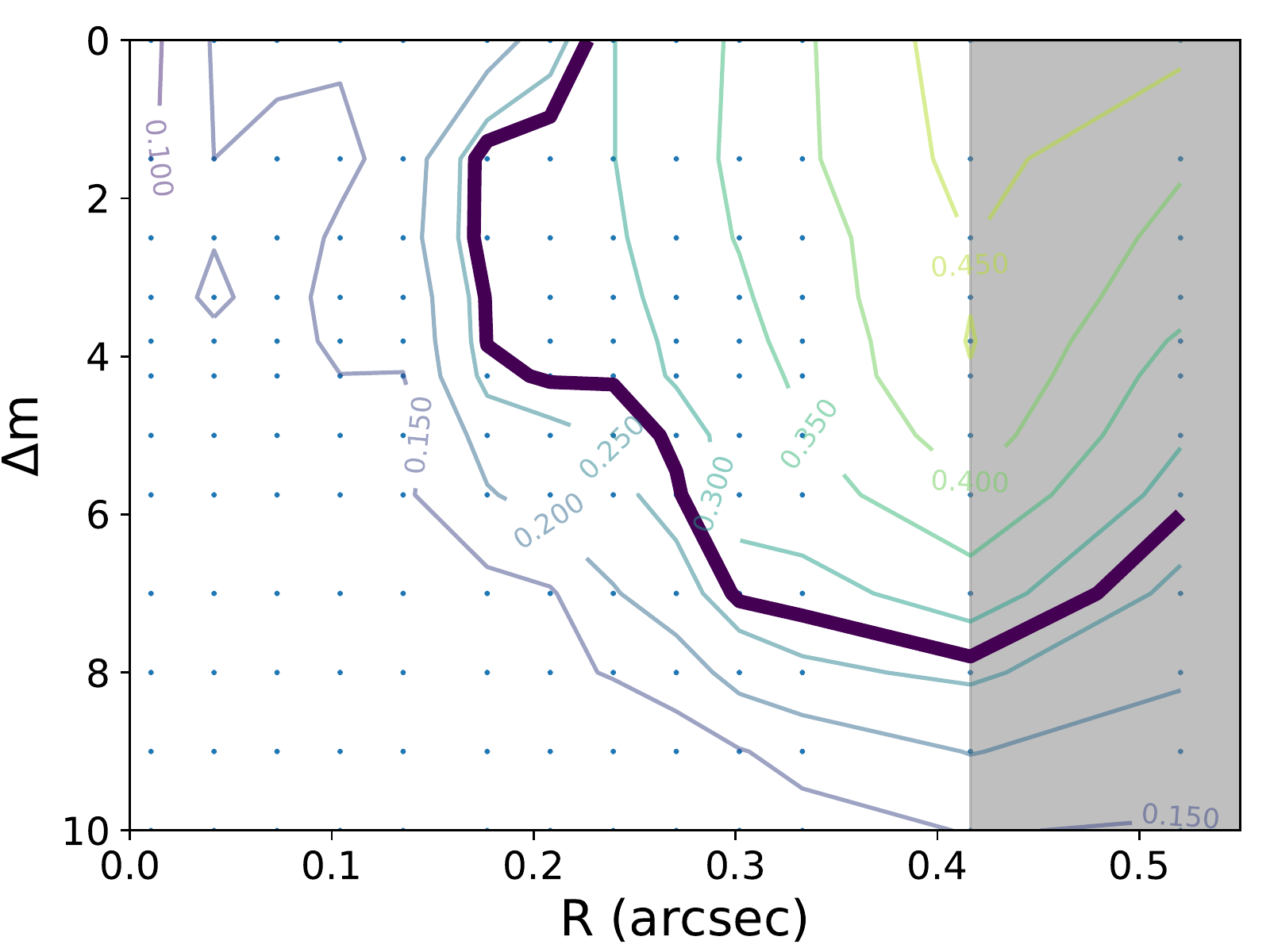} 
\caption{An example of KS statistic interpolations in the $\lambda/B_{EE}$ regime, for the `S' half-strips (i.e., those pointed most closely to the direction South after derotation). Dots show the grid of tested companion contrasts ($0 < \Delta m < 10$) and distances from the host star. The bold curve is at a KS value of 0.2716, which represents the boundary between companions that are from a different parent population, to a confidence of 95\% (See Appendix \ref{subsec:lambdaBregime}). The grey region is the region where the injected companions are 1 FWHM or less from the edge of the strip. Since the PSF of the fake companion is truncated when it is this close to the edge, the KS statistic will decrease as an edge artifact.}
\label{fig:ks_landscape}
\end{figure}
\end{centering}

\begin{centering}
\begin{figure}
\includegraphics[width=0.95\linewidth,trim=0.3cm 0cm 0cm 0cm, clip=True]{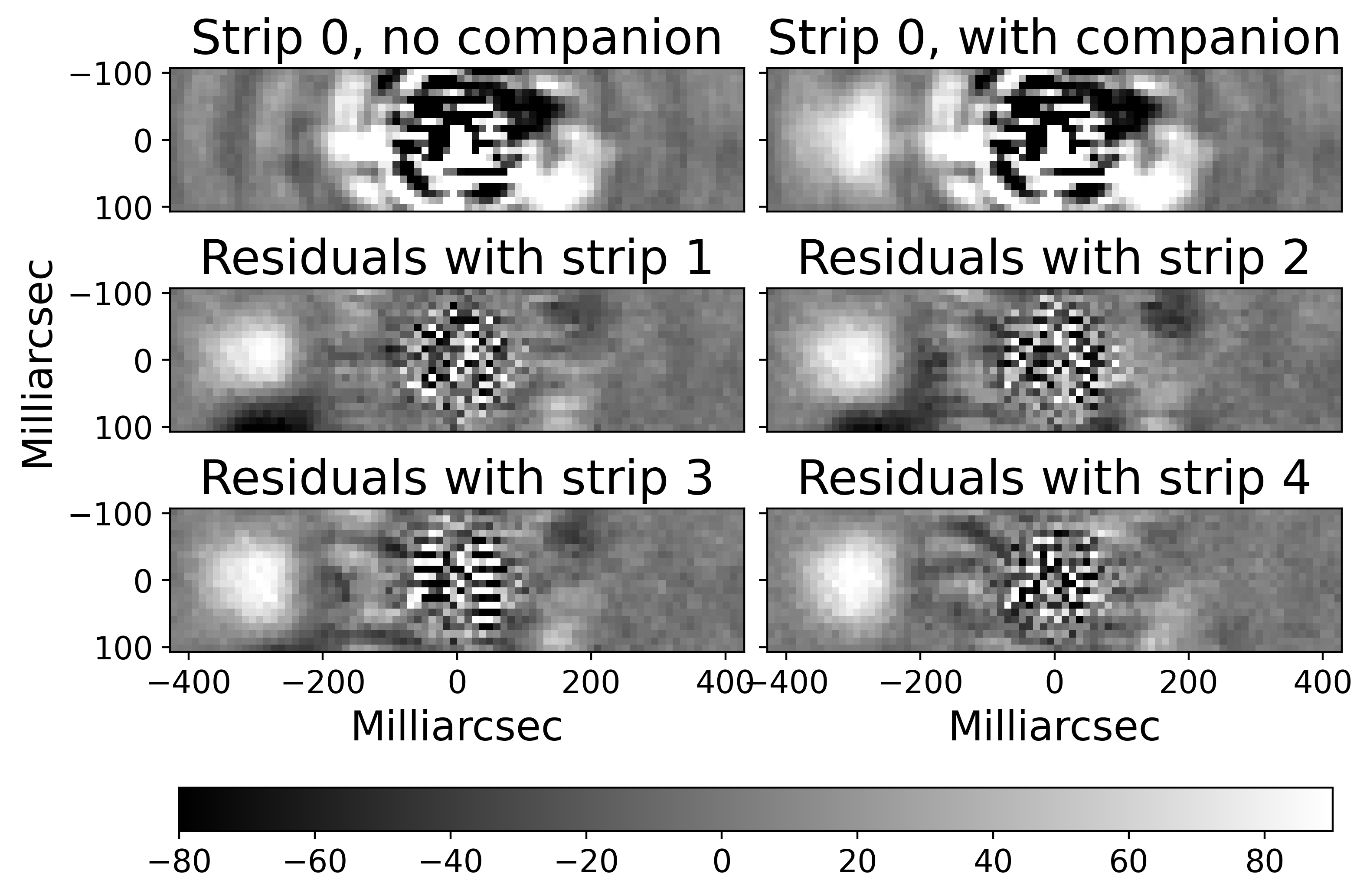}
\caption{Top row: reduced frames of strip 0, with and without a $\Delta m=7$, $\rho=0.3$'' companion. Middle and bottom rows show residuals between strip 0 with a companion, and the other strips. Since the companion is centered along a different azimuth, the residuals are strong enough to allow detection of a flux source which does not rotate with the instrument baseline.  This detection is on the critical KS contour in Fig. \ref{fig:ks_landscape} and represents a marginal detection. The truncated greyscale is in detector counts.}
\label{fig:example_marginal_ks_detec}
\end{figure}
\end{centering}

\begin{centering}
\begin{figure}
\includegraphics[width=1\linewidth,trim=0.7cm 0cm 0cm 0cm, clip=True]{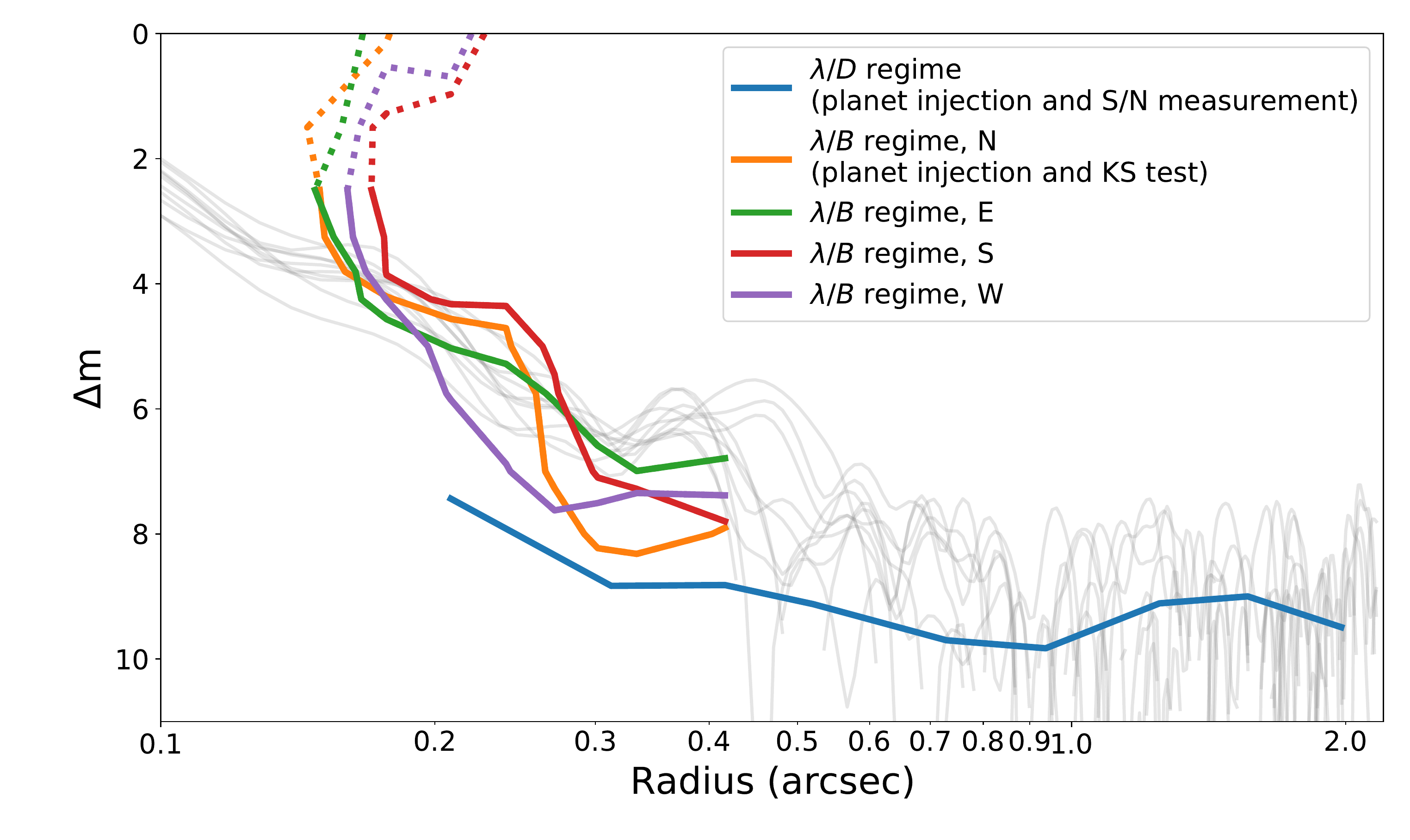}
\caption{The contrast curves. The blue line is the $\lambda/D$ contrast curve based on fake planet injections, where the fake planets have an amplitude which has converged to $S/N=5$, and with additional constraints described in Appendix \ref{appendix:contrast_curves}. The other lines show the directional averages of the critical value of the KS statistic based on fake planet injections in the $\lambda/B_{EE}$ analysis, from the cross-sections of the half-strips in Fig. \ref{fig:strips}. (For example, KS curves from 0N, 1N, 2N, 3N, and 4N are averaged to produce the ``N'' curve.) Companions to the right of these lines induce an azimuthal change in the cross-sections along the long LBT baseline that is significant enough that the null hypothesis $H_{0}$ can be rejected with confidence $\geq$95\%. The dotted regions of lines indicate regimes for which the companions are too bright for a mass solution from all four evolutionary models. The inversion at the brightest regimes ($\Delta m \lesssim 1$) is addressed in Sec. \ref{sec:results}. Grey lines are a random sampling of PSF profiles to guide the eye, and normalized to be zero on the magnitude scale at the peak.}
\label{fig:cont_curve_1d_fizeau}
\end{figure}
\end{centering}

For the $\lambda/B_{EE}$ regime, Fig. \ref{fig:ks_landscape} shows the landscape of the KS statistic. From these results, we cannot reject the null hypothesis for companions at radii smaller than $\approx0.15$'', or within 1 AU. Fig. \ref{fig:example_marginal_ks_detec} shows an example detection from along the critical KS contour at $\Delta m=7$. If, however, a fake planet amplitude is comparable to that of the host star at small angles, the changing mode pattern used to subtract the host star exacerbates the residuals at those angles. This leads to the sensitivity inversion seen at very bright fake companions ($0<\Delta m<2$) in Figs. \ref{fig:ks_landscape} and \ref{fig:cont_curve_1d_fizeau}.

\begin{figure}
\includegraphics[width=0.49\linewidth]{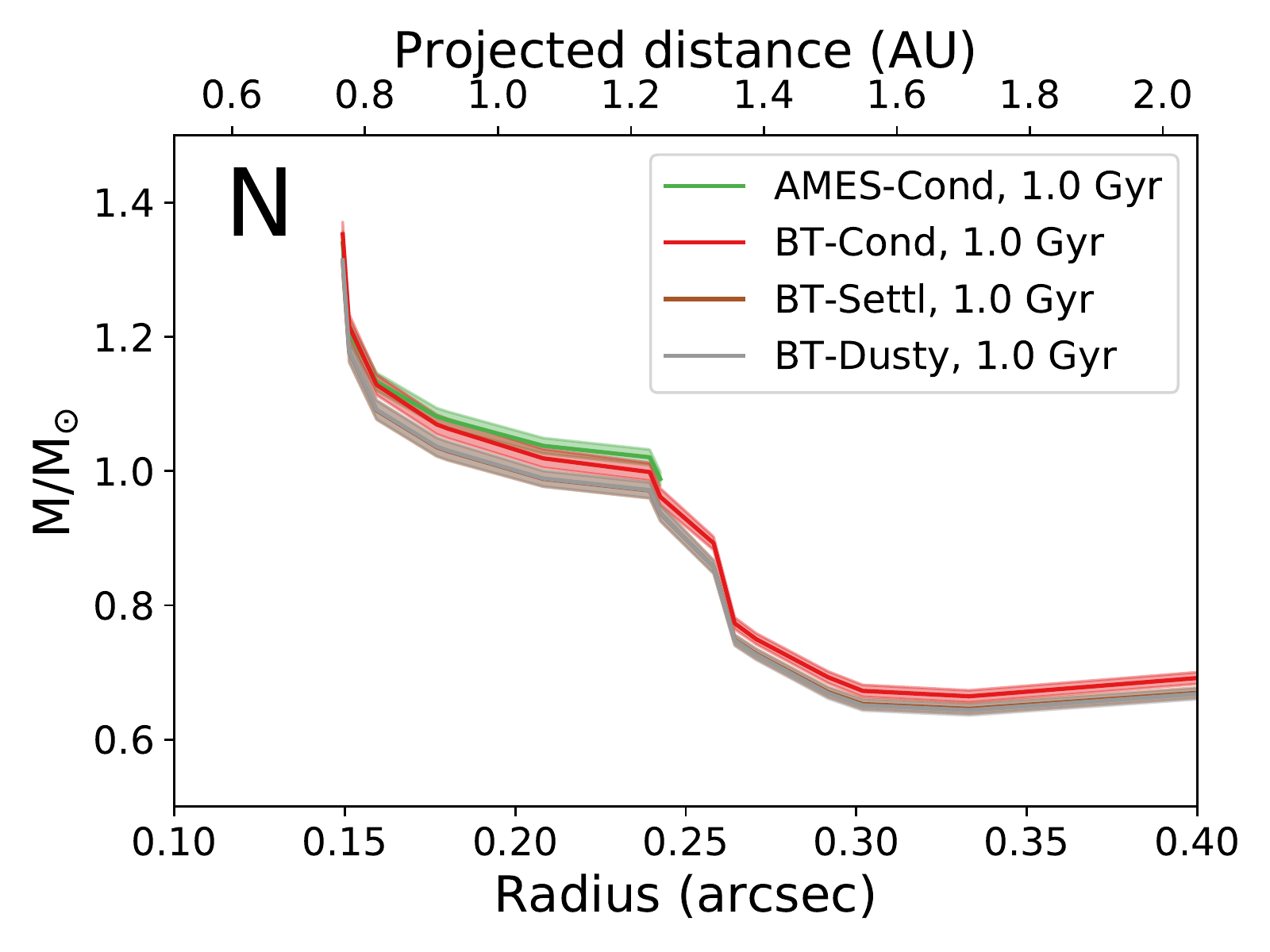}
\includegraphics[width=0.49\linewidth]{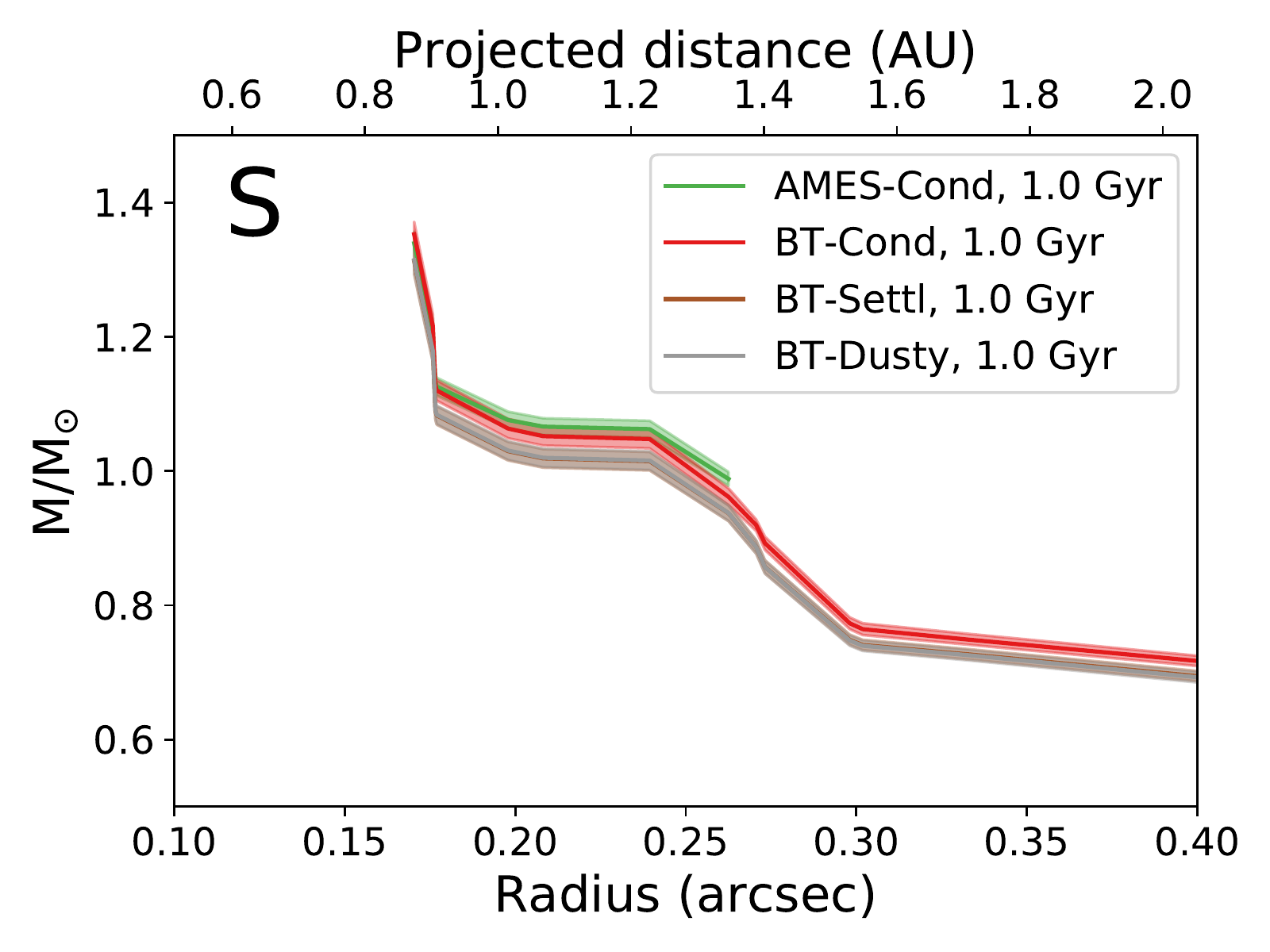} \\
\includegraphics[width=0.49\linewidth]{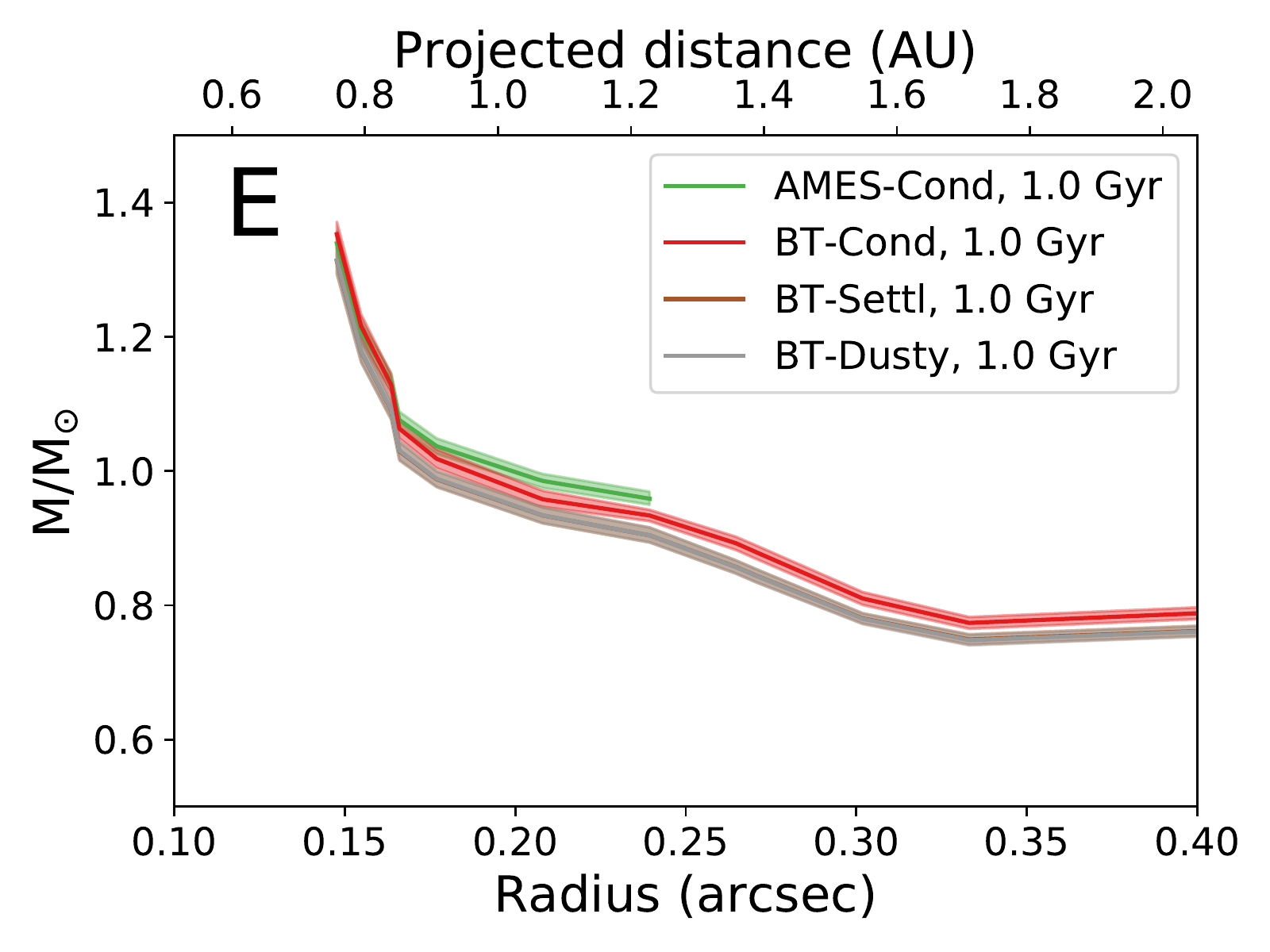}
\includegraphics[width=0.49\linewidth]{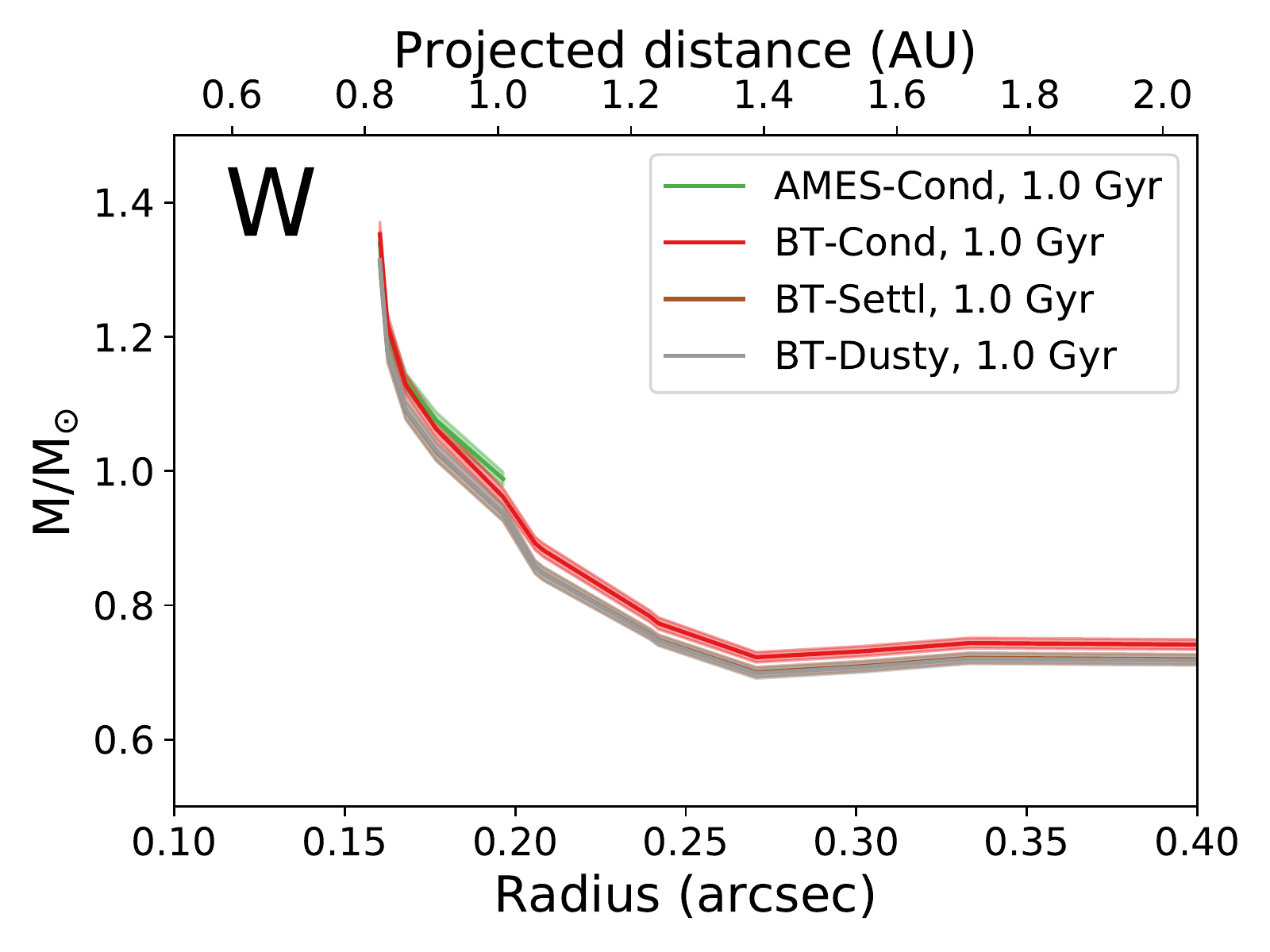}
\caption{The mass sensitivity curves from the $\lambda/B_{EE}$ analysis, based on the KS test statistic. Top left: Mass sensitivity curves for the half-strips pointed along the directions closest to cardinal direction North; i.e., comprising of sampled baselines 0N, 1N, 2N, 3N, and 4N as shown in Fig. \ref{fig:hist_angles}). Other subfigures, clockwise from top right: the same, for cardinal directions South, West, and East. The directions North and South are along the short LBT baseline, and East and West are along the long baseline.}
\label{fig:mass_sensitivity_curve_1d_lambda_B}
\end{figure}

Fig. \ref{fig:mass_sensitivity_curve_1d_lambda_B} shows that in the regime of $\lambda/B_{EE}$, the constraints based on fake companion injections and the KS test have sensitivities at the smallest angles corresponding to the highest masses which can be interpolated from the model grids, $\approx$1.3 M$_{\odot}$.

\section{Discussion} 
\label{sec:discussion}

We now contextualize our results within related work. \citet{stone2018leech} conducted a direct, filled-aperture direct imaging survey of 98 B- to M-type stars using either or both of the 8.4-m telescopes of the LBT. The LEECH observations did not involve coherent combination of the beams: when both telescopes were being used simultaneously, the PSFs were physically separate on the LMIRCam detector. Altair itself was observed only using the left-side LBT sub-telescope, and the LEECH contrast curve for that target is shown in Fig. \ref{fig:cont_curve_1d}.

This Fizeau dataset does not outperform constraints from LEECH around Altair in the $\lambda/D$ regime, but it also only has $\approx$15\% of the integration time in the innermost arcsecond. One source of lost efficiency is the time overhead from each of the detector readouts. Another source is the highly time-dependent nature of the optical aberrations in the Fizeau PSF, which is mostly due to repeated openings of the phase loop, and the need to manually realign and re-close the loop. 

Fig. \ref{fig:cont_curve_1d} shows comparisons of this Fizeau $\lambda/D$ contrast curve with that of the LEECH survey, based on simple scalings of the observation setup. These involve rescaling the collecting area, removing the neutral density filter, breaking up the observation into separate epochs to decorrelate the noise, and to allow more rotation. If these modifications are combined and speckles are sufficiently uncorrelated \citep{marois2006angular,males2021mysterious}, it would appear that the Fizeau mode can be competitive with single-aperture imaging at small angles.

In the $\lambda/B_{EE}$ regime analysis, constraints on companions to Altair are as close in as $\approx$0.15''. This inner working angle is closer in than the published constraints specific to Altair in any of the direct, non-interferometric imaging surveys mentioned in Sec. \ref{sec:intro}, which do not go closer in than 0.4'' \citep{kuchner2000search,leconte2010lyot,janson2011high,dieterich2012solar}, and the unpublished Altair-specific curve in the analysis of \citet{stone2018leech} goes as close in as 0.28'', albeit the mass ranges of interest in these surveys were substellar. Among OLBI observations of Altair  mentioned in Sec. \ref{sec:intro}, there has been no convincing evidence of a visibility profile perturbed by a very-close-in companion point source.  (The X-ray emission of Altair was once thought to suggest chromospheric activity of a companion, but the emission has been found to be consistent with Altair's own corona \citep{robrade2009altair}.)

\begin{centering}
\begin{figure}
\includegraphics[width=1\linewidth,trim=1.7cm 0cm 1.2cm 0cm, clip=True]{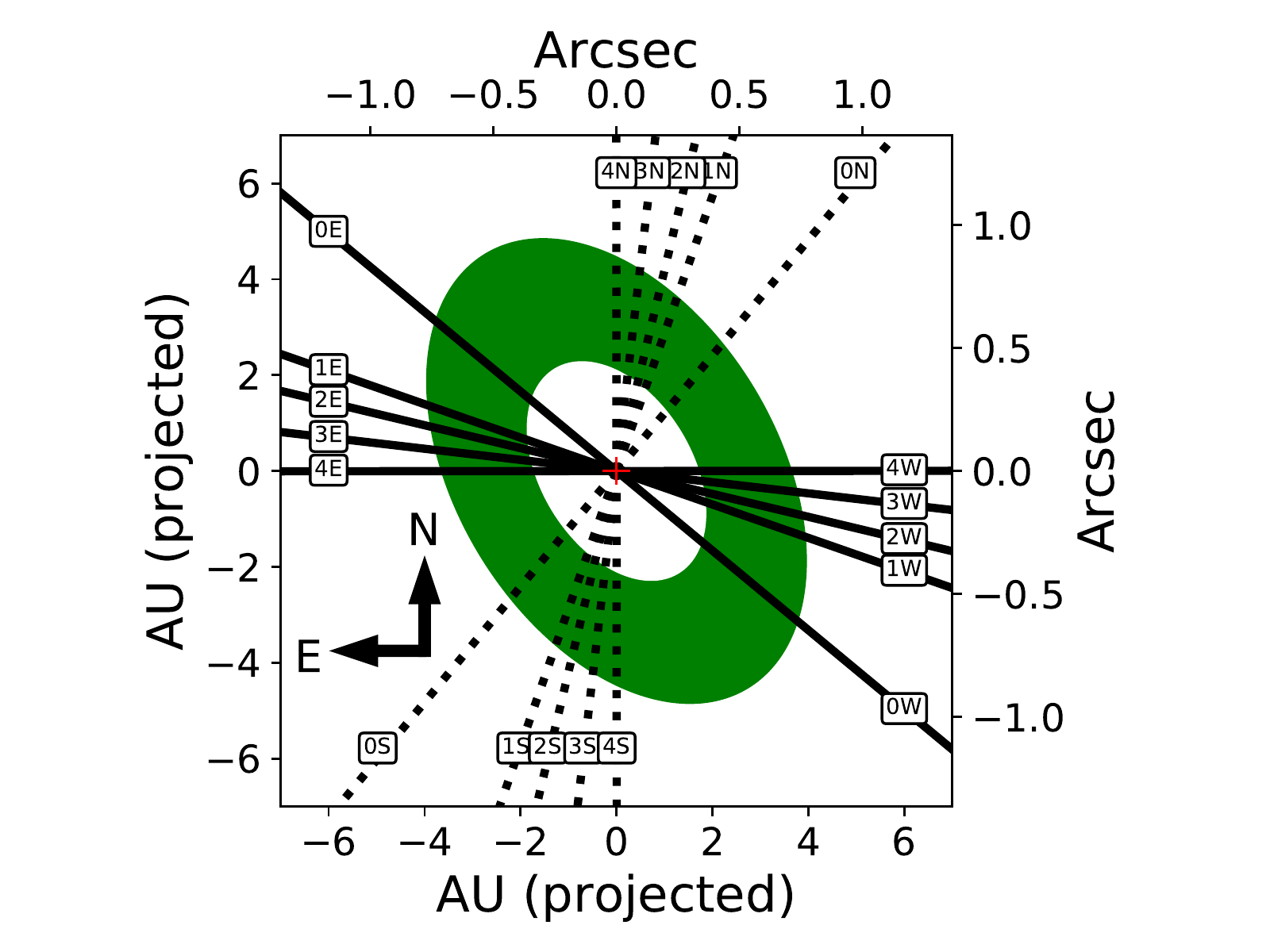}
\caption{The projected habitable zone (green) as defined by \citet{cantrell2013solar} around Altair (red cross), for planets that are spin-orbit aligned with the stellar spin axis. Long LBT baselines (see Fig. \ref{fig:strips}) are indicated by solid lines, and short LBT baselines with dashed lines.}
\label{fig:hist_angles}
\end{figure}
\end{centering}

The most recent constraints on Altair's position angle and inclination are $\textrm{PA}=301.1\pm0.3^{\circ}$ and $i = 50.7\pm1.2^{\circ}$ \citep{bouchaud2020realistic}. If  planets in the Altair system orbit in the plane perpendicular to Altair's spin axis, then the contrast curves in this work sample all of the projected habitable zone, which lies at $0.3''\leq \rho \leq 1.0''$. Fig. \ref{fig:hist_angles} shows a projection of this zone, with lines overplotted to show the sampled long LBT baselines.

It should be noted that our innermost constraints are set by the KS test applied to data with fake injected planets. In this context the KS test is weaker than fake planet recovery in the $\lambda/D$ regime, as it effectively asks ``are these strips along a given baseline drawn from different distributions?'' and assumes that variations between strips at different angles will be introduced by the PSFs of companions. The subsets of frames in the $\lambda/B_{EE}$ analysis each represent less integration time than that available for the $\lambda/D$ reduction, by virtue of the fact that only small amounts of rotation can be tolerated to preserve the narrow Fizeau fringes. Averages of the contrast curves across subsets, which have a cumulative integration time that is the same as the data in the $\lambda/D$ regime, are shown in Fig. \ref{fig:cont_curve_1d_fizeau}. Though these curves are still in the stellar mass regime, we place the closest-in direct imaging constraints on companions to Altair, at a sensitivity of 1.3 $M_{\odot}$ at $\approx$0.15''. In this analysis we have also used the highest percentage of science frames from an LBT Fizeau dataset, 57\%, in both the $\lambda/D$ and $\lambda/B_{EE}$ regimes.

\section{Future Improvements}
\label{sec:future_improvements}

The nature of the Fizeau PSF, and its stability during these observations, suggests a variety of  strategies for future improvement. 

We chose integration times and filter combinations to avoid hard saturation of the entire Airy core and the structure of Fizeau fringes contained therein. However, this leaves low S/N in the Fizeau PSF at radii corresponding to the $\lambda/D$ regime, which could yet prove valuable for putting constraints on low-mass companions, particularly in the darkest regions of the Fizeau PSF \citep{patru2017_i}. 

If the purpose of the observation is to examine both the $\lambda/D$ and $\lambda/B_{EE}$ regimes, we suggest taking future observations in a bifurcated way: a sequence of frames with hard saturation of the Airy core and high S/N in the larger surrounding Fizeau PSF structure; and a sequence of frames with only slight saturation of the innermost bright Fizeau fringes, to retain good visibility of the innermost dark Fizeau fringes. (In addition, entirely unsaturated frames should be taken to reconstruct the core of the host star PSF, as we have done in this work.) It may also be adviseable to have a greater separation of the two regimes during the reduction process. For example, separate PCA basis sets could be generated for different tesselation patterns, and may minimize effects such as the inversion seen in Fig. \ref{fig:cont_curve_1d_fizeau}.

It is always important to consider weather conditions, but it is particularly important for delicate observations such as these. To increase the S/N in the $\lambda/D$ regime, and to access angles in the $\lambda/B_{EE}$ regime, clearly one of the most critical aspects is the stability of the phase loop---and that requires good seeing. Poor seeing scrambles the coherence of interfered light, which degrades the visibility of the fringes on the phase detector to below the level required by the phase software loop. How ``good'', then, does good seeing have to be to keep the phase loop closed?

\begin{centering}
\begin{figure}
\includegraphics[width=1\linewidth]{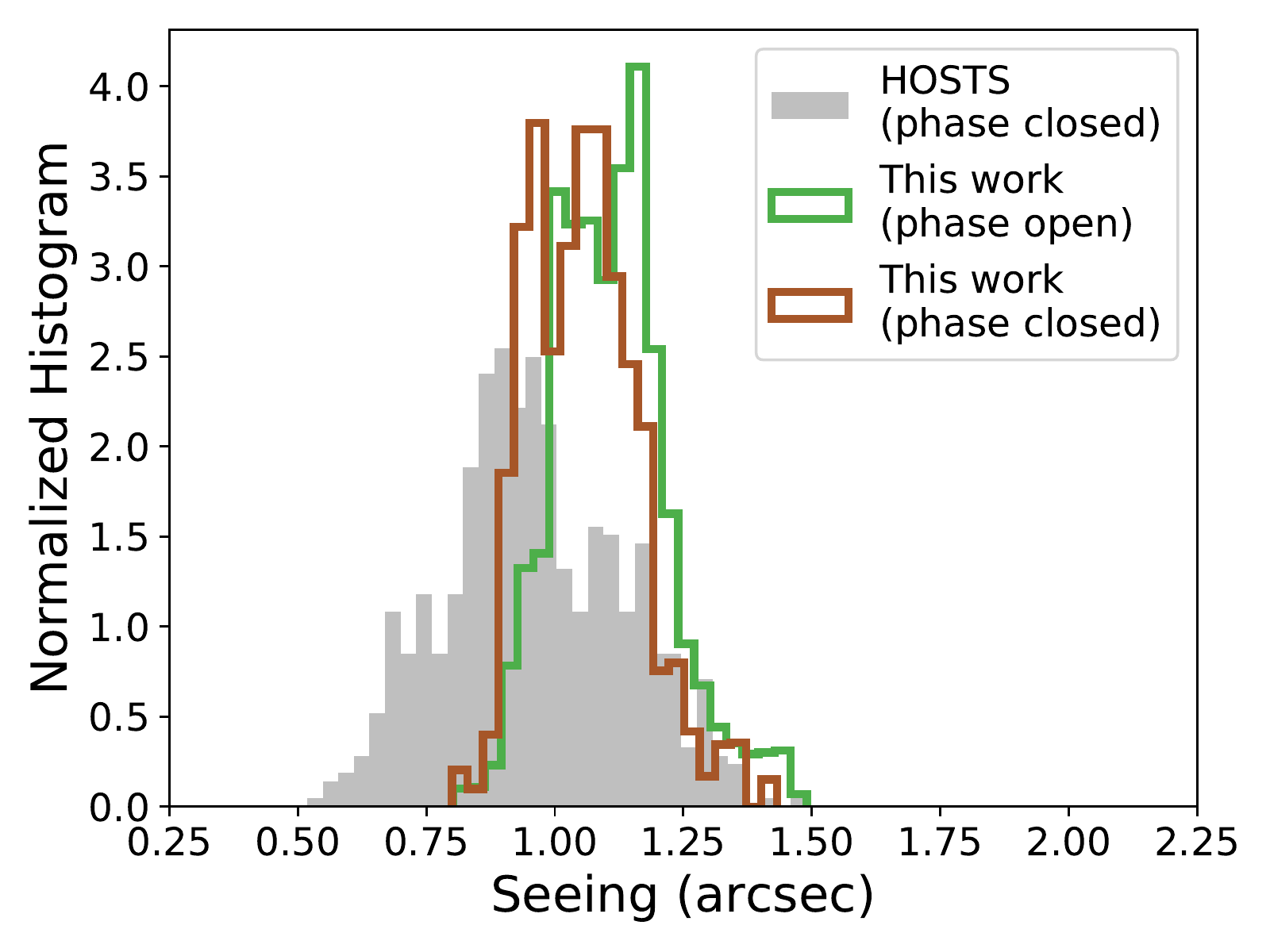} \\
\caption{Seeing values, as compared between this observation and the HOSTS survey. Seeing values from this work are from each individual frame. Seeing values from HOSTS correspond to blocks of frames, each spanning a few minutes of observing time.} 
\label{fig:seeing_hist}
\end{figure}
\end{centering}

Thus far, the great majority of LBTI observations taken with a closed phase loop have been for the HOSTS survey of exozodiacal dust disks \citep{ertel2020hosts}. In Fig. \ref{fig:seeing_hist} we compare seeing values during HOSTS observations\footnote{Available at \url{http://lbti.ipac.caltech.edu/}} with those taken in this work. HOSTS observations were either scheduled classically, or in queue mode when seeing was generally below 1.2''. Regardless of how a stretch of closed-loop HOSTS observations began, the stochastic nature of weather conditions on Mt. Graham conspired to add considerable diversity in the seeing values. No closed-loop HOSTS sequences were successful with seeing of $\gtrsim1.5$'', and this should be considered a prohibitive level of turbulence for phase control with the current instrument configuration.

\begin{centering}
\begin{figure*}
\includegraphics[width=1\textwidth,trim=5cm 0cm 5cm 0cm,clip=True]{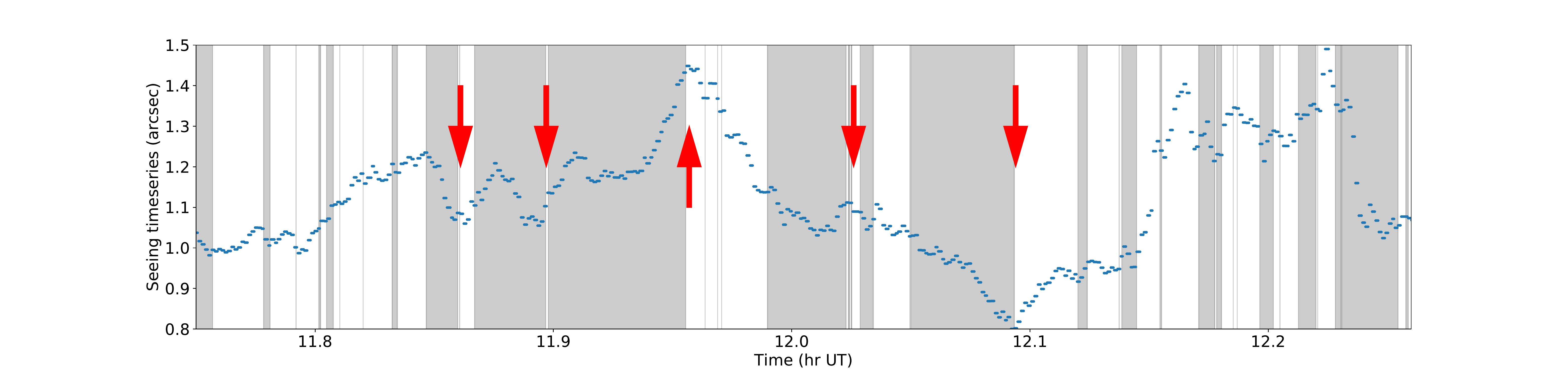} \\
\caption{A detail of the seeing. Grey regions indicate time during which the phase loop was closed. White regions indicate an open phase loop, but do not necessarily indicate that closing the phase loop was impossible---some time must be spent to perform optical realignments, and some time is spent waiting for the seeing to improve before the next re-attempt. Of significance here are the moments when the phase loop opened after the phase loop had been closed for at least several seconds, and coincident with a upward or downward change in the seeing (arrows). Note this can occur even when the seeing begins to decrease, or when the seeing is below 0.9''.}
\label{fig:seeing_arrows}
\end{figure*}
\end{centering}

However, Fig. \ref{fig:seeing_hist} also shows that the majority of our Altair observation was not during prohibitively poor seeing, and only indicates a weak dependence of the phase loop on the seeing value alone. Some other source of instability seemed to be present which forced the phase loop to open. In these observations of Altair, the seeing zigzagged with a typical amplitude of 0.1-0.2'' on a timescale of 1-3 min. Sometimes the phase loop opened at sudden changes in the seeing---at jumps \textit{and} drops (see Fig. \ref{fig:seeing_arrows}). The phase loop openings coinciding with shifts in the \textit{trend} of the seeing might mark the passage of boundaries between large turbulence pockets. It may be advantageous to restrict phase-controlled observations to periods in which seeing is smooth as a function of time, in addition to being low. The astronomer cannot control the seeing, but the Italian LBT partner Istituto Nazionale di Astrofisica (INAF) is developing the online Advanced LBT
Turbulence and Atmosphere (ALTA) Center,\footnote{\url{http://alta.arcetri.inaf.it/}} which features a tool for predicting the nightly temporal evolution of various LBT-specific environmental parameters including the seeing \citep{masciadri2019new}. The delicacy of our phase loop closure demonstrates the importance of a reliable predictive tool for the summit of Mt. Graham, and we encourage the use of such a tool in the future.

Since the LBT is an ALT-AZ telescope with sub-telescopes always perpendicular to the elevation mount, the LBT cannot freeze the angle of the long LBT baseline relative to an arbitrary object for long integrations. An observer planning to sample certain baselines around an object will have to consider the time-dependent parallactic angle, and the heaviest sampling will occur when the angular change is slow. Accessible elevations are bounded by a hard lower telescope elevation limit of $30^{\circ}$, and a higher limit of $\gtrsim80^{\circ}$ due to AO tip-tilt instability as the mount swings around rapidly during the meridian transit of the object (A. Vaz, pers. comm.).

The analysis by \citet{patru2017_i} offers important guidance for considering how to sample the object as it rotates, given the morphology of LBT Fizeau PSF. For example, along the long LBT baseline is the highest-frequency information, but for a given parallatic angle rotation, planet PSFs will drift through more of the dark host star fringes if they lie in the direction along the short baseline. Observing programs could be designed such that the LBT could move from target to target and back, so as to take snapshots of the long baseline at substantially different angles over multiple objects. For a given snapshot, the closest-in Fizeau fringes would offer the best contrasts at those angles, but the greatest amount of ``drift'' of a companion through dark fringes would happen if it moves perpendicular to them, at a wider angle from the star. 

There are also mechanical considerations or upgrades to be made to LBTI. During the observations in this work, we nodded the Fizeau PSF on the detector a single time to avoid incurring any more PSF instability or manual realignments. After this dataset was taken, sets of lenses were installed at staggered radii in a filter wheel upstream of the Phasecam detector. This could facilitate rapid side-to-side nodding in Fizeau mode by shifting the illumination on Phasecam with a wheel movement, and then nodding the telescope mount to shift the PSF on the detector and recovering the co-alignment. This could enable more dithering on the detector. Some on-sky tests of this technique have been performed, but it remains to be fully commissioned. 

Phasecam itself was originally designed to target the bright stars of the HOSTS survey, and the detector readouts have noise at levels that can be prohibitive for dim targets. Furthermore, it had to read out at a rate that was pegged to the cadence of corrections from the telescope's vibration-monitoring system \citep{bohm2016ovms} at ${\geq}500$ Hz, and the only ``ground truth'' measurement of path length changes was the illumination on the Phasecam detector. The absence of any independent measurement of mirror movements between the snapshots of the phase added phase noise, in addition to the phase noise already inflicted by the atmosphere and telescope vibrations. These factors conspire to limit phase control to bright stars of $m_{K}\lesssim4.7$. However, there are plans to upgrade Phasecam in the near future with a SAPHIRA array \citep{goebel2018overview} with sub-e$
^{-}$ read noise, to install capacitive sensors behind the pathlength corrector mirrors to measure mirror movements, and to develop the software to enable decoupling of the cadences of Phasecam and the telescope vibration-monitoring system. These upgrades are predicted to enable the phase loop to close on stars $m_{K}<9$ (J. Stone, pers. comm.). It remains to be seen how an upgraded phase-sensing detector will perform in unstable seeing as in Fig. \ref{fig:seeing_arrows}, but the target list for phase-controlled Fizeau observations will increase dramatically, and will include dimmer targets for which other improvements described in this section will be important.

The science camera LMIRcam has already been upgraded in the time since the observations taken for this work. The detector remains the same, but the ISDEC controller has been replaced with a MACIE interface card \citep{loose2018acadia}. By enabling a switch from USB 2 to 3, this has led to an order-of-magnitude increase in the data transfer speed, to 4.8 Gbps. The new MACIE also has a programming interface that allows more readout mode flexibility.

The MACIE electronics enables both of the native pixel readout rates of the LMIRcam detector: ``Slow'' (100 to 500 kHz) and ``Fast'' (5 MHz). Trade-offs between these modes include frame time, read noise, dark current, ADC bit depth, and available subarray regions. The USB3 connection allows image data to be streamed continuously to the host computer's memory buffer and to be saved to disk without any data buffering overheads, significantly increasing on-source efficiency over the previous implementation. In addition, multiple non-destructive read frames can be acquired ``up-the-ramp'' for each integration to allow slope fits and reduce read noise. This also enables a substantial increase in the proportion of time spent integrating on a target relative to the total time the detector is active. In the 5 MHz pixel rate readout mode, for example, frame times are 28 msec/frame, and integrations and readouts of the full 2048$\times$2048 array, with 10 nondestructive readouts per frame, would have an integration time efficiency of above 90\%. Even after slope fitting effectively brings this down to $\approx$80\%, this is much more favorable than the $\approx$10\% efficiency in this work (see Table \ref{table:strip_frame_subsets}), which only used one-quarter of the readout array.

It may be even more economical to use subarrays which are smaller than the full detector, though a sufficiently large footprint is required around the host star to determine the background level around bright stars (see Appendix \ref{subsec:prepping}). As a general guideline, we encourage observers to generate and reduce simulated datasets as they design their observing strategy, so as to make quantitative studies of the tradeoffs of various LMIRCam or NOMIC readout modes for their particular science program, and to make maximum use of precious observing time.

On the software side, the phase loop can also be improved to increase the number of frames with good phase control. Phasecam currently uses only $Ks$-band light for the phase correction. As mentioned in Sec. \ref{subsec:observations}, there are occasional phase jumps where the unwrapped phase skips an integer multiple of a $Ks$-wavelength in path length, and introduces a phase error on the Fizeau PSF on the science detector.  This happens frequently in unstable seeing. Until now, phase jump corrections have been made manually by the observers, but efforts are underway to supplement the phase loop with $H$-band light to swiftly and automatically correct these jumps \citep{maier2018two,maier2020implementing}. 

Work has also been performed to develop a ``correction loop'' to supplement the phase loop \citep{spalding2018fizeau,spalding2019status}. The correction loop would automate alignments and use cutouts of the science detector readouts in real time to remove non-common path aberrations (NCPA) between the phase and science detectors. Some initial on-sky tests of alignment algorithms have been performed, but again, this loop remains to be fully commissioned. In this work we have tried to account for phase and other aberrations in the science PSF by PCA-decomposing each frame, but sensitive future observations in the $\lambda/B_{EE}$ regime will require the Fizeau fringes to remain as frozen in place as possible, for as many frames as possible. Development of a correction loop to remove NCPA should continue.

Complimentary to improvements of the phase control would be a comparative study of post-processing techniques that could cope with loosened observing constraints. For example, in the $\lambda/D$ regime, phase closure may be an overly restrictive requirement in the reduction. The phase jitter of the Fizeau fringes will add some noise to the background, and the study by \citet{patru2017_ii} of the effects on the Fizeau PSF structure and a number of merit functions is an important step. Given the spatial heterogeneity of the potential contrast gain in the Fizeau PSF pointed out by \citet{patru2017_i}, what is needed next is fake planet recovery in synthetic datasets with very dense ``seeding'' of fake planets throughout the stellar PSF structure.

\section{Conclusion} 
\label{sec:consclusions}

In this work we have performed high-contrast imaging with the LBT Fizeau PSF on the star Altair, one of the nearest early-type stars. The inclination of the star is low enough that our contrast curves sample all phases of circular orbits in the habitable zone normal to the stellar spin axis.  

The data reduction was separated into two regimes: one at radii $\sim \lambda/D$ from the host star, as is traditionally done in high-contrast imaging with AO-corrected PSFs; and one at radii $\sim \lambda/B_{EE}$, which is unique to the LBT.

Even with AO correction, the Fizeau PSF exhibits unique aberrations of differential tip, tilt, and phase. We carried out an analyses which accommodates those additional degrees of freedom and treats every detector readout individually. In this manner we have been able to use 57\% of the science frames. 

These observations have a sensitivity for 1.0 Gyr down to $\approx0.5\textrm{ M}_{\odot}$ at large radii, roughly at the K- to M-type transition. More significantly, we place constraints on companions of masses $\approx$1.3 $M_{\odot}$ at 0.15'' along the long baseline of the LBT. Constraints are similar along the short baseline, with a slightly better performance on the long baseline at most angles $\lesssim0.25$''.

Future users of the LBT Fizeau mode should take into account the strong trade-off in sensitivity between the two radial regimes: signal-to-noise at radii $\sim \lambda/D$ will be maximized with hard saturation of the Airy core, but the innermost Fizeau fringes at radii $\sim \lambda/B_{EE}$ will be preserved with shorter integration times.

In either regime, the most obvious mechanical need is for the ability to close the phase loop with minimal interruptions or down time, and thereby maximize the number of useable readouts. This would also allow integration times longer than the atmospheric coherence timescale. It is therefore important to take these observations in good and stable seeing. 

When the PSF stability is good enough and a more exacting selection can be made of the data, further improvement may be sought by using the merit criteria of \citet{patru2017_ii}. When the sensitivity becomes competitive with pre-existing filled-aperture datasets, data reductions with fake planet PSFs dispersed in more locations around the host star will also allow testing of the contrast gain maps of \citet{patru2017_i}.

\acknowledgments

We acknowledge Jordan Stone's assistance during the LBTI observations that went into this work. We also thank the dedicated LBT support personnel for all their assistance to make an operation like the LBT possible, and to Steve Ertel, Tomas Stolker, Kevin Wagner, and George Rieke for insightful discussions. The LBTI itself is funded by the National Aeronautics and Space Administration as part of its Exoplanet Exploration Program, and LMIRCam was funded by NSF grant 0705296.

This material is based upon High Performance Computing (HPC) resources supported by the University of Arizona TRIF, UITS, and Research, Innovation, and Impact (RII) and maintained by the UArizona Research Technologies department. We also gratefully acknowledge a special project allocation from the UA Research Computing HPC.

The material in this paper is also based upon work supported by CyVerse cyberinfrastructure, which is funded by the National Science Foundation under Award Numbers DBI-0735191,  DBI-1265383, and DBI-1743442. URL: \url{www.cyverse.org}. We also thank CyVerse for support to attend the 2018 CyVerse Container Camp. 

Thanks go to Blake Joyce and Julian Pistorius for their assistance, which was made possible through the University of Arizona Research Technologies Collaborative Support program, and Blake and Julian's support of PhTea and Hacky Hour.

This made use of the SIMBAD database, operated at CDS, Strasbourg, France \citep{wenger2000simbad}; and the SVO Filter Profile Service (\url{http://svo2.cab.inta-csic.es/theory/fps/}) and VOSA, developed under the Spanish Virtual Observatory project, supported by the Spanish MINECO through grant AYA2017-84089. VOSA has been partially updated by using funding from the European Union's Horizon 2020 Research and Innovation Programme, under Grant Agreement nº 776403 (EXOPLANETS-A).

This publication also made use of data products from the Two Micron All Sky Survey, which is a joint project of the University of Massachusetts and the Infrared Processing and Analysis Center/California Institute of Technology, funded by the National Aeronautics and Space Administration and the National Science Foundation.

The authors also wish to recognize and acknowledge the significant cultural role and reverence that the summit of Mt. Graham (in Apache \textit{Dzi{\l} Nchaa Si'an}, ``big seated mountain'') is held within the indigenous White Mountain and San Carlos Apache communities.

Part of this work has been carried out within the framework of the National Centre of Competence in Research PlanetS supported by the Swiss National Science Foundation. SPQ acknowledges the financial support of the SNSF.

KM's work is supported by the NASA Exoplanets Research Program (XRP) by cooperative agreement NNX16AD44G.

ES was supported by the Ed and Jill Bessey Scholarship in Astrobiology, and is currently supported by Jeff Chilcote under NSF MRI grant 1920180.

\section{ORCID iDs}

\begin{footnotesize}
\noindent
Eckhart Spalding {\includegraphics[scale=0.15]{orcid_64x64.png}} \url{https://orcid.org/0000-0003-3819-0076} \\
Phil Hinz {\includegraphics[scale=0.15]{orcid_64x64.png}} \url{https://orcid.org/0000-0002-1954-4564} \\
Katie Morzinski {\includegraphics[scale=0.15]{orcid_64x64.png}} \url{https://orcid.org/0000-0002-1384-0063} \\
Jared Males {\includegraphics[scale=0.15]{orcid_64x64.png}} \url{https://orcid.org/0000-0002-2346-3441} \\
Michael Meyer {\includegraphics[scale=0.15]{orcid_64x64.png}} \url{https://orcid.org/0000-0003-1227-3084} \\
Sascha Quanz {\includegraphics[scale=0.15]{orcid_64x64.png}} \url{https://orcid.org/0000-0003-3829-7412} \\
Jarron Leisenring  {\includegraphics[scale=0.15]{orcid_64x64.png}} \url{https://orcid.org/0000-0002-0834-6140} \\
\end{footnotesize}

\facility{LBT/LBTI}

Other software: \texttt{Python} \citep{van1995python,oliphant2007python}, \texttt{matplotlib} \citep{hunter2007matplotlib}, \texttt{numpy} \citep{walt2011numpy}, \texttt{astropy} \citep{price2018astropy}, \texttt{scipy} \citep{2020SciPy-NMeth}, Singularity \citep{kurtzer2017singularity}, Singularity Hub \citep{sochat2017enhancing}, Binder \citep{jupyter2018binder}.

\appendix
\section{The reduction pipeline}
\label{sec:app_pca}

Compared to an Airy PSF, the Fizeau PSF has additional degrees of freedom in its aberrations. The amplitudes of these aberrations were also highly time-variable during this observation, which required the host star PSF to be modeled on a frame-by-frame basis in the reduction pipeline which we describe here. The functionality of the pipeline can be split into three main sections corresponding to those numbered in Fig. \ref{fig:kiboko_flowchart}.

\begin{center}
\begin{centering}
\begin{figure}
\begin{center}
\includegraphics[width=0.95\linewidth, trim= 0cm 11cm 7cm 4cm, clip=True]{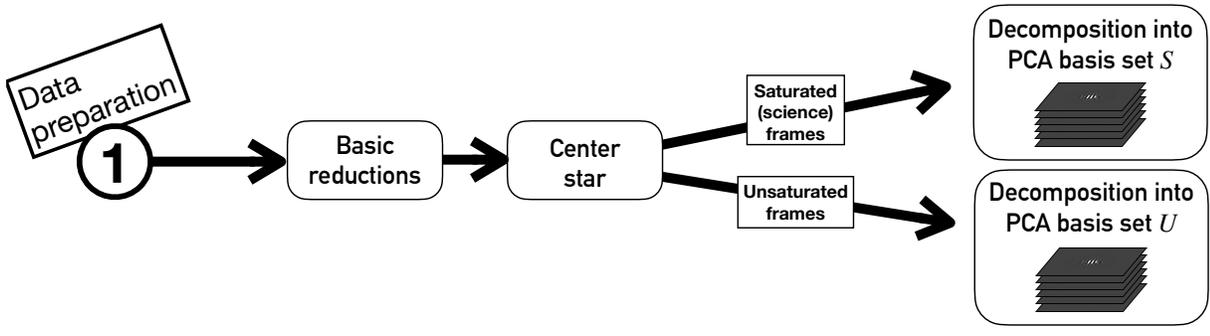}\\
\includegraphics[width=0.95\linewidth, trim= 0cm 5cm 7cm 5cm, clip=True]{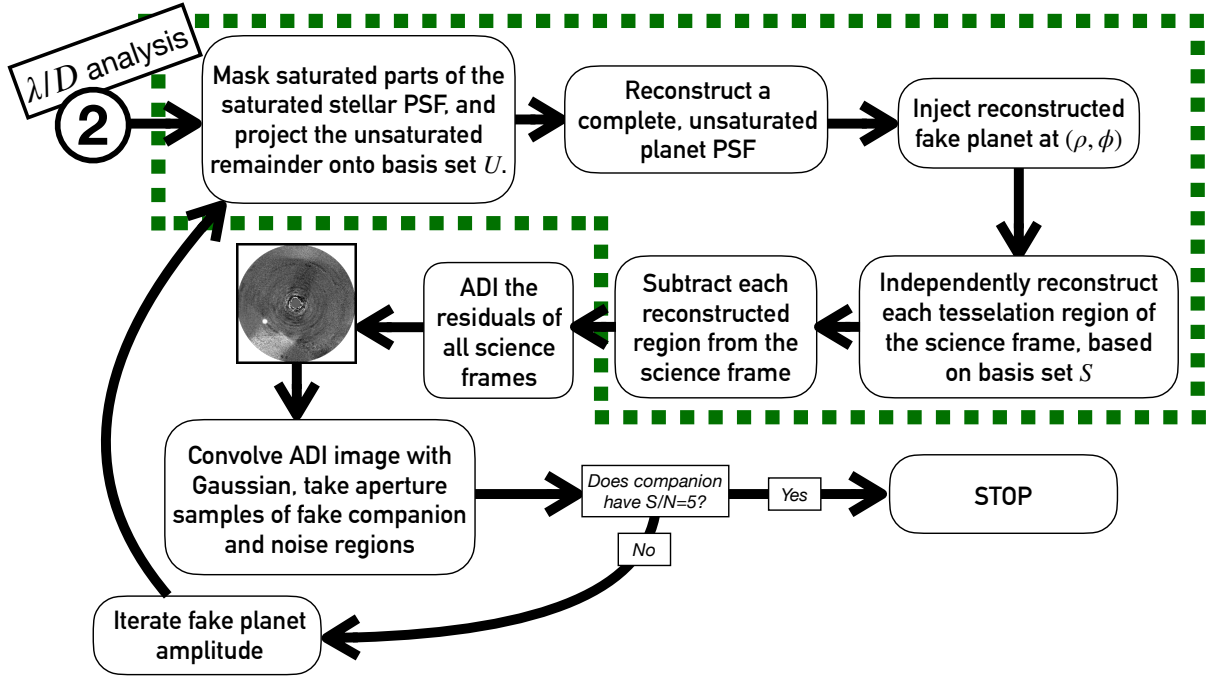}\\
\includegraphics[width=0.95\linewidth, trim= 0cm 4.5cm 7cm 9.5cm, clip=True]{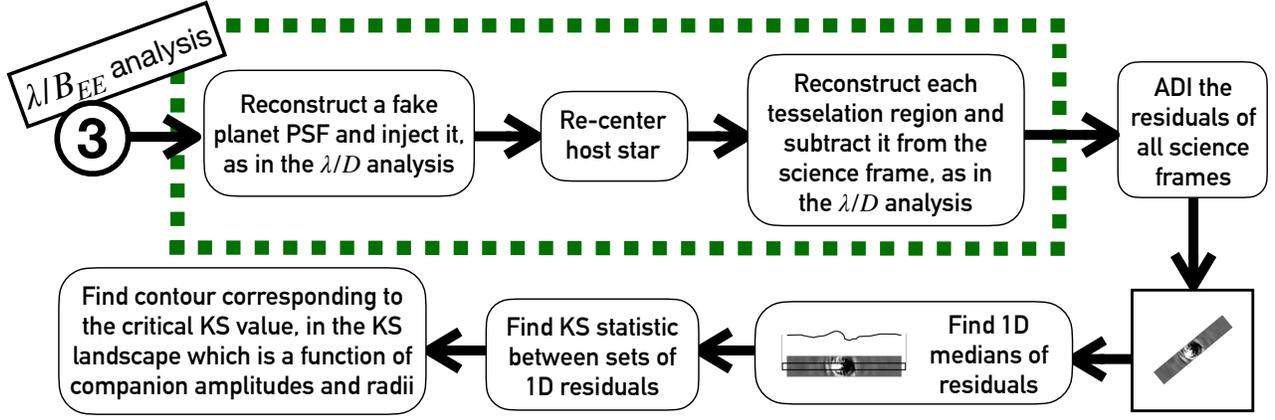}
\end{center}
\caption{The high-level functionality of the pipeline, as split into three sections which correspond to the text in Secs. \ref{subsec:prepping}, \ref{subsec:lambdaDregime}, and \ref{subsec:lambdaBregime}. Parts outlined in green dashed line are run on a frame-by-frame basis.} 
\label{fig:kiboko_flowchart}
\end{figure}
\end{centering}
\end{center}

\subsection{Section 1: Data preparation}
\label{subsec:prepping}

The standard reduction tasks included the subtraction of darks, the application of a bad pixel mask, and the scraping of meta-data from the FITS headers. The pipeline also subtracted out a spurious but consistent gradient in \texttt{y} on the detector, which may be connected to finite voltage settling times after the detector is reset following a readout.

The LMIRCam detector has 32 64-pixel-wide channels which independently and continually move up and down in bias due to tiny voltage changes. To subtract the changing bias and the thermal infrared background, we experimented with PCA decompositions of the background in the style of \citet{hunziker2018pca}. The basis set included modes from the background decomposition as well as 32 concatenated PCA modes corresponding to variations in each of the 32 channels. The latter modes were binary maps, with a value of 1 in a given channel and zero in the other channels. When reconstructing the background for subtraction, the star illumination was masked out, and the background underneath the star was reconstructed using the PCA basis set. We found that the best background subtraction simply involved decomposing the background with a 32-mode basis set consisting of only the binary channel modes, while masking the star with a circle of radius 70 pixels (7.4$\lambda/D$).

Next, the rough centers of the Fizeau/Airy PSFs were found by finding the location of the pixel with the maximum value after applying a Gaussian smoothing filter. From each detector readout a subarray equivalent to 4.3''$\times$4.3'' is cut out, and these are centered by fitting a Gaussian profile and shifting and spline-interpolating the images to a sub-pixel level.  (As mentioned in Table \ref{table:strip_frame_subsets}, for $\approx$70\% of the science frames, the cutout region stretched beyond one side of the readout image by 1.0 arcsec. The pipeline replaced that ``overshoot'' region with NaNs in the PSF cutouts.)

Cutouts of saturated (science) frames with filter configurations corresponding to Blocks A and D (as listed in Table \ref{table:blocks}) were used as a training set to generate a PCA basis set $S$ for saturated PSFs, with 100 modes. The same was done with cutouts of unsaturated frames from Blocks B and C to generate the basis set $U$ for unsaturated PSFs.

\subsection{Section 2: $\lambda/D$ analysis}
\label{subsec:lambdaDregime}

To begin the process of subtracting the host star PSF and injecting fake planets, a median science frame was subtracted from the stack of science frames. The host star was then decomposed twice: with PCA basis set $S$, corresponding to the residuals of a saturated PSF to enable best host star subtraction; and basis set $U$ to generate fake planet PSFs. For the latter, the saturated regions of the science frame (defined as pixels where detector counts ${>}55$k) are masked during the decomposition, so as to reconstruct the full PSF within the saturated regions.  (However, the saturation only involved the innermost bright lobes of the Fizeau PSF so as to leave the innermost dark Fizeau fringes unsaturated.) The full frames of the residuals are then projected onto the basis set $U$. The host star amplitude was determined by taking an ADI image of the reconstructed host star PSFs, convolving the image with a Gaussian kernel, then taking the maximum value. A fake planet generated from the reconstruction $U$ was then injected at a given starting amplitude and radius and angle from True North ($\rho,\phi$). It is worth emphasizing that as the host star PSF changes from one frame to the next, the shifting of the Fizeau fringes and other PSF distortions are stamped into the fake planet PSF for that frame, just as the aberrations between a host star and true planet PSF would be correlated.

\begin{centering}
\begin{figure}
\includegraphics[width=0.45\linewidth, trim=2cm 0.5cm 2cm 0cm, clip=True]{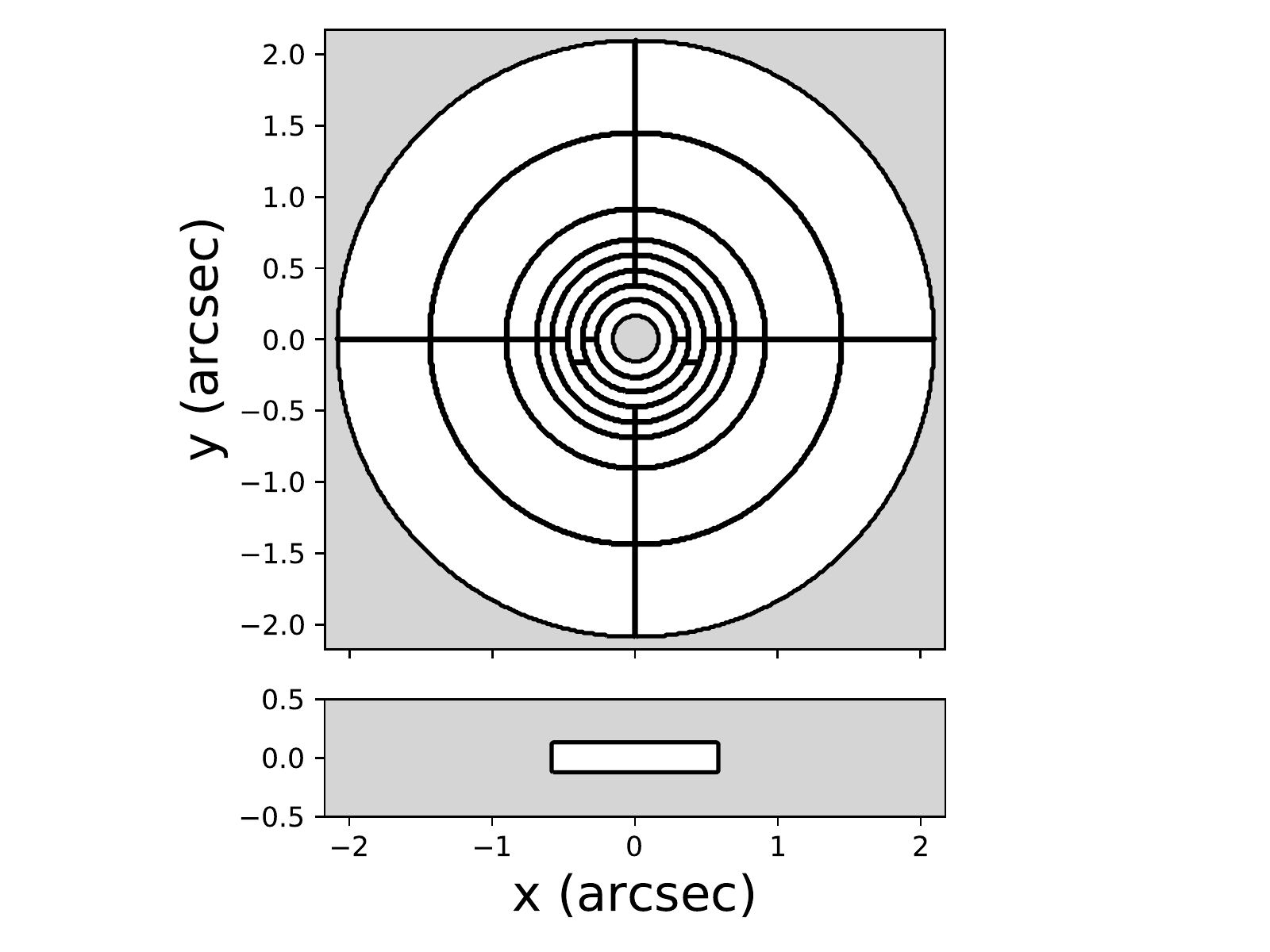} \includegraphics[width=0.45\linewidth]{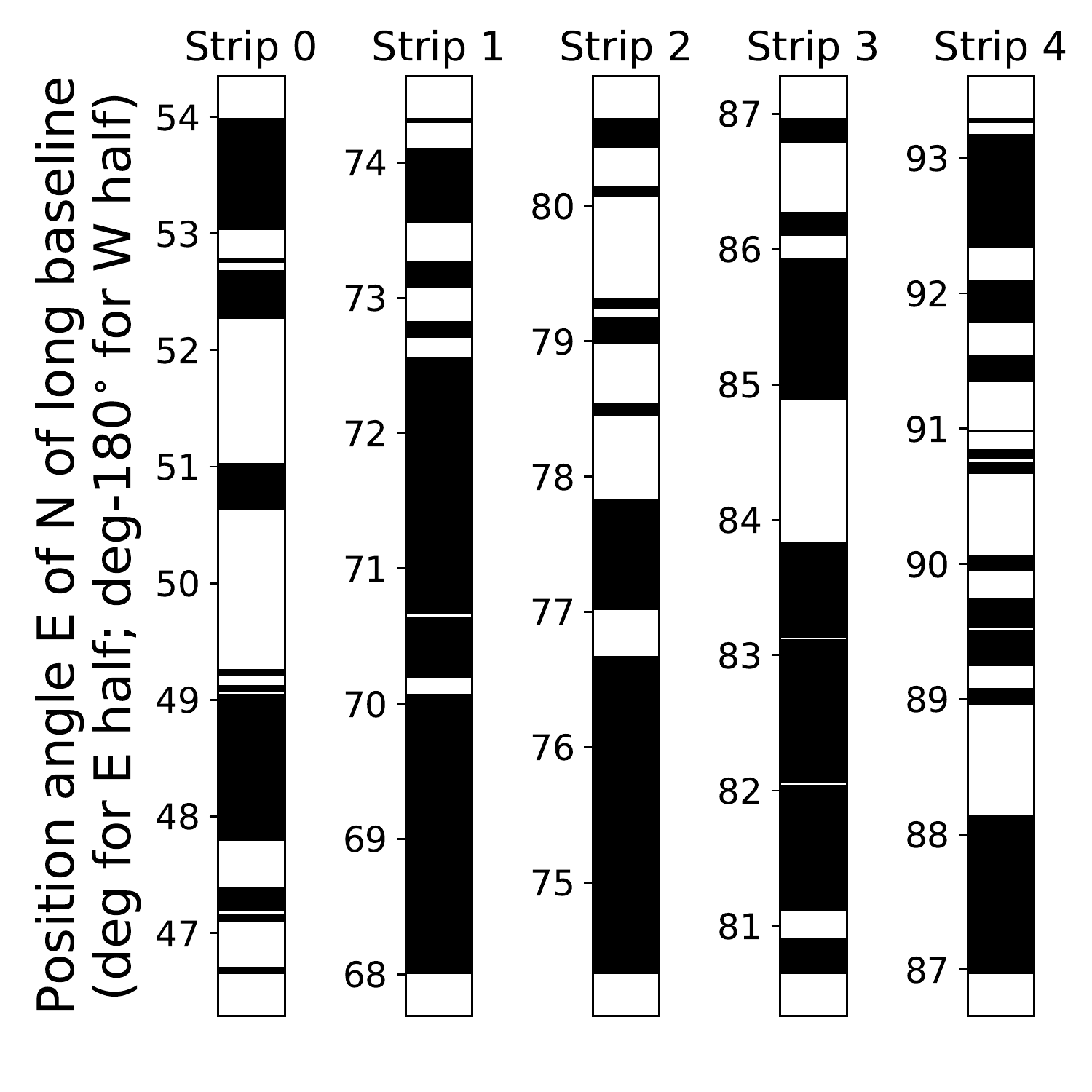}
\caption{Left, top: The tesselation pattern for PCA-decomposing individual regions in the $\lambda /D$ regime for subtracting the host star. Grey areas are not involved in the reduction. Left, bottom: One of two tesselation patterns for the $\lambda/B_{EE}$ regime, consisting of a single strip across the long Fizeau baseline and includes the host star PSF. (The other pattern is the same, except rotated by 90$^{\circ}$ so as to place the long axis along the short LBT baseline.) Right: Sampled position angles (black regions) swept around the object by the long LBT baseline, for each of the five strips used in the $\lambda/B_{EE}$ analysis. White regions correspond to where no data was taken. For example, for the data that comprised Strip 0, there is continuous Fizeau coverage around Altair of position angles between 53$^{\circ}$ to 54$^{\circ}$ East of North (and by symmetry, 233$^{\circ}$ to 234$^{\circ}$ East of North), and no data at all from 51$^{\circ}$ to 52$^{\circ}$ (nor 231$^{\circ}$ to 232$^{\circ}$).} 
\label{fig:tesselation}
\end{figure}
\end{centering}

All of the preceding describes the process of preparing frames and injecting a first round of fake planets for both the $\lambda/D$ and $\lambda/B_{EE}$ analyses, but it is at this point that the two analyses split into two strictly separate treatments. To carry out the PCA decompositions for subtracting the host star, each science frame for the $\lambda/D$ analysis was split into a tesselation pattern, where regions were sized to be at least 10$\times$ the area $\pi (\textrm{FWHM}/2)^{2}\approx74$ pixels$^{2}$, the area of the footprint of an Airy disk (Fig. \ref{fig:tesselation}). This was done to optimize the subtraction of the host star PSF in individual regions, but without overfitting to the basis set \citep{lafreniere2007new}. Each region was independently projected onto the basis set $S$ and the resulting reconstruction was subtracted from that region. The outside edges and the centermost circle do not correspond to regions for decomposition in the $\lambda/D$ regime and are ignored. Following the subtraction of every region from each frame, the frames were derotated according to their parallactic angles, and a median was taken across them to yield a single ADI frame. (Since the LBT is an ALT-AZ telescope and LBTI has no derotator, the PSF does not rotate with the sky. Thus the PSF decomposition must be done before derotating the frames.) 

For a final ADI frame, the signal-to-noise of a fake planet was calculated as the maximum counts within a FWHM-sized aperture around the fake companion, divided by the empirical standard deviation of counts sampled from a string of ``necklace-bead'' patches along an annulus at the same radius of the fake companion. The centers of these patches were separated from each other by a circumferential distance of FWHM, and excluded the companion itself (see \citet{mawet2014fundamental}). Each noise patch was defined to contain the pixels within a radius of 0.75 pixel of the calculated center of the patch. The value of counts from the patch was taken to be the median value of counts among the pixels in that patch. The pipeline iterated the fake planet amplitudes and repeated reductions of each dataset until the signal-to-noise of a companion converged to 5. 

\subsection{Section 3: $\lambda/B_{EE}$ analysis}
\label{subsec:lambdaBregime}

At radii corresponding to the innermost Fizeau fringes ($\lambda/D\gtrsim\rho\gtrsim\lambda/B_{EE}$), information is washed out if the PSFs are all derotated and medianed. The degrees of freedom for placing a companion around the host star also diminish, to the point where radii from the star become small enough that the circumference at that radius is $\sim$1 FWHM.

To treat this regime, we injected fake planets in the frames as in Sec. \ref{subsec:lambdaDregime}, but with a number of key differences. Firstly, fake planets are injected corresponding to a pre-set grid of (amplitude, radius) space, without converging on an amplitude iteratively. Secondly, reductions are performed with subsets of frames which span narrow ranges of parallactic angle. Thirdly, the tessellation of the science frame, which is used to project regions onto basis sets, consisted of a single rectangle with a long axis along the long LBTI baseline, and a second identical region along the short LBT baseline. (See Fig. \ref{fig:tesselation} and Sec. \ref{subsubsec:regime_fizeau}.) By design there was not much angular rotation in each reduction, so the residuals of the resulting ADI frame had a footprint similar to that of the tessellation region.

Fourthly, we chose a different way of quantifying the residuals after host star subtraction. We generate 1D residuals by taking the median along the 7-pixel-wide short axis of the tesselation region (the strip), to generate a 1D set of residuals corresponding to the long axis. The 7-pixel width was chosen to be wide enough so as to take a median over several pixels and produce a 1-D set of residuals as a function of radius which does not exhibit pixel-to-pixel noise; but still narrow enough to be unaffected by the loss of effective integration time along the edges of the strip following ADI. Sets of 1D residuals are compared with each other as long as they are not immediately adjacent to each other (unless the strip of interest is 0E, which is separated from the other strips by a larger position angle). We do not compare half-strips pointed in different cardinal directions. This reduces the parallactic angle diversity among the compared strips, but avoids the misidentification of quasistatic asymmetries in the residuals around the Fizeau PSF as a ``companion.''

For example, the strips along the baseline pointing closest to East are denoted 0E, 1E, 2E, 3E, and 4E. If a fake planet was injected along the position angle of half-strip 3E, then comparisons are made between the 1D residuals of 3E and each of the half-strips 0E and 1E. The half-strips 2E and 4E are excluded, because they are immediately adjacent to 3E and contain a significant amount of the fake planet PSF centered in strip 3E that it would be comparable to making a comparison between the same 1D residuals---which would act to decrease sensitivity to companions along strip 3E. If, on the other hand, a fake planet was injected along the position angle of half-strip 0E, then the residuals are compared with each of 1E, 2E, 3E, and 4E.

Finally, we use a different statistical test than in the $\lambda/D$ regime for deciding whether a companion has been detected. For the $\lambda/B_{EE}$ regime we used the Kolmogorov-Smirnov (KS) two-sample test to compare pairs of the 1D residuals. This KS two-sample test either accepts or rejects the null hypothesis $H_{0}$ that the residuals do not come from different parent distributions \citep{conover1971statistics}. The KS statistic is the maximum difference between the two cumulative distribution functions (CDFs), or

\begin{equation}
D = \textrm{max}\left|S_{n}(x)-S_{m}(x)\right|
\end{equation}

\noindent
where we use $S_{n}$ (of length $n$) and $S_{m}$ (of length $m$) to denote the two CDFs of the sets of empirical residuals.

The probability of this value being less than a critical value converges to the Kolmogorov-Smirnov cumulative distribution function $L(z)$ as in \citet{feller1948},

\begin{equation}
\textrm{Prob}\left\{D \leq z \sqrt{\frac{m+n}{mn}}\right\} \longrightarrow L(z) \equiv 1-2 \sum_{\nu=1}^{\infty}(-1)^{\nu-1}e^{-\nu^{2}z^{2}}
\end{equation}

\noindent
We reject $H_{0}$---that the distributions are indeed from the same population---with a confidence of 95\% at a value of $z=1.358$ (e.g., \citet{smirnov1948}).

We compare the residuals of all the half-strips with each other to generate KS statistic distributions across the grid of injected planet parameters (i.e., the grid of amplitude and radius from the host star). A median is taken across the distributions from all half-strip comparisons, and the critical value contour of the median landscape is considered to be the contrast curve. Beyond that contour, where $D>z\sqrt{(m+n)/mn}$ from a comparison of two strips (one of which has a planet injected along its median angle), the hypothesis $H_{0}$ is rejected. In this analysis, we compare samples of the values of pixels along radii of $n=m=50$ pixels around the host star, so the critical value is $D_{crit}=z\sqrt{(m+n)/mn}=1.358\sqrt{2/50}=0.2716$

\section{Synthetic photometry of Altair}
\label{sec:abs_mag}

To find the magnitude of Altair at wavelengths relevant for this work, we write the apparent magnitude of a star on the Vega scale in terms of the mean flux and a zero point:

\begin{equation}
m_{\textrm{star}} = -2.5\textrm{log}_{10}(\langle f_{\lambda}\rangle_{P}) + 2.5\textrm{log}_{10}(\textrm{zp}_{P}(f_{\lambda,\textrm{Vega}}^{(0)}))
\end{equation}

\noindent
where the $P$ subscript makes it explicit that these quantities are to be calculated for a photon detector like LMIRcam, not an energy detector. The zero point for an energy detector (subscript $E$) for the Paranal NACO NB405 filter is defined by the Virtual Observatory SED Analyzer (VOSA) \citep{bayo2008vosa} as

\begin{equation}
\textrm{zp}_{E}(f_{\lambda,\textrm{Vega}}^{(0)})\equiv \frac{ \int d\lambda R_{\lambda}(\lambda)f_{\lambda,\textrm{Vega}}^{(0)}(\lambda)}{\int d\lambda R_{\lambda}(\lambda)}
\label{eqn:zpE}
\end{equation}

\noindent
with units of $\textrm{erg}\cdot\textrm{cm}^{-2}\textrm{s}^{-1}\angstrom^{-1}$. We calculated this zero point using the Vega spectrum from the python package \texttt{pysynphot}, checked the result with the SVO value, and then calculated our own photon detector zero point as

\begin{equation}
\textrm{zp}_{P}(f_{\lambda,\textrm{Vega}}^{(0)})\equiv \frac{ \int d\lambda R_{\lambda} \lambda f_{\lambda,\textrm{Vega}}^{(0)}}{\int d\lambda R_{\lambda} \lambda}
\end{equation}

\noindent
where we have dropped the $(\lambda)$ arguments to compactify the notation. The average photon flux of a star at the top of the Earth's atmosphere is the average flux at the surface of the star, scaled for distance:

\begin{equation}
\langle f_{\lambda} \rangle _{P} = \frac{\int d\lambda R_{\lambda} \lambda  f_{\lambda,\textrm{star}}^{(0)} }{\int d\lambda R_{\lambda} \lambda } = \left( \frac{R_{\textrm{star}}}{D} \right)^{2}\frac{\int d\lambda R_{\lambda} \lambda  f_{\lambda,\textrm{star}}^{(\textrm{surf)}} }{\int d\lambda R_{\lambda} \lambda }  
\end{equation}

For the stellar spectrum, we used a Kurucz synthetic spectrum with specifications T\textsubscript{eff}=7750 K, log($g$)=4.0, and [Fe/H]=0 \citep{castelli1997notes}. This is to mimic Altair with T\textsubscript{eff}=7550 K, log($g$)=4.13, and [Fe/H]=-0.24 \citep{erspamer2003automated}, though the science wavelengths are well into the Rayleigh-Jeans regime for an star of this temperature.

Putting everything together, we find

\begin{equation}
M_{ \textrm{star}} = -2.5 \textrm{log}_{10} \left\{ \left( \frac{R_{\textrm{star}}}{D} \right)^{2}\frac{\int d\lambda R_{\lambda} \lambda f_{\lambda,\textrm{star}}^{(\textrm{surf)}} }{\int d\lambda R_{\lambda} \lambda f_{\lambda,\textrm{Vega}}^{(0)}}  \right\}   - 5\textrm{log}_{10}\left( \frac{\textrm{d}}{\textrm{10 pc}} \right) = 1.87
\label{eqn:final_M}
\end{equation}

\begin{centering}
\begin{figure}
\includegraphics[width=\linewidth]{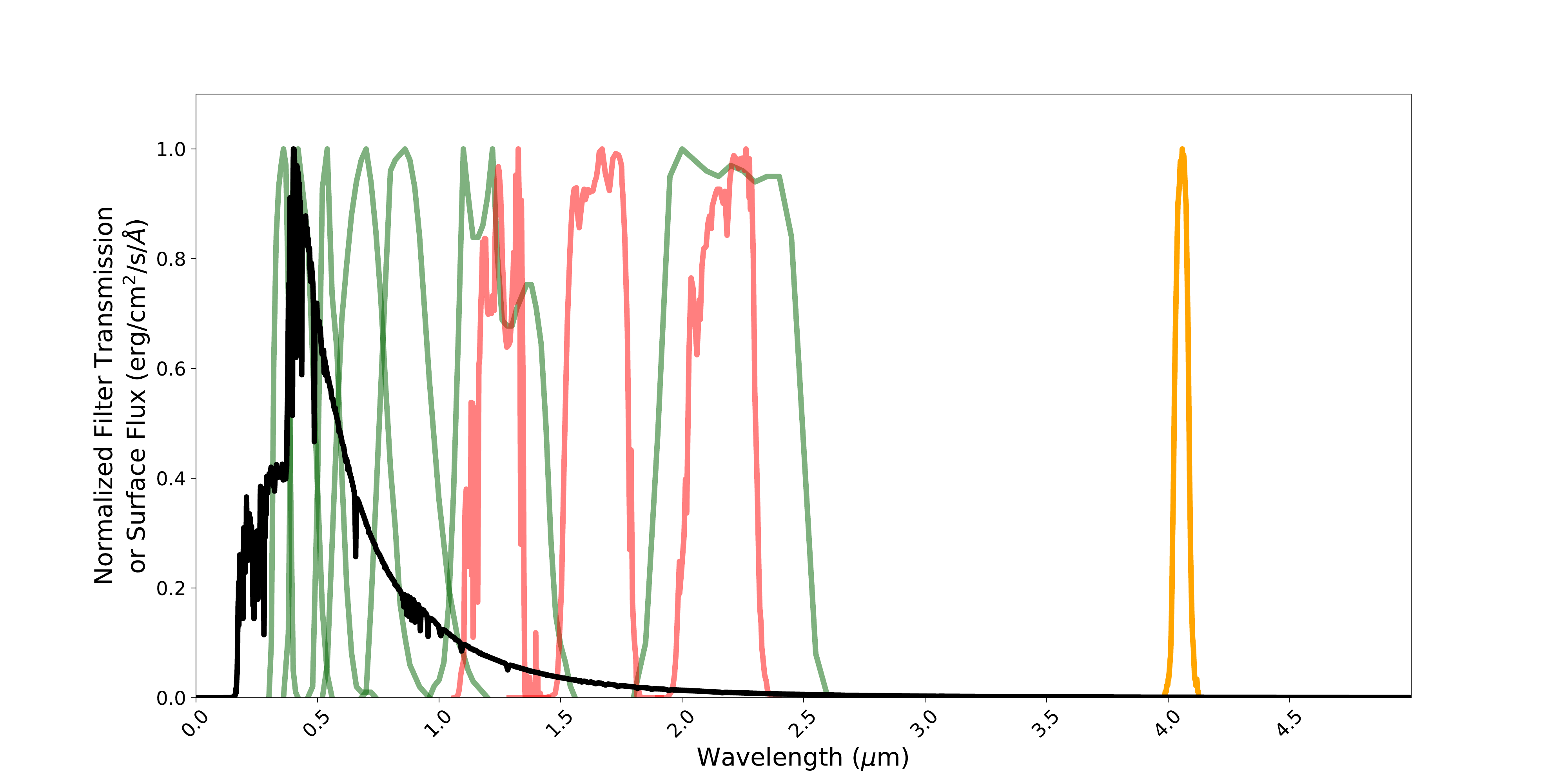}
\caption{The Johnson $UBVRIJK$ (green), 2MASS $JHKs$ (red), and NB4.05 (yellow) filters and the model Altair spectrum} 
\label{fig:filters_2_check}
\end{figure}
\end{centering}

\begin{figure}
\begin{floatrow}
\ffigbox{%
 \includegraphics[width=1\linewidth]{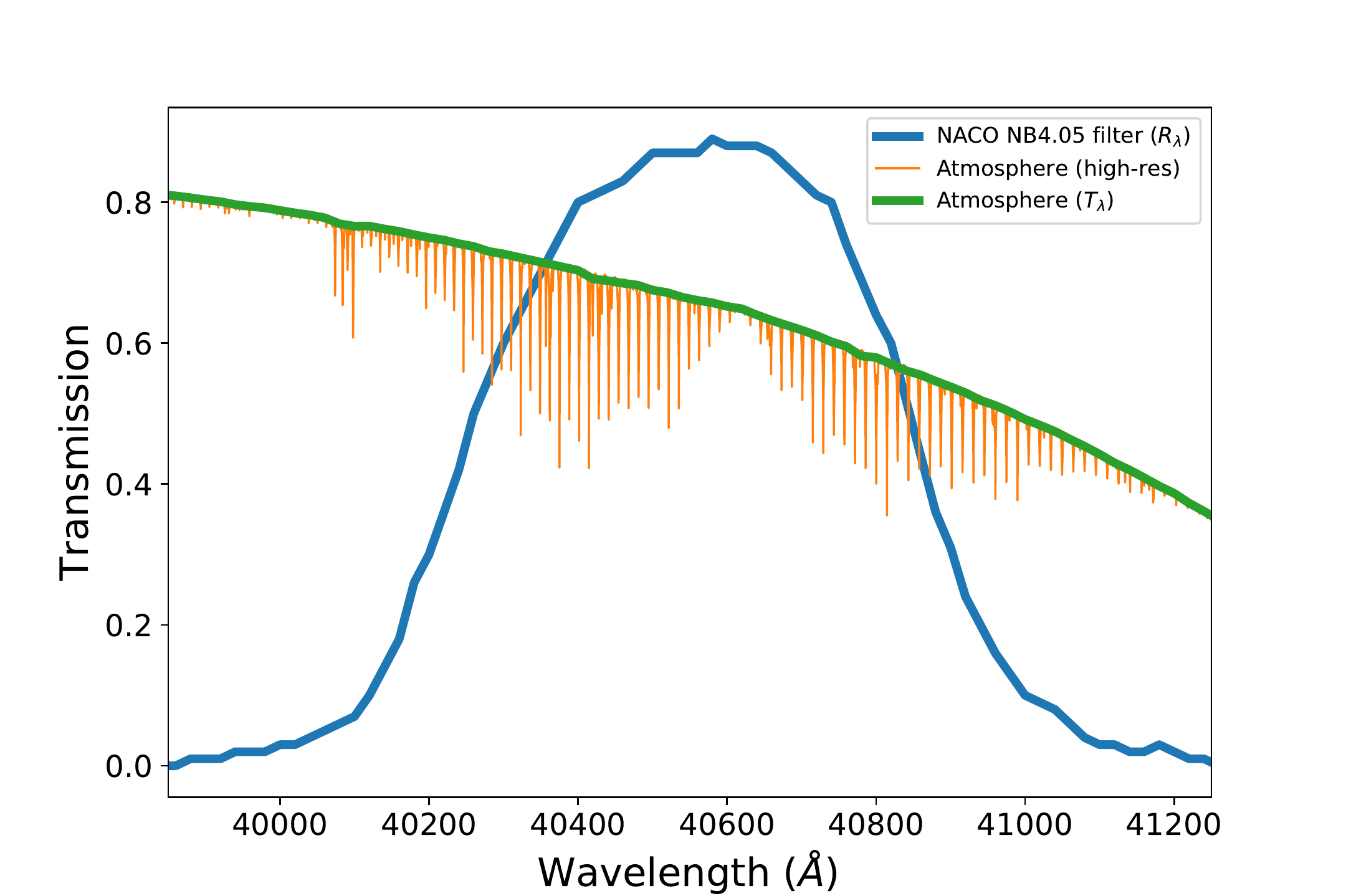}%
}{%
  \caption{Atmospheric transmission over the bandpass.}%
  \label{fig:atm_trans_band}
}
\capbtabbox{%
  \begin{tabular}{lcc} \hline
  Filter & $m_{calc}$ & $m_{meas}$ \\ \hline
  Johnson $U$ & 1.12 & 1.07  \\ \hline
  Johnson $B$ & 1.02 & 0.98  \\ \hline
  Johnson $V$ & 0.88 & 0.76  \\ \hline
  Johnson $R$ & 0.73 & 0.62  \\ \hline
  Johnson $I$ & 0.64 & 0.49  \\ \hline
  Johnson $J$ & 0.53 & 0.35  \\ \hline
  Johnson $K$ & 0.43 & 0.24  \\ \hline
  2MASS $J$ & 0.52 & 0.313  \\ \hline
  2MASS $H$ & 0.44 & 0.102 \\ \hline
  2MASS $Ks$ & 0.43 & 0.102  \\ \hline
  \end{tabular}
}{%
  \caption{Comparison of apparent magnitudes (on the Vega scale) $m_{meas}$ of Altair from the literature, and $m_{calc}$ calculated using a Kurucz spectrum. Johnson magnitude values are from \citep{ducati2002catalogue}, and 2MASS magnitudes from \citep{skrutskie2006two}.}%
}
\label{table:lit_mags}
\end{floatrow}
\end{figure}

To this precision, the answer is the same if the calculation is performed again for an energy detector with the zero point provided by VOSA. We used the zero points provided by VOSA (as defined in Eqn. \ref{eqn:zpE}) to calculate synthetic photometry for Altair for standard filters used in the literature (Table \ref{table:lit_mags} and Fig. \ref{fig:filters_2_check}), in the energy-detector approximation. We find the average of the difference between our synthetic magnitudes and literature magnitudes to be $\langle m_{syn}-m_{lit} \rangle=+0.17$, with a standard deviation of 0.10. We subtract this offset from the result in Eqn. \ref{eqn:final_M}, and consider the standard deviation to be the error in the magnitude. This yields our adopted value of the absolute magnitude in this bandpass:

\begin{equation}
M_{ \textrm{Altair}} = 1.70 \pm 0.10
\end{equation}

We also considered the possibility of color error in the measurement of the magnitude difference $\Delta m$ between a host star and a companion, because we measure this difference between objects with different spectra from behind a terrestrial atmosphere with a transmission is a strong function of wavelength.

To quantify the error we should expect, we used an atmospheric transmission model from ATRAN \citep{lord1992nasa}, corresponding to a latitude and altitude of the LBT, water vapor of 11 mm H$_{2}$O, and a zenith angle of 30$^{\circ}$ (Fig. \ref{fig:atm_trans_band}). Together with the Kurucz model spectrum for Altair, we tested blackbodies ranging from $\textrm{T}=200$ K to $\textrm{T}=2800$ K to mimic companion planet spectra, and found the magnitude error in $\Delta m$ to always have an absolute value $<0.01$ mag, which is negligible for our purposes.

\section{Calculation of contrast curves in $\lambda/D$ regime}
\label{appendix:contrast_curves}

\begin{deluxetable}{l | l | l}
\tabletypesize{\scriptsize}
\tablecaption{Defined quantities for calculating the $\lambda/D$ contrast curve}
\tablewidth{0pt}
\tablehead{
\colhead{Parameter} & \colhead{Description} & \colhead{Units}
}
\startdata
$\bar{x}_{1}=x_{1}$  & Fake companion amplitude, ``averaged'' over one value & Counts   \\
\hline
$\bar{x}_{2}\approx 0$  & \begin{tabular}{@{}l@{}}Amplitude of the noise ring, averaged over $N_{FWHM,r}-1$,\\the rounded floor number of FWHM which can fit in an annulus at a given \\ radius of the host star, minus 1 to remove the location of the fake companion\end{tabular} & Counts \\
\hline
$s_{1,2}=s_{2}$ & \begin{tabular}{@{}l@{}}Empirical standard deviation of pooled samples 1 and 2 (See Eqn. ).\\Since sample 1 just has 1 sample, this is the standard deviation\\of the amplitudes of the noise necklace beads\end{tabular} & Counts \\
\hline
$s_{2}$ & Empirical standard deviation of the amplitudes of the noise necklace beads & Counts \\
\hline
 $A_{5}$ & \begin{tabular}{@{}l@{}}Amplitudes of fake injected planets, following smoothing by convolution with\\a Gaussian\end{tabular} & Counts \\
\enddata

\tablenotetext{a}{This is stopped down within the instrument from a physical diameter of 8.4 m.}
\label{table:2sample_params}
\end{deluxetable}

In the past, much of the high-contrast literature has made use of ``5-sigma'' contrast curves which are based on taking the level of the standard deviation of the noise at a given radius, and multiplying it by five to obtain the amplitude of a signal where the false positive fraction (FPF) is expected to be $\sim3\times10^{-7}$ for Gaussian-distributed noise. However, this measure implies that the amplitude of the true positive fraction (TPF) is fixed at 0.5, since noise will conspire to make the other half of true signals fall below the threshold \citep{jensen2017new}.

Furthermore, the decreasing number of degrees of freedom at small radii---due to the decreasing number of FWHMs that can be fit onto a given annulus, like beads on a shrinking necklace---make empirical approximations of the parent noise distribution increasingly coarse \citep{mawet2014fundamental}. If we are willing to be flexible with the FPF, we can generate a more informative contrast curve that takes into account shrinking degrees of freedom at smaller angles from the host star. 

The following is based on some of the methodologies and notation conventions from \citet{mawet2014fundamental}, \citet{ruane2017deep} and \citet{stone2018leech}. We start by making the approximating assumption that at each radius from the host star the amplitudes of the parent speckle population are indeed Gaussian distributed $\mathcal{N}(\mu,\sigma)$. In that case, we still do now know what the true parameters $\mu$, $\sigma$ of the Gaussian are, based on empirical measurements. We must resort to modeling the probability density distribution of the rescaled variable $t\equiv\frac{\bar{x}-\mu}{\sigma/\sqrt{n}}$ as a $t$-distribution, where $n$ is the empirical sample size and $\bar{x}$ is the empirical average \citep{student1908probable}:

\begin{equation}
    P_{t}(t) = \frac{\Gamma (n/2)}{\sqrt{\pi (n-1)} \Gamma (\frac{n-1}{2})}\left( 1+\frac{t^{2}}{n-1} \right)^{-n/2}
\label{eqn:tdistrib}
\end{equation}

\noindent
This distribution has degrees of freedom $\nu=n-1$, and the gamma function $\Gamma (z)\equiv\int_{0}^{\infty}e^{-t}t^{z-1}dt$ for real arguments $z>0$. When comparing the means of two populations where the true variances are assumed to be the same, the $t$-statistic becomes 

\begin{equation}
    t=\frac{\bar{x}_{1}-\bar{x}_{2}}{s_{1,2}\sqrt{\frac{1}{n_{1}}+\frac{1}{n_{2}}}}=\frac{x_{1}-\bar{x}_{2}}{s_{2}\sqrt{1+\frac{1}{n_{2}}}}
\label{eqn:tstat}
\end{equation}

\noindent
with variables defined in Table \ref{table:2sample_params}. (See expanded versions of this expression in, for example, Sec. 11.3.2 in \citet{martin1971statistics}.) The amplitude of a companion which fulfills our TPF and FPF criteria is $\mu_{c}=x_{1}-\bar{x}_{2}\approx x_{1}$.

We apply the constraint that we are only willing to tolerate a number of total false positive detections $N_{FP}^{(tot)}=0.01$ in a dataset that ranges in angle out to a maximum integer radius of $R_{max}=20$ in units of FWHM. Then at each radius there is a constant number of false positives $N_{FP,r}=N_{FP}^{(tot)}/R_{max}=5\times10^{-4}$, and the FPF is a function of radius:

\begin{equation}
    FPF(r)=\frac{N_{FP,r}}{2\pi r}
\label{eqn:fpf_r}
\end{equation}

We want to find the threshold $t=\tau$ at each radius from the host star which satisfies the following constraints on the FPF and TPF of detections under the hypotheses $H_{0}$ (there is no companion) and $H_{1}$ (there is a companion): 

\begin{equation}
    FPF(r) = \int_{\tau}^{\infty}P_{t}(t|H_{0}) dt \textrm{, \ \ \ } TPF = 0.95 = \int_{\mu_{c}-\tau}^{\infty}P_{t}(t|H_{1}) dt 
\label{eqn:fpf_tpf_invertible}
\end{equation}

\noindent
To make the expressions simultaneously invertible to find the companion amplitude $\mu_{c}$ that satisfies both expressions, the lower bound of the TPF integral is the offset $\mu_{c}-\tau$, so that we can find that bound by using the CDF of the $t$-distribution at $\mu_{c}-\tau$. We then solve for the bounds by using the inverse of the cumulative distribution function, the percent point function, of the $t$-distribution:

\begin{equation}
    \mu_{c}=\textrm{ppf}_{t,\tau}(1-FPF(r)) + \textrm{ppf}_{t,\mu_{c}-\tau}(TPF=0.95)
\end{equation}

This threshold is still in $t$-space and needs to be scaled to contrast, based on the results of the fake planet injections. Those injected planets converged---following a smoothing to remove pixel-to-pixel noise---on amplitudes $A_{5}$ corresponding to $S/N=5$. Convergence on this value folds in throughput effects. The empirical noise is found from those rescaled injected companions as $s_{2}=A_{5}/5$.

From Eqn. \ref{eqn:tstat} we have, for a given radius from the host star, the linear contrast $F$ of a companion

\begin{equation}
    F=\mu_{c} s_{2} \sqrt{1+\frac{1}{n_{2}}}
\end{equation}

\noindent
This corresponds to a contrast curve corrected for diminishing degrees of freedom at small angles, and which satisfies our constraints on the FPF and TPF. The $\lambda/D$ contrast curve in Fig. \ref{fig:cont_curve_1d_fizeau} is the magnitude equivalent of this, viz. $\Delta m = -2.5\textrm{log}_{10}(F)$. 
\bibstyle{apj}
\bibliography{report}

\begin{thebibliography}{}
\expandafter\ifx\csname natexlab\endcsname\relax\def\natexlab#1{#1}\fi
\providecommand{\url}[1]{\href{#1}{#1}}
\providecommand{\dodoi}[1]{doi:~\href{http://doi.org/#1}{\nolinkurl{#1}}}
\providecommand{\doeprint}[1]{\href{http://ascl.net/#1}{\nolinkurl{http://ascl.net/#1}}}
\providecommand{\doarXiv}[1]{\href{https://arxiv.org/abs/#1}{\nolinkurl{https://arxiv.org/abs/#1}}}

\bibitem[{Absil {et~al.}(2013)Absil, Defr{\`e}re, Du~Foresto, Di~Folco,
  M{\'e}rand, Augereau, Ertel, Hanot, Kervella, Mollier,
  {et~al.}}]{absil2013near}
Absil, O., Defr{\`e}re, D., Du~Foresto, V.~C., {et~al.} 2013, Astronomy \&
  Astrophysics, 555, A104

\bibitem[{Allard {et~al.}(2001)Allard, Hauschildt, Alexander, Tamanai, \&
  Schweitzer}]{allard2001limiting}
Allard, F., Hauschildt, P.~H., Alexander, D.~R., Tamanai, A., \& Schweitzer, A.
  2001, The Astrophysical Journal, 556, 357

\bibitem[{Allard {et~al.}(2012)Allard, Homeier, \& Freytag}]{allard2012models}
Allard, F., Homeier, D., \& Freytag, B. 2012, Philosophical Transactions of the
  Royal Society A: Mathematical, Physical and Engineering Sciences, 370, 2765

\bibitem[{Bailey {et~al.}(2014)Bailey, Hinz, Puglisi, Esposito, Vaitheeswaran,
  Skemer, Defr{\`e}re, Vaz, \& Leisenring}]{bailey2014large}
Bailey, V.~P., Hinz, P.~M., Puglisi, A.~T., {et~al.} 2014, in SPIE Proceedings,
  914803

\bibitem[{Baines {et~al.}(2017)Baines, Armstrong, Schmitt, Zavala, Benson,
  Hutter, Tycner, \& van Belle}]{baines2017fundamental}
Baines, E.~K., Armstrong, J.~T., Schmitt, H.~R., {et~al.} 2017, The
  Astronomical Journal, 155, 30

\bibitem[{Baraffe {et~al.}(2003)Baraffe, Chabrier, Barman, Allard, \&
  Hauschildt}]{baraffe2003evolutionary}
Baraffe, I., Chabrier, G., Barman, T.~S., Allard, F., \& Hauschildt, P. 2003,
  Astronomy \& Astrophysics, 402, 701

\bibitem[{Barber {et~al.}(2006)Barber, Tennyson, Harris, \&
  Tolchenov}]{barber2006high}
Barber, R., Tennyson, J., Harris, G.~J., \& Tolchenov, R. 2006, Monthly Notices
  of the Royal Astronomical Society, 368, 1087

\bibitem[{{Baron} {et~al.}(2010){Baron}, {Chen}, \& {Hauschildt}}]{phoenix}
{Baron}, E., {Chen}, B., \& {Hauschildt}, P. 2010, PHOENIX: A General-purpose
  State-of-the-art Stellar and Planetary Atmosphere Code, Astrophysics Source
  Code Library.
\newblock \doeprint{1010.056}

\bibitem[{Bayo {et~al.}(2008)Bayo, Rodrigo, y~Navascu{\'e}s, Solano,
  Guti{\'e}rrez, Morales-Calder{\'o}n, \& Allard}]{bayo2008vosa}
Bayo, A., Rodrigo, C., y~Navascu{\'e}s, D.~B., {et~al.} 2008, Astronomy \&
  Astrophysics, 492, 277

\bibitem[{Beichman {et~al.}(2014)Beichman, Benneke, Knutson, Smith, Lagage,
  Dressing, Latham, Lunine, Birkmann, Ferruit,
  {et~al.}}]{beichman2014observations}
Beichman, C., Benneke, B., Knutson, H., {et~al.} 2014, Publications of the
  Astronomical Society of the Pacific, 126, 1134

\bibitem[{Boehle {et~al.}(2019)Boehle, Quanz, Lovis, S{\'e}gransan, Udry, \&
  Apai}]{boehle2019combining}
Boehle, A., Quanz, S.~P., Lovis, C., {et~al.} 2019, Astronomy \& Astrophysics,
  630, A50

\bibitem[{B{\"o}hm {et~al.}(2016)B{\"o}hm, Pott, Borelli, Hinz, Defr{\`e}re,
  Downey, Hill, Summers, Conrad, K{\"u}rster, {et~al.}}]{bohm2016ovms}
B{\"o}hm, M., Pott, J.-U., Borelli, J., {et~al.} 2016, in {SPIE Proceedings},
  99062R

\bibitem[{Bouchaud {et~al.}(2020)Bouchaud, de~Souza, Rieutord, Reese, \&
  Kervella}]{bouchaud2020realistic}
Bouchaud, K., de~Souza, A.~D., Rieutord, M., Reese, D., \& Kervella, P. 2020,
  Astronomy \& Astrophysics, 633, A78

\bibitem[{Bowens {et~al.}(2021)Bowens, Meyer, Delacroix, Absil, van Boekel,
  Quanz, Shinde, Kenworthy, Carlomagno, de~Xivry,
  {et~al.}}]{bowens2021exoplanets}
Bowens, R., Meyer, M.~R., Delacroix, C., {et~al.} 2021, Astronomy \&
  Astrophysics, 653, A8

\bibitem[{Bowler(2016)}]{bowler2016imaging}
Bowler, B.~P. 2016, Publications of the Astronomical Society of the Pacific,
  128, 102001

\bibitem[{Burse {et~al.}(2016)Burse, Ramaprakash, Chordia, Punnadi, Chillal,
  Mestri, Bharti, Sinha, \& Kohok}]{burse2016isdec}
Burse, M., Ramaprakash, A., Chordia, P., {et~al.} 2016, in SPIE Proceedings,
  991526

\bibitem[{Buzasi {et~al.}(2005)Buzasi, Bruntt, Bedding, Retter, Kjeldsen,
  Preston, Mandeville, Suarez, Catanzarite, Conrow,
  {et~al.}}]{buzasi2005altair}
Buzasi, D., Bruntt, H., Bedding, T., {et~al.} 2005, The Astrophysical Journal,
  619, 1072

\bibitem[{Cantrell {et~al.}(2013)Cantrell, Henry, \& White}]{cantrell2013solar}
Cantrell, J.~R., Henry, T.~J., \& White, R.~J. 2013, The Astronomical Journal,
  146, 99

\bibitem[{Castelli {et~al.}(1997)Castelli, Gratton, \&
  Kurucz}]{castelli1997notes}
Castelli, F., Gratton, R., \& Kurucz, R. 1997, Astronomy \& Astrophysics, 318,
  841

\bibitem[{Conover(1971)}]{conover1971statistics}
Conover, W. 1971, Practical {N}onparametric {S}tatistics (New York, Wiley)

\bibitem[{Conrad {et~al.}(2015)Conrad, De~Kleer, Leisenring, La~Camera,
  Arcidiacono, Bertero, Boccacci, Defr{\`e}re, De~Pater, Hinz,
  {et~al.}}]{conrad2015spatially}
Conrad, A., De~Kleer, K., Leisenring, J., {et~al.} 2015, The Astronomical
  Journal, 149, 175

\bibitem[{Conrad(2016)}]{conrad2016role}
Conrad, A.~R. 2016, in SPIE Proceedings, 99070L

\bibitem[{de~Kleer {et~al.}(2017)de~Kleer, Skrutskie, Leisenring, Davies,
  Conrad, De~Pater, Resnick, Bailey, Defrere, Hinz, {et~al.}}]{de2017multi}
de~Kleer, K., Skrutskie, M., Leisenring, J., {et~al.} 2017, Nature, 545, 199

\bibitem[{de~Kleer {et~al.}(2021)de~Kleer, Skrutskie, Leisenring, Davies,
  Conrad, de~Pater, Resnick, Bailey, Defr{\`e}re, Hinz,
  {et~al.}}]{de2021resolving}
---. 2021, The Planetary Science Journal, 2, 227

\bibitem[{de~Souza {et~al.}(2005)de~Souza, Kervella, Jankov, Vakili, Ohishi,
  Nordgren, \& Abe}]{kervella2005gravitational}
de~Souza, A., Kervella, P., Jankov, S., {et~al.} 2005, Astronomy \&
  Astrophysics, 442, 567

\bibitem[{Defr\`{e}re {et~al.}(2014)Defr\`{e}re, Hinz, Downey, Ashby, Bailey,
  Brusa, Christou, Danchi, Grenz, Hill, {et~al.}}]{defrere2014co}
Defr\`{e}re, D., Hinz, P., Downey, E., {et~al.} 2014, in SPIE Proceedings,
  914609

\bibitem[{Dieterich {et~al.}(2012)Dieterich, Henry, Golimowski, Krist, \&
  Tanner}]{dieterich2012solar}
Dieterich, S.~B., Henry, T.~J., Golimowski, D.~A., Krist, J.~E., \& Tanner,
  A.~M. 2012, The Astronomical Journal, 144, 64

\bibitem[{Ducati(2002)}]{ducati2002catalogue}
Ducati, J. 2002, CDS/ADC Collection of Electronic Catalogues, 2237

\bibitem[{Erspamer \& North(2003)}]{erspamer2003automated}
Erspamer, D., \& North, P. 2003, Astronomy \& Astrophysics, 398, 1121

\bibitem[{Ertel {et~al.}(2018)Ertel, Defr{\`e}re, Hinz, Mennesson, Kennedy,
  Danchi, Gelino, Hill, Hoffmann, Rieke, {et~al.}}]{ertel2018hosts}
Ertel, S., Defr{\`e}re, D., Hinz, P., {et~al.} 2018, The Astronomical Journal,
  155, 194

\bibitem[{Ertel {et~al.}(2020)Ertel, Defr{\`e}re, Hinz, Mennesson, Kennedy,
  Danchi, Gelino, Hill, Hoffmann, Mazoyer, {et~al.}}]{ertel2020hosts}
---. 2020, The Astronomical Journal, 159, 177

\bibitem[{Feller(1948)}]{feller1948}
Feller, W. 1948, Annals of Mathematical Statistics, 19, 177

\bibitem[{Ferguson {et~al.}(2005)Ferguson, Alexander, Allard, Barman, Bodnarik,
  Hauschildt, Heffner-Wong, \& Tamanai}]{ferguson2005low}
Ferguson, J.~W., Alexander, D.~R., Allard, F., {et~al.} 2005, The Astrophysical
  Journal, 623, 585

\bibitem[{G{\'a}sp{\'a}r {et~al.}(2013)G{\'a}sp{\'a}r, Rieke, \&
  Balog}]{gaspar2013collisional}
G{\'a}sp{\'a}r, A., Rieke, G.~H., \& Balog, Z. 2013, The Astrophysical Journal,
  768, 25

\bibitem[{Gatewood \& de~Jonge(1995)}]{gatewood1995map}
Gatewood, G., \& de~Jonge, J.~K. 1995, The Astrophysical Journal, 450, 364

\bibitem[{Goebel {et~al.}(2018)Goebel, Hall, Guyon, Warmbier, \&
  Jacobson}]{goebel2018overview}
Goebel, S.~B., Hall, D.~N., Guyon, O., Warmbier, E., \& Jacobson, S.~M. 2018,
  Journal of Astronomical Telescopes, Instruments, and Systems, 4, 026001

\bibitem[{Greenbaum {et~al.}(2018)Greenbaum, Pueyo, Ruffio, Wang, De~Rosa,
  Aguilar, Rameau, Barman, Marois, Marley, {et~al.}}]{greenbaum2018gpi}
Greenbaum, A.~Z., Pueyo, L., Ruffio, J.-B., {et~al.} 2018, The Astronomical
  Journal, 155, 226

\bibitem[{Hadjara {et~al.}(2014)Hadjara, de~Souza, Vakili, Jankov, Millour,
  Meilland, Khorrami, Chelli, Baffa, Hofmann, {et~al.}}]{hadjara2014beyond}
Hadjara, M., de~Souza, A.~D., Vakili, F., {et~al.} 2014, Astronomy \&
  Astrophysics, 569, A45

\bibitem[{Hill(2010)}]{hill2010large}
Hill, J.~M. 2010, Applied Optics, 49, D115

\bibitem[{Hinz {et~al.}(2004)Hinz, Connors, McMahon, Cheng, Peng, Hoffmann,
  McCarthy, \& Angel}]{hinz2004large}
Hinz, P.~M., Connors, T., McMahon, T., {et~al.} 2004, in SPIE Proceedings, Vol.
  5491, 787

\bibitem[{Hinz {et~al.}(2008)Hinz, Solheid, Durney, \& Hoffmann}]{hinz2008nic}
Hinz, P.~M., Solheid, E., Durney, O., \& Hoffmann, W.~F. 2008, in SPIE
  Proceedings, 701339

\bibitem[{Hinz {et~al.}(2016)Hinz, Defr{\`e}re, Skemer, Bailey, Stone,
  Spalding, Vaz, Pinna, Puglisi, Esposito, {et~al.}}]{hinz2016overview}
Hinz, P.~M., Defr{\`e}re, D., Skemer, A., {et~al.} 2016, in SPIE Proceedings,
  990704

\bibitem[{Hoffleit \& Warren~Jr(1995)}]{hoffleit1995vizier}
Hoffleit, D., \& Warren~Jr, W. 1995, VizieR Online Data Catalog, 5050

\bibitem[{Hoffmann {et~al.}(2014)Hoffmann, Hinz, Defr{\`e}re, Leisenring,
  Skemer, Arbo, Montoya, \& Mennesson}]{hoffmann2014operation}
Hoffmann, W.~F., Hinz, P.~M., Defr{\`e}re, D., {et~al.} 2014, in SPIE
  Proceedings, 91471O

\bibitem[{Howard \& Fulton(2016)}]{howard2016limits}
Howard, A.~W., \& Fulton, B.~J. 2016, Publications of the Astronomical Society
  of the Pacific, 128, 114401

\bibitem[{Hunter(2007)}]{hunter2007matplotlib}
Hunter, J.~D. 2007, Computing in Science \& Engineering, 9, 90

\bibitem[{Hunziker {et~al.}(2018)Hunziker, Quanz, Amara, \&
  Meyer}]{hunziker2018pca}
Hunziker, S., Quanz, S.~P., Amara, A., \& Meyer, M.~R. 2018, Astronomy \&
  Astrophysics, 611, A23

\bibitem[{Hunziker {et~al.}(2020)Hunziker, Schmid, Mouillet, Milli, Zurlo,
  Delorme, Abe, Avenhaus, Baruffolo, Bazzon, {et~al.}}]{hunziker2020refplanets}
Hunziker, S., Schmid, H.~M., Mouillet, D., {et~al.} 2020, Astronomy \&
  Astrophysics, 634, A69

\bibitem[{Janson {et~al.}(2011)Janson, Bonavita, Klahr, Lafreni{\`e}re,
  Jayawardhana, \& Zinnecker}]{janson2011high}
Janson, M., Bonavita, M., Klahr, H., {et~al.} 2011, The Astrophysical Journal,
  736, 89

\bibitem[{Jensen-Clem {et~al.}(2017)Jensen-Clem, Mawet, Gonzalez, Absil,
  Belikov, Currie, Kenworthy, Marois, Mazoyer, Ruane, {et~al.}}]{jensen2017new}
Jensen-Clem, R., Mawet, D., Gonzalez, C. A.~G., {et~al.} 2017, The Astronomical
  Journal, 155, 19

\bibitem[{Kirchschlager {et~al.}(2017)Kirchschlager, Wolf, Krivov, Mutschke, \&
  Brunngr{\"a}ber}]{kirchschlager2017constraints}
Kirchschlager, F., Wolf, S., Krivov, A.~V., Mutschke, H., \& Brunngr{\"a}ber,
  R. 2017, Monthly Notices of the Royal Astronomical Society, 467, 1614

\bibitem[{Kuchner \& Brown(2000)}]{kuchner2000search}
Kuchner, M.~J., \& Brown, M.~E. 2000, Publications of the Astronomical Society
  of the Pacific, 112, 827

\bibitem[{Kuchner {et~al.}(1998)Kuchner, Brown, \& KORESkO}]{kuchner199811}
Kuchner, M.~J., Brown, M.~E., \& KORESkO, C.~D. 1998, Publications of the
  Astronomical Society of the Pacific, 110, 1336

\bibitem[{Kurtzer {et~al.}(2017)Kurtzer, Sochat, \&
  Bauer}]{kurtzer2017singularity}
Kurtzer, G.~M., Sochat, V., \& Bauer, M.~W. 2017, PLOS One, 12, e0177459

\bibitem[{Lafreniere {et~al.}(2007)Lafreniere, Marois, Doyon, Nadeau, \&
  Artigau}]{lafreniere2007new}
Lafreniere, D., Marois, C., Doyon, R., Nadeau, D., \& Artigau, {\'E}. 2007, The
  Astrophysical Journal, 660, 770

\bibitem[{Lara \& Rieutord(2011)}]{lara2011gravity}
Lara, F.~E., \& Rieutord, M. 2011, Astronomy \& Astrophysics, 533, A43

\bibitem[{Leconte {et~al.}(2010)Leconte, Soummer, Hinkley, Oppenheimer,
  Sivaramakrishnan, Brenner, Kuhn, Lloyd, Perrin, Makidon,
  {et~al.}}]{leconte2010lyot}
Leconte, J., Soummer, R., Hinkley, S., {et~al.} 2010, The Astrophysical
  Journal, 716, 1551

\bibitem[{Leisenring {et~al.}(2014)Leisenring, Hinz, Skrutskie, Skemer,
  Woodward, Veillet, Arcidiacono, Bailey, Bertero, Boccacci,
  {et~al.}}]{leisenring2014fizeau}
Leisenring, J., Hinz, P.~M., Skrutskie, M., {et~al.} 2014, in {SPIE
  Proceedings}, 91462S

\bibitem[{Leisenring {et~al.}(2012)Leisenring, Skrutskie, Hinz, Skemer, Bailey,
  Eisner, Garnavich, Hoffmann, Jones, Kenworthy, {et~al.}}]{leisenring2012sky}
Leisenring, J.~M., Skrutskie, M., Hinz, P., {et~al.} 2012, in SPIE Proceedings,
  84464F

\bibitem[{Loose {et~al.}(2018)Loose, Smith, Alkire, Joshi, Kelly, Siskind,
  Mann, Chen, Askarov, Fox-Rabinovitz, {et~al.}}]{loose2018acadia}
Loose, M., Smith, B., Alkire, G., {et~al.} 2018, in SPIE Proceedings, Vol.
  10709

\bibitem[{Lord(1992)}]{lord1992nasa}
Lord, S. 1992, Nasa technical memorandum 103957, Ames Research Center, Moffett
  Field, CA

\bibitem[{Maier {et~al.}(2018)Maier, Hinz, Defr{\`e}re, Ertel, \&
  Downey}]{maier2018two}
Maier, E., Hinz, P., Defr{\`e}re, D., Ertel, S., \& Downey, E. 2018, in SPIE
  Proceedings, 107011M

\bibitem[{Maier {et~al.}(2020)Maier, Hinz, Defr{\`e}re, Grenz, Downey, Ertel,
  Morzinski, \& Douglas}]{maier2020implementing}
Maier, E.~R., Hinz, P., Defr{\`e}re, D., {et~al.} 2020, Journal of Astronomical
  Telescopes, Instruments, and Systems, 6, 035001

\bibitem[{Males {et~al.}(2021)Males, Fitzgerald, Belikov, \&
  Guyon}]{males2021mysterious}
Males, J.~R., Fitzgerald, M.~P., Belikov, R., \& Guyon, O. 2021, Accepted to
  PASP. arXiv:2107.04604

\bibitem[{Males {et~al.}(2014)Males, Close, Guyon, Morzinski, Puglisi, Hinz,
  Follette, Monnier, Tolls, Rodigas, {et~al.}}]{males2014direct}
Males, J.~R., Close, L.~M., Guyon, O., {et~al.} 2014, in SPIE Proceedings,
  914820

\bibitem[{Marleau \& Cumming(2013)}]{marleau2013constraining}
Marleau, G.-D., \& Cumming, A. 2013, Monthly Notices of the Royal Astronomical
  Society, 437, 1378

\bibitem[{Marois {et~al.}(2006)Marois, Lafreniere, Doyon, Macintosh, \&
  Nadeau}]{marois2006angular}
Marois, C., Lafreniere, D., Doyon, R., Macintosh, B., \& Nadeau, D. 2006, The
  Astrophysical Journal, 641, 556

\bibitem[{Martin(1971)}]{martin1971statistics}
Martin, B. R.~C. 1971, Statistics for {P}hysicists (New York, Academic Press)

\bibitem[{Masciadri {et~al.}(2019)Masciadri, Turchi, \&
  Martelloni}]{masciadri2019new}
Masciadri, E., Turchi, A., \& Martelloni, G. 2019, Proceedings of AO4ELT6
  Conference, 9-14 June 2019, \texttt{arXiv:1911.02819}

\bibitem[{Mawet {et~al.}(2014)Mawet, Milli, Wahhaj, Pelat, Absil, Delacroix,
  Boccaletti, Kasper, Kenworthy, Marois, {et~al.}}]{mawet2014fundamental}
Mawet, D., Milli, J., Wahhaj, Z., {et~al.} 2014, The Astrophysical Journal,
  792, 97

\bibitem[{Mennesson {et~al.}(2014)Mennesson, Millan-Gabet, Serabyn, Colavita,
  Absil, Bryden, Wyatt, Danchi, Defrere, Dor{\'e},
  {et~al.}}]{mennesson2014constraining}
Mennesson, B., Millan-Gabet, R., Serabyn, E., {et~al.} 2014, The Astrophysical
  Journal, 797, 119

\bibitem[{Meyer {et~al.}(2018)Meyer, Currie, Guyon, Hasegawa, Kasper, Marois,
  Monnier, Morzinski, Packham, \& Quanz}]{meyer2018finding}
Meyer, M.~R., Currie, T., Guyon, O., {et~al.} 2018, White paper;
  arXiv:1804.03218

\bibitem[{Millan-Gabet {et~al.}(2011)Millan-Gabet, Serabyn, Mennesson, Traub,
  Barry, Danchi, Kuchner, Stark, Ragland, Hrynevych,
  {et~al.}}]{millan2011exozodiacal}
Millan-Gabet, R., Serabyn, E., Mennesson, B., {et~al.} 2011, The Astrophysical
  Journal, 734, 67

\bibitem[{Monnier {et~al.}(2007)Monnier, Zhao, Pedretti, Thureau, Ireland,
  Muirhead, Berger, Millan-Gabet, Van~Belle, Ten~Brummelaar,
  {et~al.}}]{monnier2007imaging}
Monnier, J.~D., Zhao, M., Pedretti, E., {et~al.} 2007, Science, 317, 342

\bibitem[{Mordasini {et~al.}(2012)Mordasini, Alibert, Klahr, \&
  Henning}]{mordasini2012characterization}
Mordasini, C., Alibert, Y., Klahr, H., \& Henning, T. 2012, Astronomy \&
  Astrophysics, 547, A111

\bibitem[{Nielsen {et~al.}(2019)Nielsen, De~Rosa, Macintosh, Wang, Ruffio,
  Chiang, Marley, Saumon, Savransky, Ammons, {et~al.}}]{nielsen2019gemini}
Nielsen, E.~L., De~Rosa, R.~J., Macintosh, B., {et~al.} 2019, The Astronomical
  Journal, 158, 13

\bibitem[{Nunez {et~al.}(2017)Nunez, Scott, Mennesson, Absil, Augereau, Bryden,
  ten Brummelaar, Ertel, du~Foresto, Ridgway, {et~al.}}]{nunez2017near}
Nunez, P.~D., Scott, N., Mennesson, B., {et~al.} 2017, Astronomy \&
  Astrophysics, 608, A113

\bibitem[{Ohishi {et~al.}(2004)Ohishi, Nordgren, \&
  Hutter}]{ohishi2004asymmetric}
Ohishi, N., Nordgren, T.~E., \& Hutter, D.~J. 2004, The Astrophysical Journal,
  612, 463

\bibitem[{Oliphant(2007)}]{oliphant2007python}
Oliphant, T.~E. 2007, Computing in Science \& Engineering, 9, 10

\bibitem[{Oppenheimer {et~al.}(2001)Oppenheimer, Golimowski, Kulkarni,
  Matthews, Nakajima, Creech-Eakman, \&
  Durrance}]{oppenheimer2001coronagraphic}
Oppenheimer, B., Golimowski, D., Kulkarni, S., {et~al.} 2001, The Astronomical
  Journal, 121, 2189

\bibitem[{Patru {et~al.}(2017{\natexlab{a}})Patru, Esposito, Puglisi, Riccardi,
  Pinna, Arcidiacono, Antichi, Mennesson, Defr{\`e}re, Hinz,
  {et~al.}}]{patru2017_i}
Patru, F., Esposito, S., Puglisi, A., {et~al.} 2017{\natexlab{a}}, Monthly
  Notices of the Royal Astronomical Society, 472, 2544

\bibitem[{Patru {et~al.}(2017{\natexlab{b}})Patru, Esposito, Puglisi, Riccardi,
  Pinna, Arcidiacono, Antichi, Mennesson, Defr{\`e}re, Hinz,
  {et~al.}}]{patru2017_ii}
---. 2017{\natexlab{b}}, Monthly Notices of the Royal Astronomical Society,
  472, 3288

\bibitem[{{Perryman} {et~al.}(1997){Perryman}, {Lindegren}, {Kovalevsky},
  {Hog}, {Bastian}, {Bernacca}, {Creze}, {Donati}, {Grenon}, {Grewing}, {van
  Leeuwen}, {van der Marel}, {Mignard}, {Murray}, {Le Poole}, {Schrijver},
  {Turon}, {Arenou}, {Froeschle}, \& {Petersen}}]{1997perrymannhipparcos}
{Perryman}, M.~A.~C., {Lindegren}, L., {Kovalevsky}, J., {et~al.} 1997,
  Astronomy \& Astrophysics, 500, 501

\bibitem[{Peterson {et~al.}(2006)Peterson, Hummel, Pauls, Armstrong, Benson,
  Gilbreath, Hindsley, Hutter, Johnston, Mozurkewich,
  {et~al.}}]{peterson2006resolving}
Peterson, D.~M., Hummel, C., Pauls, T., {et~al.} 2006, The Astrophysical
  Journal, 636, 1087

\bibitem[{Plez(1998)}]{plez1998new}
Plez, B. 1998, Astronomy \& Astrophysics, 337, 495

\bibitem[{Price-Whelan {et~al.}(2018)Price-Whelan, Sip{\H{o}}cz, G{\"u}nther,
  Lim, Crawford, Conseil, Shupe, Craig, Dencheva, Ginsburg,
  {et~al.}}]{price2018astropy}
Price-Whelan, A., Sip{\H{o}}cz, B., G{\"u}nther, H., {et~al.} 2018, The
  Astronomical Journal, 156, 123

\bibitem[{Prieto {et~al.}(2004)Prieto, Barklem, Lambert, \&
  Cunha}]{prieto2004sn}
Prieto, C.~A., Barklem, P.~S., Lambert, D.~L., \& Cunha, K. 2004, Astronomy \&
  Astrophysics, 420, 183

\bibitem[{{Project Jupyter} {et~al.}(2018){Project Jupyter}, Bussonnier, Forde,
  Freeman, Granger, Head, Holdgraf, Kelley, Nalvarte, Osheroff,
  {et~al.}}]{jupyter2018binder}
{Project Jupyter}, Bussonnier, M., Forde, J., {et~al.} 2018, in Proceedings of
  the 17th Python in Science Conference, 113--120

\bibitem[{Quanz {et~al.}(2015)Quanz, Amara, Meyer, Girard, Kenworthy, \&
  Kasper}]{quanz2015confirmation}
Quanz, S.~P., Amara, A., Meyer, M.~R., {et~al.} 2015, The Astrophysical
  Journal, 807, 64

\bibitem[{Quanz {et~al.}(2012)Quanz, Crepp, Janson, Avenhaus, Meyer, \&
  Hillenbrand}]{quanz2012searching}
Quanz, S.~P., Crepp, J.~R., Janson, M., {et~al.} 2012, The Astrophysical
  Journal, 754, 127

\bibitem[{Rajan {et~al.}(2017)Rajan, Rameau, De~Rosa, Marley, Graham,
  Macintosh, Marois, Morley, Patience, Pueyo,
  {et~al.}}]{rajan2017characterizing}
Rajan, A., Rameau, J., De~Rosa, R.~J., {et~al.} 2017, The Astronomical Journal,
  154, 10

\bibitem[{Reiners \& Royer(2004)}]{reiners2004altair}
Reiners, A., \& Royer, F. 2004, Astronomy \& Astrophysics, 428, 199

\bibitem[{Reyl{\'e} {et~al.}(2021)Reyl{\'e}, Jardine, Fouqu{\'e}, Caballero,
  Smart, \& Sozzetti}]{reyle202110}
Reyl{\'e}, C., Jardine, K., Fouqu{\'e}, P., {et~al.} 2021, Astronomy \&
  Astrophysics

\bibitem[{Richichi {et~al.}(2009)Richichi, Percheron, \&
  Davis}]{richichi2009list}
Richichi, A., Percheron, I., \& Davis, J. 2009, Monthly Notices of the Royal
  Astronomical Society, 399, 399

\bibitem[{Roberts(2011)}]{roberts2011astrometric}
Roberts, L.~C. 2011, Monthly Notices of the Royal Astronomical Society, 413,
  1200

\bibitem[{Robrade \& Schmitt(2009)}]{robrade2009altair}
Robrade, J., \& Schmitt, J. 2009, Astronomy \& Astrophysics, 497, 511

\bibitem[{Roddier(1999)}]{roddier1999adaptive}
Roddier, F. 1999, Adaptive Optics in Astronomy (Cambridge University Press)

\bibitem[{Rodrigo {et~al.}(2012)Rodrigo, Solano, \& Bayo}]{rodrigo2012svo}
Rodrigo, C., Solano, E., \& Bayo, A. 2012, IVOA Rep, 1015

\bibitem[{Romero-Wolf {et~al.}(2021)Romero-Wolf, Bryden, Agnes, Arenberg,
  Bradford, D’Amico, Debes, Greenhouse, Hu, Matousek,
  {et~al.}}]{romero2021starshade}
Romero-Wolf, A., Bryden, G., Agnes, G., {et~al.} 2021, Journal of Astronomical
  Telescopes, Instruments, and Systems, 7, 021219

\bibitem[{Rothberg {et~al.}(2018)Rothberg, Kuhn, Power, Hill, Veillet, Edwards,
  Thompson, \& Wagner}]{rothberg2018current}
Rothberg, B., Kuhn, O., Power, J., {et~al.} 2018, in SPIE Proceedings, 1070205

\bibitem[{Ruane {et~al.}(2017)Ruane, Mawet, Kastner, Meshkat, Bottom,
  Castell{\'a}, Absil, Gonzalez, Huby, Zhu, {et~al.}}]{ruane2017deep}
Ruane, G., Mawet, D., Kastner, J., {et~al.} 2017, The Astronomical Journal,
  154, 73

\bibitem[{Schroeder {et~al.}(2000)Schroeder, Golimowski, Brukardt, Burrows,
  Caldwell, Fastie, Ford, Hesman, Kletskin, Krist,
  {et~al.}}]{schroeder2000search}
Schroeder, D.~J., Golimowski, D.~A., Brukardt, R.~A., {et~al.} 2000, The
  Astronomical Journal, 119, 906

\bibitem[{Skrutskie {et~al.}(2006)Skrutskie, Cutri, Stiening, Weinberg,
  Schneider, Carpenter, Beichman, Capps, Chester, Elias,
  {et~al.}}]{skrutskie2006two}
Skrutskie, M., Cutri, R., Stiening, R., {et~al.} 2006, The Astronomical
  Journal, 131, 1163

\bibitem[{Skrutskie {et~al.}(2010)Skrutskie, Jones, Hinz, Garnavich, Wilson,
  Nelson, Solheid, Durney, Hoffmann, Vaitheeswaran,
  {et~al.}}]{skrutskie2010large}
Skrutskie, M., Jones, T., Hinz, P., {et~al.} 2010, in SPIE Proceedings, 77353H

\bibitem[{Smirnov(1948)}]{smirnov1948}
Smirnov, N. 1948, The Annals of Mathematical Statistics, 19, 279

\bibitem[{Sochat {et~al.}(2017)Sochat, Prybol, \&
  Kurtzer}]{sochat2017enhancing}
Sochat, V.~V., Prybol, C.~J., \& Kurtzer, G.~M. 2017, PloS One, 12, e0188511

\bibitem[{Spalding {et~al.}(2018)Spalding, Hinz, Ertel, Maier, \&
  Stone}]{spalding2018fizeau}
Spalding, E., Hinz, P., Ertel, S., Maier, E., \& Stone, J. 2018, in SPIE
  Proceedings, 107010J

\bibitem[{Spalding {et~al.}(2019)Spalding, Hinz, Morzinksi, Ertel, Grenz,
  Maier, Stone, \& Vaz}]{spalding2019status}
Spalding, E., Hinz, P., Morzinksi, K., {et~al.} 2019, in SPIE Proceedings,
  111171S

\bibitem[{Spiegel {et~al.}(2011)Spiegel, Burrows, \&
  Milsom}]{spiegel2011deuterium}
Spiegel, D.~S., Burrows, A., \& Milsom, J.~A. 2011, The Astrophysical Journal,
  727, 57

\bibitem[{Stone {et~al.}(2018)Stone, Skemer, Hinz, Bonavita, Kratter, Maire,
  Defrere, Bailey, Spalding, Leisenring, {et~al.}}]{stone2018leech}
Stone, J.~M., Skemer, A.~J., Hinz, P.~M., {et~al.} 2018, The Astronomical
  Journal, 156, 286

\bibitem[{Student(1908)}]{student1908probable}
Student. 1908, Biometrika, 1

\bibitem[{Suarez {et~al.}(2005)Suarez, Bruntt, \& Buzasi}]{suarez2005modelling}
Suarez, J.~C., Bruntt, H., \& Buzasi, D. 2005, Astronomy \& Astrophysics, 438,
  633

\bibitem[{Talens {et~al.}(2017)Talens, Spronck, Lesage, Otten, Stuik, Pollacco,
  \& Snellen}]{talens2017multi}
Talens, G., Spronck, J., Lesage, A.-L., {et~al.} 2017, Astronomy \&
  Astrophysics, 601, A11

\bibitem[{Thalmann {et~al.}(2011)Thalmann, Usuda, Kenworthy, Janson, Mamajek,
  Brandner, Dominik, Goto, Hayano, Henning, {et~al.}}]{thalmann2011piercing}
Thalmann, C., Usuda, T., Kenworthy, M., {et~al.} 2011, The Astrophysical
  Journal Letters, 732, L34

\bibitem[{Thureau {et~al.}(2014)Thureau, Greaves, Matthews, Kennedy, Phillips,
  Booth, Duch{\^e}ne, Horner, Rodriguez, Sibthorpe,
  {et~al.}}]{thureau2014unbiased}
Thureau, N., Greaves, J., Matthews, B., {et~al.} 2014, Monthly Notices of the
  Royal Astronomical Society, 445, 2558

\bibitem[{Turchi {et~al.}(2016)Turchi, Masciadri, \&
  Fini}]{turchi2016forecasts}
Turchi, A., Masciadri, E., \& Fini, L. 2016, in SPIE Proceedings, 990938

\bibitem[{van Belle(2012)}]{van2012interferometric}
van Belle, G.~T. 2012, The Astronomy \& Astrophysics Review, 20, 51

\bibitem[{van Belle {et~al.}(2001)van Belle, Ciardi, Thompson, Akeson, \&
  Lada}]{van2001altair}
van Belle, G.~T., Ciardi, D.~R., Thompson, R.~R., Akeson, R.~L., \& Lada, E.~A.
  2001, The Astrophysical Journal, 559, 1155

\bibitem[{Van~Leeuwen(2007)}]{van2007validation}
Van~Leeuwen, F. 2007, Astronomy \& Astrophysics, 474, 653

\bibitem[{van Lieshout {et~al.}(2014)van Lieshout, Dominik, Kama, \&
  Min}]{van2014near}
van Lieshout, R., Dominik, C., Kama, M., \& Min, M. 2014, Astronomy \&
  Astrophysics, 571, A51

\bibitem[{Van~Rossum \& Drake~Jr(1995)}]{van1995python}
Van~Rossum, G., \& Drake~Jr, F.~L. 1995, {The Python Language Reference Manual}
  (Centrum voor Wiskunde en Informatica Amsterdam)

\bibitem[{Vican(2012)}]{vican2012age}
Vican, L. 2012, The Astronomical Journal, 143, 135

\bibitem[{Vigan {et~al.}(2015)Vigan, Gry, Salter, Mesa, Homeier, Moutou, \&
  Allard}]{vigan2015high}
Vigan, A., Gry, C., Salter, G., {et~al.} 2015, Monthly Notices of the Royal
  Astronomical Society, 454, 129

\bibitem[{Vigan {et~al.}(2021)Vigan, Fontanive, Meyer, Biller, Bonavita, Feldt,
  Desidera, Marleau, Emsenhuber, Galicher, {et~al.}}]{vigan2021sphere}
Vigan, A., Fontanive, C., Meyer, M., {et~al.} 2021, Astronomy \& Astrophysics,
  651, A72

\bibitem[{{Virtanen} {et~al.}(2020){Virtanen}, {Gommers}, {Oliphant},
  {Haberland}, {Reddy}, {Cournapeau}, {Burovski}, {Peterson}, {Weckesser},
  {Bright}, {van der Walt}, {Brett}, {Wilson}, {Jarrod Millman}, {Mayorov},
  {Nelson}, {Jones}, {Kern}, {Larson}, {Carey}, {Polat}, {Feng}, {Moore}, {Vand
  erPlas}, {Laxalde}, {Perktold}, {Cimrman}, {Henriksen}, {Quintero}, {Harris},
  {Archibald}, {Ribeiro}, {Pedregosa}, {van Mulbregt}, \&
  {Contributors}}]{2020SciPy-NMeth}
{Virtanen}, P., {Gommers}, R., {Oliphant}, T.~E., {et~al.} 2020, Nature
  Methods, 17, 261

\bibitem[{Walt {et~al.}(2011)Walt, Colbert, \& Varoquaux}]{walt2011numpy}
Walt, S. v.~d., Colbert, S.~C., \& Varoquaux, G. 2011, Computing in Science \&
  Engineering, 13, 22

\bibitem[{Wang {et~al.}(2019)Wang, Meyer, Boss, Close, Currie, Dragomir,
  Fortney, Gaidos, Hasegawa, Kitiashvili, {et~al.}}]{wang2019new}
Wang, J., Meyer, M.~R., Boss, A., {et~al.} 2019, White paper; arXiv:1903.07556

\bibitem[{Wenger {et~al.}(2000)Wenger, Ochsenbein, Egret, Dubois, Bonnarel,
  Borde, Genova, Jasniewicz, Lalo{\"e}, Lesteven, {et~al.}}]{wenger2000simbad}
Wenger, M., Ochsenbein, F., Egret, D., {et~al.} 2000, Astronomy \& Astrophysics
  Supplement Series, 143, 9

\end{thebibliography}

\end{document}